\documentclass[a4paper,fleqn]{cas-dc}

\usepackage[justification=centering]{caption}
\usepackage[numbers]{natbib}
\usepackage{amsmath,amsfonts}
\usepackage{array}
\usepackage{url}
\usepackage{verbatim}
\usepackage{graphicx}
\usepackage{booktabs}
\usepackage{multirow}
\usepackage{makecell}
\usepackage{tabularx}
\usepackage[dvipsnames]{xcolor}
\usepackage{tikz}
\usetikzlibrary{trees,positioning,shapes,shadows,arrows,calc}
\graphicspath{{images/}}
\usepackage{longtable}
\usepackage{multicol}

\newcolumntype{Y}{>{\hsize=.1\hsize \centering\arraybackslash}X}
\usepackage{parskip}
\usepackage[pagewise, mathlines]{lineno}

\usepackage{caption}
\usepackage{subcaption}

\definecolor{se}{RGB}{78,113,190}
\definecolor{acc}{RGB}{222,131,68}
\definecolor{dsc}{RGB}{165,165,165}
\definecolor{sp}{RGB}{245,194,66}
\definecolor{auroc}{RGB}{106,154,208}
\definecolor{fpr}{RGB}{126,171,85}
\definecolor{tld}{RGB}{45,67,116}
\definecolor{pre}{RGB}{148,77,32}
\definecolor{visual}{RGB}{99,99,99}
\definecolor{bd}{RGB}{148,117,36}
\definecolor{gc}{RGB}{52,93,141}
\definecolor{others}{RGB}{75,103,51}

\def\tsc#1{\csdef{#1}{\textsc{\lowercase{#1}}\xspace}}
\tsc{WGM}
\tsc{QE}

\begin{document}
\let\WriteBookmarks\relax
\def\floatpagepagefraction{1}
\def\textpagefraction{.001}

\shorttitle{Human Treelike Tubular Structure Segmentation}    

\shortauthors{H. Li, Z. Tang, Y. Nan, and G. Yang}  

\title [mode = title]{Human Treelike Tubular Structure Segmentation: A Comprehensive Review and Future Perspectives}  

\tnotemark[1] 

\tnotetext[1]{This study was supported in part by the BHF (TG/18/5/34111, PG/16/78/32402), the ERC IMI (101005122), the H2020 (952172), the MRC (MC/PC/21013), the Royal Society (IEC/NSFC/211235), the SABER project supported by Boehringer Ingelheim Ltd, the Imperial College Undergraduate Research Opportunities Programme (UROP), the NVIDIA Academic Hardware Grant Program, and the UKRI Future Leaders Fellowship (MR/V023799/1).} 

%

\author[1,2]{Hao Li}[orcid=0000-0003-2476-0664]

\fnmark[1]

\credit{Methodology, Formal analysis, Investigation, Writing - Original Draft, Visualization}

\address[1]{National Heart and Lung Institute, Faculty of Medicine, Imperial College London,
            London,
            United Kingdom}

\address[2]{Department of Bioengineering, Faculty of Engineering, Imperial College London,
            London,
            United Kingdom}

\author[1,2]{Zeyu Tang}[orcid=0000-0003-3789-2906]

\fnmark[1]

\credit{Methodology, Formal analysis, Investigation, Writing - Original Draft, Visualization}

\author[1]{Yang Nan}[orcid=0000-0002-4542-3336]

\fnmark[2]

\credit{Conceptualization, Methodology, Writing - Review and Editing, Supervision}

\author[1,3]{Guang Yang}[orcid=0000-0001-7344-7733]

\fnmark[2]

\cormark[1]

\ead{g.yang@imperial.ac.uk}

\credit{Conceptualization, Methodology, Writing - Review and Editing, Supervision, Funding acquisition}

\address[3]{Royal Brompton Hospital,
            London,
            United Kingdom}

\cortext[1]{Corresponding author}

\fntext[1]{H. Li and Z. Tang contributed equally to this review}
\fntext[2]{Y. Nan and G. Yang are the co-last senior authors}


\begin{abstract}
Various structures in human physiology follow a treelike morphology, which often expresses complexity at very fine scales. Examples of such structures are intrathoracic airways, retinal blood vessels, and hepatic blood vessels. Large collections of 2D and 3D images have been made available by medical imaging modalities such as magnetic resonance imaging (MRI),  computed tomography (CT), Optical coherence tomography (OCT) and ultrasound in which the spatial arrangement can be observed. Segmentation of these structures in medical imaging is of great importance since the analysis of the structure provides insights into disease diagnosis, treatment planning, and prognosis. Manually labelling extensive data by radiologists is often time-consuming and error-prone. As a result, automated or semi-automated computational models have become a popular research field of medical imaging in the past two decades, and many have been developed to date. In this survey, we aim to provide a comprehensive review of currently publicly available datasets, segmentation algorithms, and evaluation metrics. In addition, current challenges and future research directions are discussed.
\end{abstract}




\maketitle










\setlength{\fboxsep}{1pt}

\section{Introduction}
\begin{figure}[!ht]
    \centering
    \includegraphics[width=0.5\textwidth]{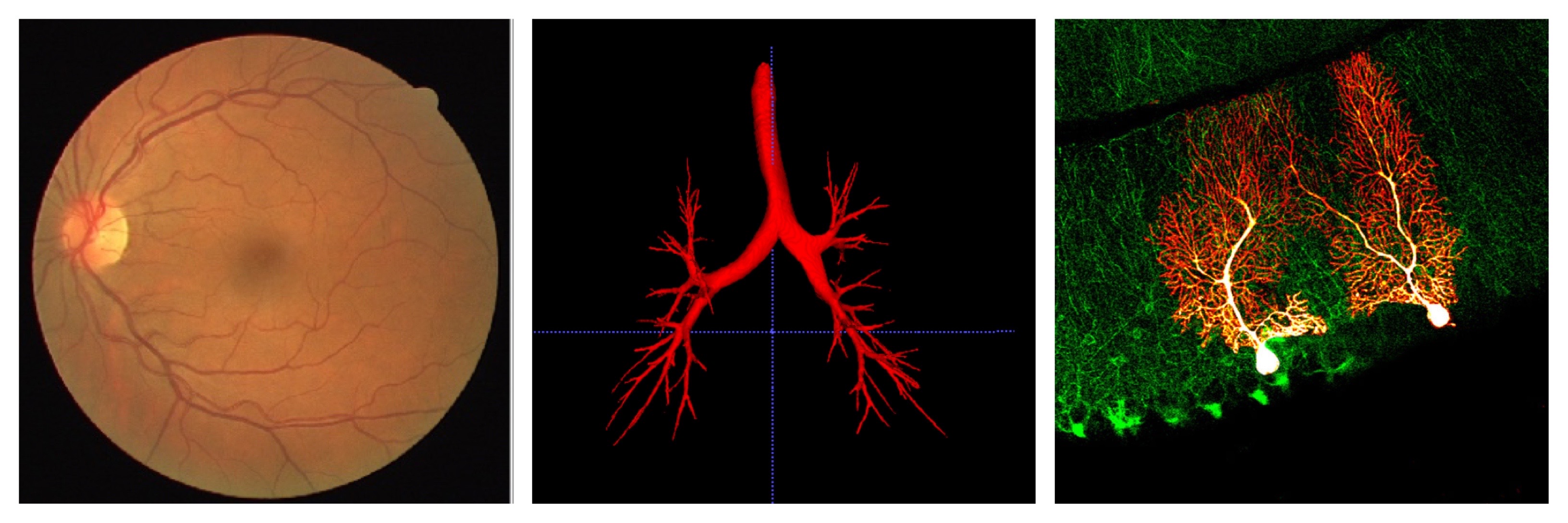}
    \caption{Examples of treelike structures in human body \cite{Staal2004,loExtractionAirwaysCT2012,neural}}
    \label{fig:trees}
\end{figure}

Anatomical structures with tree-shaped topologies are very commonly found in the human body. For example, in our transport system, blood vessels branch into networks to provide material exchange between tissues and cells while airway branches allow the exchange of gases in and out of the body. In our nervous system, tens of billions of neurons form hundreds of trillions of connections with one another (Fig \ref{fig:trees}). The analysis of such topology structure can reveal pathological conditions of the corresponding organs.

Human airways or respiratory tract start from the nasal cavity where the air is breathed in. Inhaled air then travels down through the pharynx and larynx from where it enters the trachea. The trachea is a cylindrical tube about 2.5 centimetres in diameter and 12.5 centimetres in length. It bifurcates to form two main bronchi (left and right), which then continue to branch up to 17 generations beyond which alveoli begin to develop and eventually terminate at alveolar sacs. Structural alterations of the airways often correlate with a range of chronic lung diseases, including asthma, chronic obstructive pulmonary disease (COPD), cystic fibrosis (CF), and coronavirus disease (COVID-19). Computerized tomography (CT) for the human body has been available in clinics for about 50 years. CT was not particularly suited for thorax imaging until the end of the 20th century when the resolution and speed of the scanner had been significantly improved. However, one CT examination contains a large number of CT images, usually up to hundreds \cite{Rubin2000}. Manually annotating these data is time-consuming and error-prone. Therefore, developing automated or semi-automated systems is necessary and this has led to swift progress in the research area of medical imaging.

Blood vessels are also tubular structures commonly found in our body anatomy, and are of great clinical importance as a diagnostic and prognostic factor. For example, the morphological shape of retinal blood vessels, which is imaged using fundus photography, can help diagnose hypertension, diabetes, and atherosclerosis \cite{diabeticRetinopathy, Yau2012}. Many algorithms have been proposed for the segmentation of the blood vessels as well. 
We aim to provide a comprehensive review of the algorithm for human treelike structure segmentation, particularly for airways in this article. Pu et al. \cite{Pu2012Review} has done a review on airway segmentation 10 years ago; Therefore, we will mainly focus on newly developed methods afterwards. In addition, we found the segmentation algorithms for human blood vessels are very similar and will include a few of them here for comparison and discussion. A detailed systematic review on blood vessel segmentation has already been done by \cite{mocciaBloodVesselSegmentation2018,Chen2021,Ciecholewski2021,Abdulsahib2021Review,Mookiah2021Review} 
However, none of them focused on the big picture of summarising algorithms for segmenting all treelike structures in human anatomy. 
In this review, we would like to first briefly introduce different modalities used for imaging treelike structures. Then we move on to the segmentation, summarising representative conventional and ML/DL algorithms for delineating treelike anatomical regions in the past two decades. A list of open datasets and model implementations is compiled for future research work. Finally, we discuss the trend of development and current limitations.
\subsection{Searching criteria}
The literature search was conducted mainly by Z.T. and later reviewed by Y.N. and H.L. The inclusion of an article was based on agreement by at least two of the three authors. We used the following keywords for a systematic searching on Scopus: "airway" OR "vessel" AND "segmentation" AND "deep learning" OR "network" and 2603 results were found; We further limited the results to journal articles written in English that are already published and were left with 1190 papers for the title and abstract screening. 897 papers were excluded as they are either duplicated or have completely irrelevant topics with the keywords only being mentioned few times.  During the writing of this review, a small fraction of papers that have relevant contents but were missed by our search on Scopus or excluded by an error during screening were added back, which in total gave us 72 of them for this study. 
\begin{figure}
    \centering
    \includegraphics[width=0.4\textwidth]{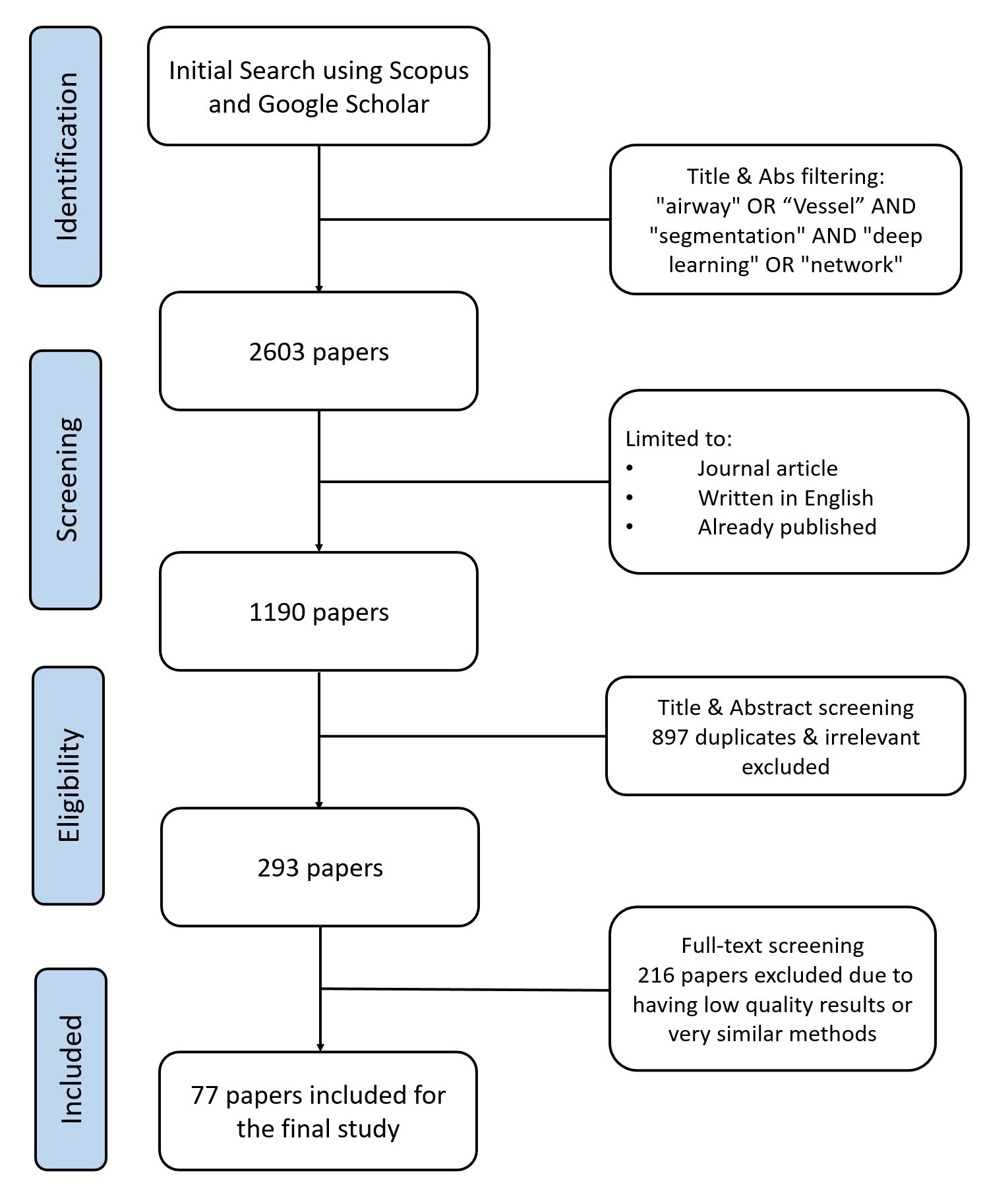}
    \caption{PRISMA Flow Chart}
    \label{fig:prisma}
\end{figure}

\section{Imaging treelike structures in human}
\subsection{CT}
\begin{figure}
    \centering
    \includegraphics[width=0.5\textwidth]{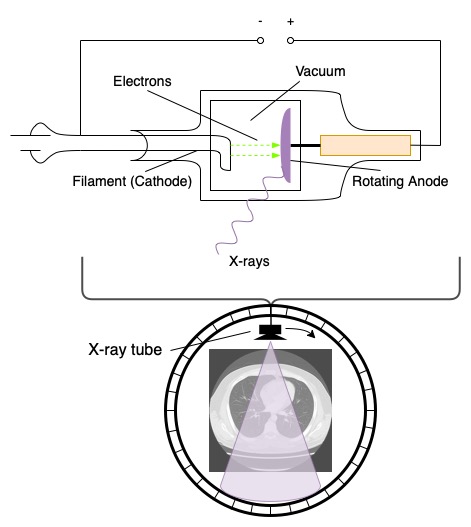}
    \caption{\textcolor{Black}{The X-ray beam is generated inside the X-ray tube which rotates around the patient lying in the centre of a CT scanner. The beam is directed to traverse the patients from various angles and is recorded by electronic detectors. Complex reconstruction algorithms help generate the image of the body's internal structure.}}
    \label{fig:ct}
\end{figure}

X-rays used in medical imaging are usually called soft X-rays because of their low energy level. X-rays are generated inside the X-ray tube as shown in Fig. \ref{fig:ct} which is a vacuum tube that utilizes high voltage to accelerate electrons released from a hot cathode. One type of X-ray, known as the characteristic X-ray, is generated when an electron in the metal atom is kicked out by the high-energy incoming electrons leaving a hole later filled by electrons from a higher energy state. The other type of X-ray, known as the Bremsstrahlung X-ray, is generated when the high-energy incoming electron passes through the atom and emits an X-ray when it loses kinetic energy. \textcolor{Black}{The generated X-ray is then directed to traverse the patients from various angles and is recorded by electronic detectors. Beer-Lambert law governs X-ray attenuation, which is influenced by a variety of factors such as beam energy, tissue type, sample thickness, and others. The higher the attenuation of the beam, the lighter the tissue appears on the reconstructed CT images; The lower the attenuation, the darker the tissue appears on the reconstructed CT images.}
\begin{figure}
    \centering
    \includegraphics[width=0.5\textwidth]{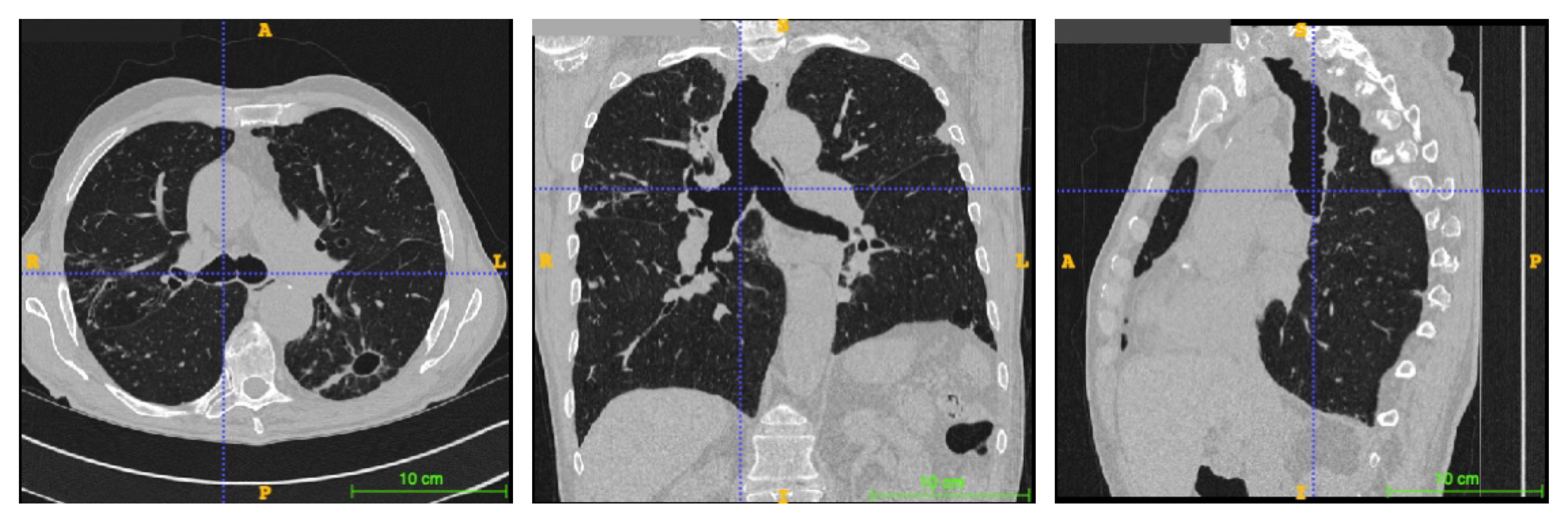}
    \caption{\textcolor{Black}{Lung CT scans from EXACT'09 dataset \cite{loExtractionAirwaysCT2012}; from left to right is axial, coronal and sagittal plane respectively.}}
    \label{fig:ctex}
\end{figure}
\subsection{MRI}
Instead of using ionizing radiation, strong external magnetic field is generated surrounding the patient lying in an MRI machine. The stronger the magnetic field, the faster the protons (nuclei of the hydrogen atom) precess. The precessing frequency lies in the range of radio frequency (RF) electromagnetic wave. The MRI machine sends RF pulses that have the same frequency of the processing and excite the proton to a higher energy state. When the RF pulse is switched off, the protons gradually relax themselves and release excess energy in the form of RF waves which can be detected. \textcolor{Black}{This signal is referred to as the free-induction decay (FID) response signal which is measured by a conductive field coil placed around the object being imaged.} By evaluating the relaxation rate (longitudinal T1 and transverse T2) of protons in different tissues, we can obtain 3-D images of deep tissues with high resolution. A schematic diagram is shown below \cite{Wallyn2019}
\begin{figure}
    \centering
    \includegraphics[width=0.5\textwidth]{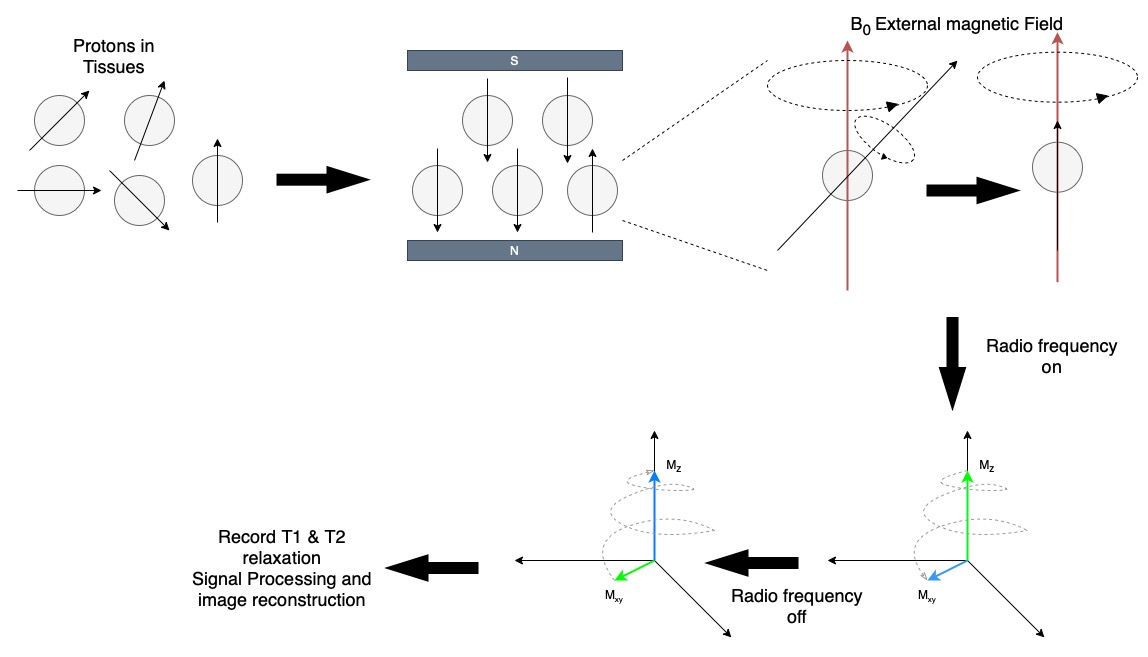}
    \caption{\textcolor{Black}{When strong external magnetic field is applied, the hydrogen nuclei precess along the direction of the field. Then a radio-frequency pulse with frequency equal to the Larmor frequency is applied perpendicular to the field. The RF pulse makes the magnetic moment of the hydrogen nuclei tilt. When the RF pulse stops, the nuclei realign themselves to the external magnetic field.}}
    \label{fig:principle_mri}
\end{figure}
\begin{figure}
    \centering
    \includegraphics[width=0.5\textwidth]{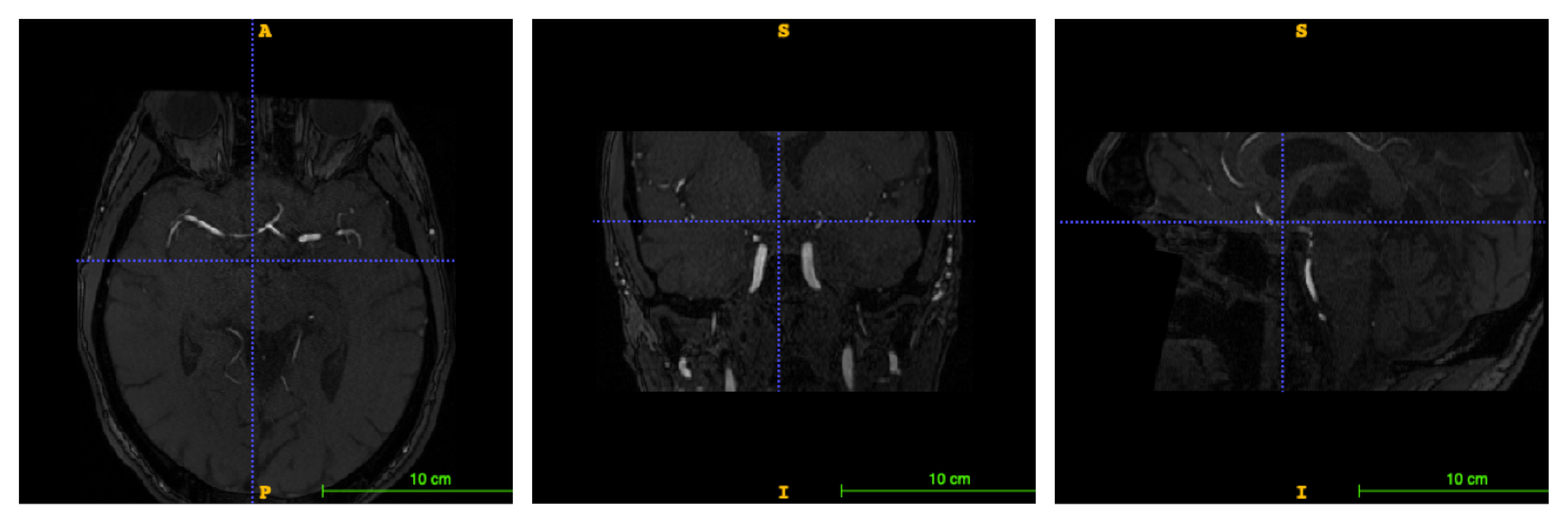}
    \caption{\textcolor{Black}{Time-of-flight(TOF) magnetic resonance angiography(MRA) of cerebral vessels \cite{brain_vessel}; from left to right is the axial, coronal and sagittal plane respectively.}}
    \label{fig:brain_vessel}
\end{figure}
\subsection{Fundus Photography}
\begin{figure}
    \centering
    \includegraphics[width=0.4\textwidth]{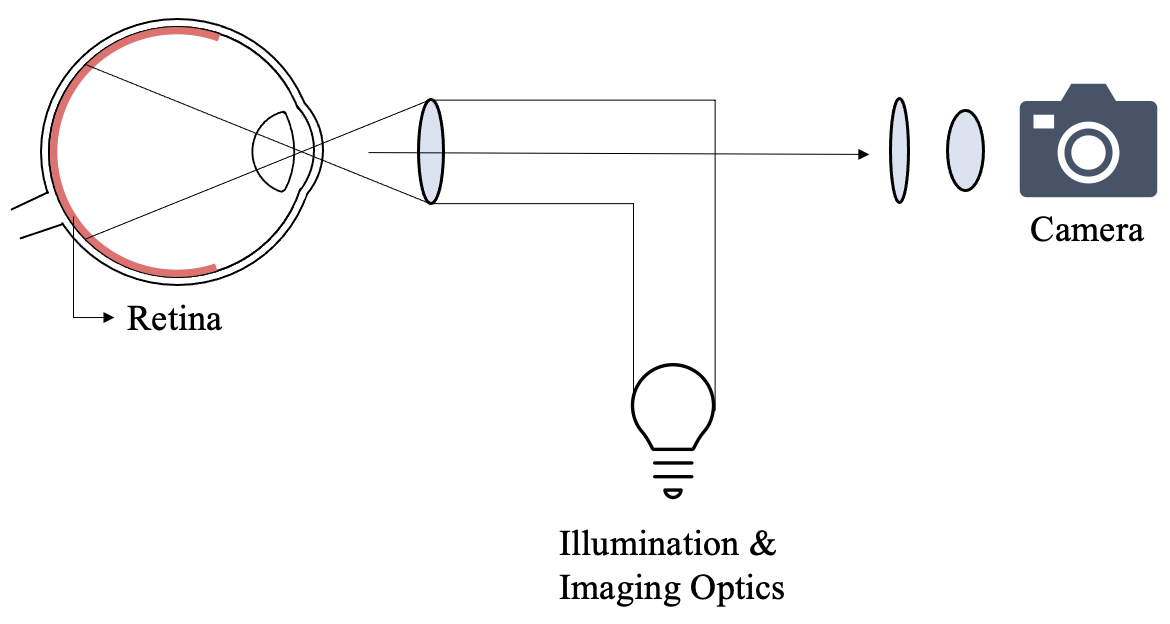}
    \caption{\textcolor{Black}{Light is projected through a set of filters and lenses, resulting in a focused doughnut-shaped form into the eye. Assuming the illumination and the object are aligned and focused, the image of the retinal exits the cornea through the central dark region of the doughnut.}}
    \label{fig:fp}
\end{figure}

Fundus retinal images allow non-invasive capture of the major anatomical structures of the retina, including the deep microvascular system. Depending on the instrument and the associated acquisition protocol, the main fundus imaging techniques are colour fundus photography (CFP) and scanning laser ophthalmoscope (SLO). The CFP utilizes a low-power microscope with a camera attachment to image the retina. The optical principle is similar to the indirect ophthalmoscope to provide a magnified view of the inner surface of the eye (see Figure~\ref{fig:fp}). The SLO uses laser scanning to provide high-contrast images of blood vessels and uses wide-field instruments.

\begin{figure}
    \centering
    \includegraphics[width=0.4\textwidth]{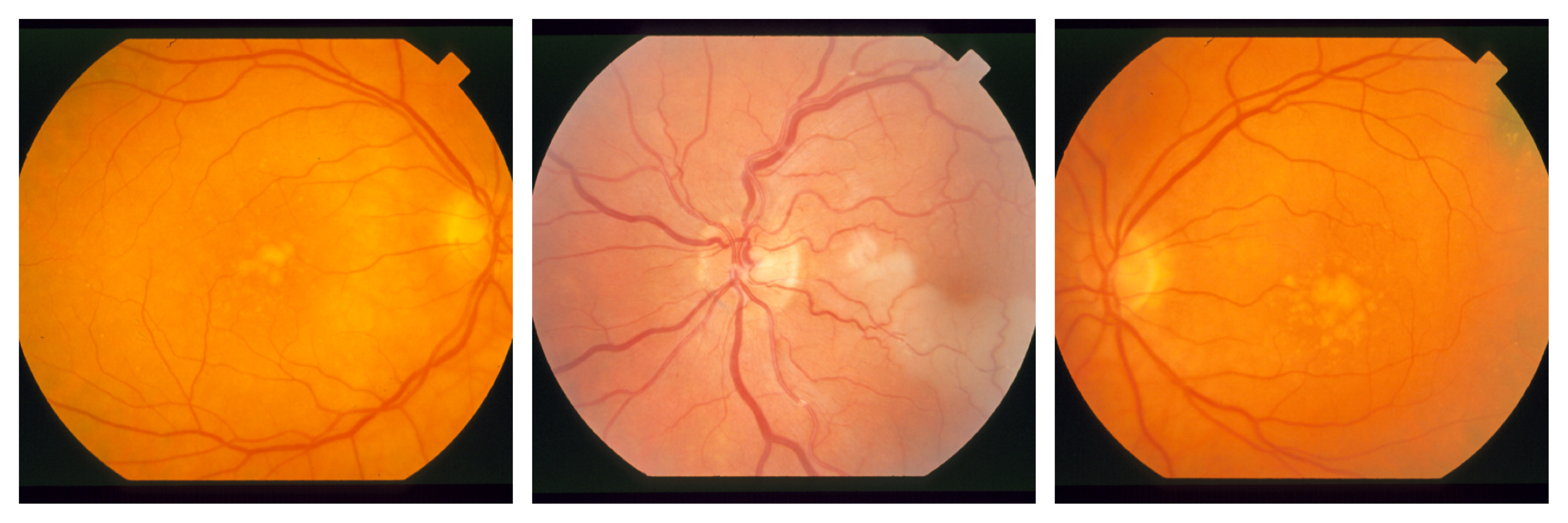}
    \caption{\textcolor{Black}{Examples of fundus images from STARE dataset \cite{STARE} captured by a TopCon TRV-50 fundus camera at 35 field of view. Each slide was digitized to produce a $605\times700$ pixel image, 24 bits per pixel (stan- dard RGB).}}
    \label{fig:fp_examples}
\end{figure}

\subsection{OCT}
OCT creates cross-sectional images by performing a series of low time-coherence interferometry (LCI) depth-scans which measures absolute distances. The low time-coherence light source travels through the beam splitter, one being directed to the sample the other being directed to the reference mirror. \textcolor{Black}{OCT bathes the sample in electromagnetic waves and uses principles of interference to determine from what depth a photon was reflected inside a sample.} Each layer in the sample has sharp change of refractive index which manifest themselves as intensity peaks in the interference pattern. The interference only occurs when the distances to the sample and reference are matched within coherence length of the light. A schematic diagram is shown below created with BioRender \cite{Tomlins2005TheoryDA}. \textcolor{Black}{OCT is mainly used  in ophthalmic and intravascular imaging. It can identify most retinal pathologies such as retinal holes and detachments, and opaques in arteries around the heart. Diseases such as diabetic retinopathy, diabetic edema and glaucoma can also be diagnosed. A more detailed description of OCT angiography can be found here \cite{OCTA}.} 
\begin{figure}
    \centering
    \includegraphics[width=0.5\textwidth]{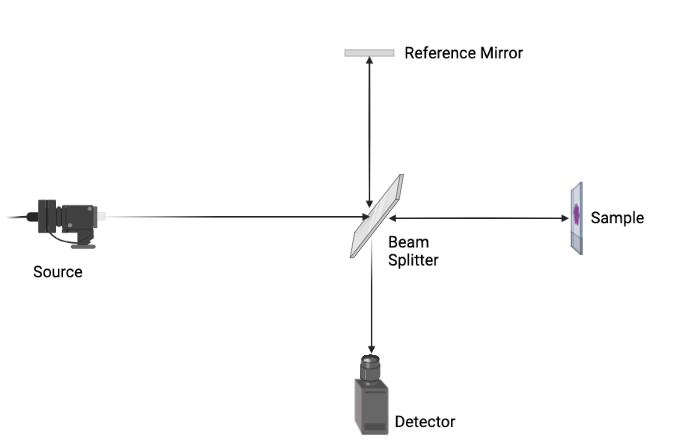}
    \caption{\textcolor{Black}{The low-coherence interferometer splits light into two paths. One path uses a reference mirror to reflect the light at a known path length. The other path leads to the sample to be imaged. The presence or absence of light reflected from a depth corresponding to the path length set by the reference mirror determines the presence or absence of detected interference.}}
    \label{fig:OCT}
\end{figure}

\begin{figure}
    \centering
    \includegraphics[width=0.4\textwidth]{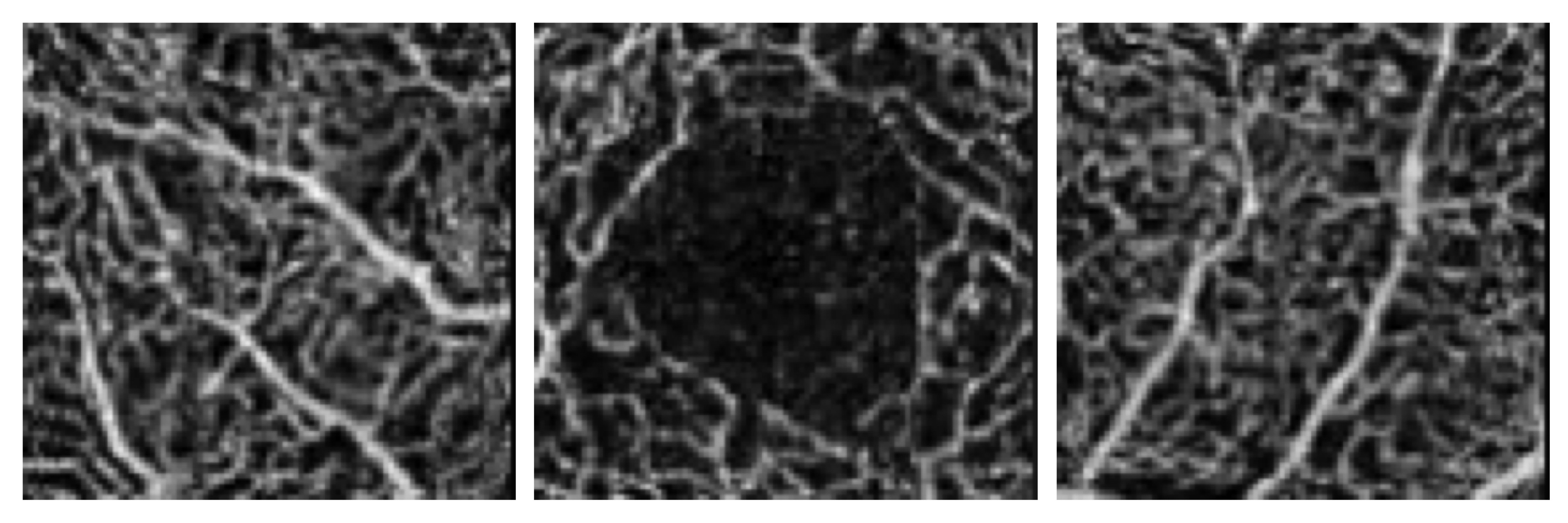}
    \caption{\textcolor{Black}{OCT images from PREVENT \cite{retinal_oct} captured using the commercial RTVue-XR
 Avanti OCT system (OptoVue, Fremont, CA). Consequent
 B-scans, each one consisting of 304×304 A-scans, were
 generated in 3×3 mm field of view centered at
 the fovea.}}
    \label{fig:retinal_OCT}
\end{figure}

\section{Segmentation Algorithms}

In this section, an exhaustive review of segmentation methods for human treelike tubular structures is presented following the taxonomy as shown in Figure \ref{fig:taxonomy}. The segmentation methods and algorithms can be categorized into three classes: conventional, machine learning (ML) based, and deep learning (DL) based. Although DL-based methods are a subset of ML-based methods that specifically utilizes artificial neural networks (ANNs) such as convolutional neural networks (CNNs), this review presents DL-based methods as a separate category for focused discussion. In addition, preprocessing and postprocessing techniques and loss functions adopted by DL-based approaches are also summarized in this section.

\tikzset{
  basic/.style  = {draw, text width=5cm, drop shadow, rectangle},
  root/.style   = {basic, rounded corners=3pt, thin, align=center, fill=white},
  level 2/.style = {basic, rounded corners=5pt, thin, align=center, fill=white, text width=4cm},
  level 3/.style = {basic, thin, align=left, fill=white, text width=5cm}
}

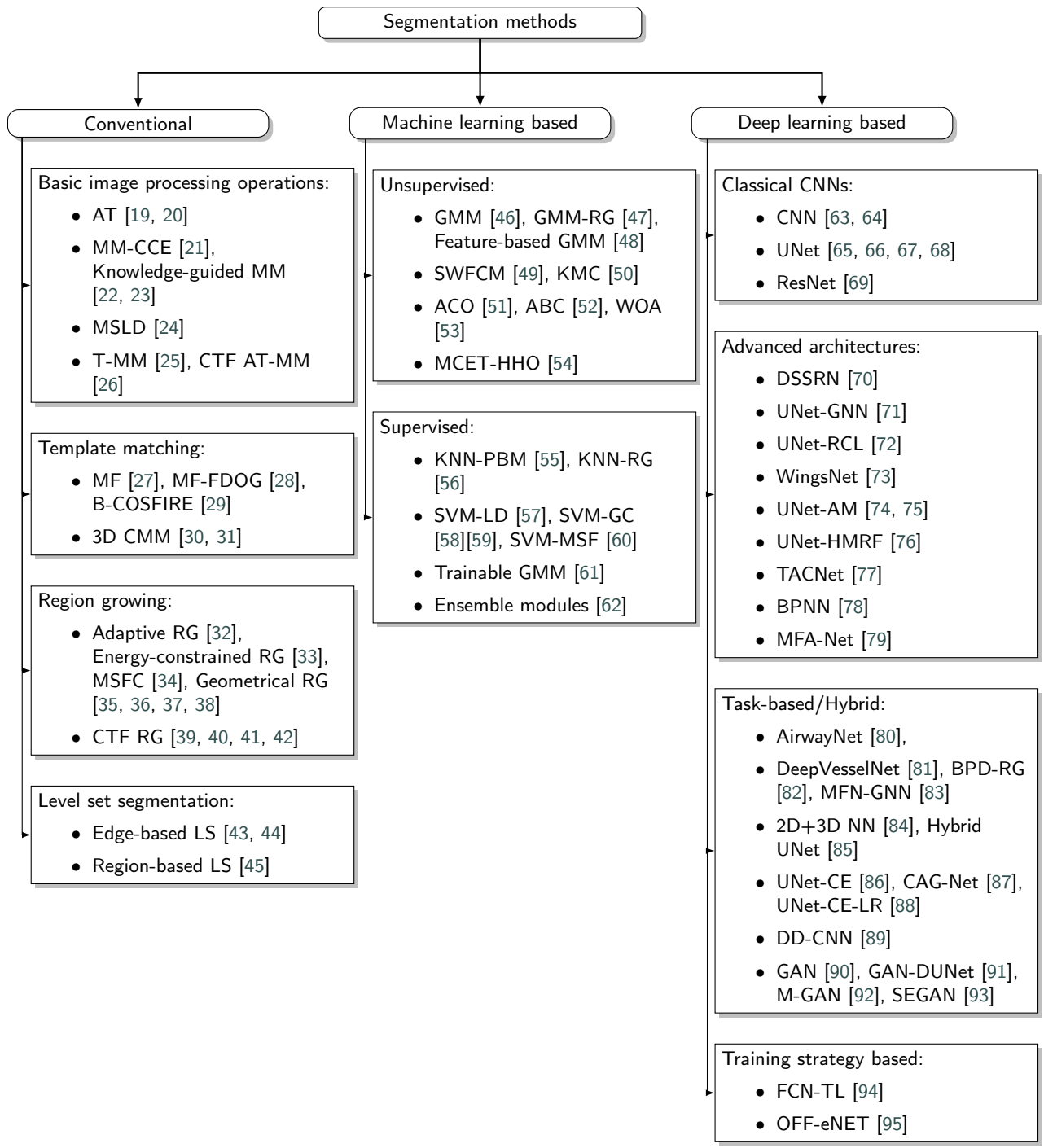
\begin{figure*}
\centering
\begin{tikzpicture}[
  level 1/.style={sibling distance=17em, level distance=5em},
  edge from parent/.style={->,solid,black,thick,draw}, 
  edge from parent path={(\tikzparentnode.south) -- (\tikzchildnode.north)},
  >=latex, node distance=1.5em, edge from parent fork down]

\node[root] {Segmentation methods}
  child {node[level 2] (c1) {Conventional}}
  child {node[level 2] (c2) {Machine learning based}}
  child {node[level 2] (c3) {Deep learning based}};

\begin{scope}[every node/.style={level 3}]
\node [below = of c1, xshift=25pt] (c11) {Basic image processing operations:
\begin{itemize}
    \item AT \cite{xiaoyijiangAdaptiveLocalThresholding2003,jiantaopuDifferentialGeometricApproach2011}
    \item MM-CCE \cite{zanaSegmentationVessellikePatterns2001}, Knowledge-guided MM \cite{passatMagneticResonanceAngiography2006,bouraoui3DSegmentationCoronary2010}
    \item MSLD \cite{nguyenEffectiveRetinalBlood2013}
    \item T-MM \cite{aykacSegmentationAnalysisHuman2003}, CTF AT-MM \cite{camaranetoUnsupervisedCoarsetofineAlgorithm2017}
\end{itemize}
};
\node [below = of c11] (c12) {Template matching:
\begin{itemize}
    \item MF \cite{hooverLocatingBloodVessels2000}, MF-FDOG \cite{zhangRetinalVesselExtraction2010}, B-COSFIRE \cite{azzopardiTrainableCOSFIREFilters2015}
    \item 3D CMM \cite{krissianModelBasedDetectionTubular2000,worzSegmentationQuantificationHuman2007}
\end{itemize}
};
\node [below = of c12] (c13) {Region growing:
\begin{itemize}
    \item Adaptive RG \cite{kiralyThreeDimensionalHumanAirway2002}, Energy-constrained RG \cite{fetitaPulmonaryAirways3D2004}, MSFC \cite{tschirrenIntrathoracicAirwayTrees2005}, Geometrical RG \cite{martinez-perezSegmentationBloodVessels2007,fabijanskaTwopassRegionGrowing2009,cetinVesselTractographyUsing2013,cetinHigherOrderTensorVessel2015}
    \item CTF RG \cite{bartzHybridSegmentationExploration2003,mayerHybridSegmentationVirtual2004,grahamRobust3DAirway2010, Mendonca2006}
\end{itemize}
};
\node [below = of c13] (c14) {Level set segmentation:
\begin{itemize}
    \item Edge-based LS \cite{hutchisonOrientedFluxSymmetry2010,lorigoCURVESCurveEvolution2001}
    \item Region-based LS \cite{klepaczkoSimulationMRAngiography2016}
\end{itemize}
};

\node [below = of c2, xshift=25pt] (c21) {Unsupervised:
\begin{itemize}
    \item GMM \cite{hassounaCerebrovascularSegmentationTOF2006}, GMM-RG \cite{oliveiraSegmentationLiverIts2011}, Feature-based GMM \cite{roychowdhuryBloodVesselSegmentation2014}
    \item SWFCM \cite{kandeUnsupervisedFuzzyBased2010}, KMC \cite{goceriVesselSegmentationAbdominal2017}
    \item ACO \cite{cinsdikiciDetectionBloodVessels2009}, ABC \cite{hassanienRetinalBloodVessel2015}, WOA \cite{hassanRetinalFundusVasculature2018}
    \item MCET-HHO \cite{ramos-sotoEfficientRetinalBlood2021}
\end{itemize}
};
\node [below = of c21] (c22) {Supervised:
\begin{itemize}
    \item KNN-PBM \cite{staalRidgeBasedVesselSegmentation2004}, KNN-RG \cite{loVesselguidedAirwayTree2010}
    \item SVM-LD \cite{ricciRetinalBloodVessel2007}, SVM-GC \cite{mengAutomaticSegmentationAirway2017}\cite{Zhai16lungvessel}, SVM-MSF \cite{leeHybridAirwaySegmentation2019}
    \item Trainable GMM \cite{soaresRetinalVesselSegmentation2006}
    \item Ensemble modules \cite{frazEnsembleClassificationBasedApproach2012}
\end{itemize}
};

\node [below = of c3, xshift=25pt] (c31) {Classical CNNs:
\begin{itemize}
    \item CNN \cite{charbonnierImprovingAirwaySegmentation2017,yunImprovementFullyAutomated2018}
    \item UNet \cite{meyerDeepNeuralNetwork2017, garcia-ucedajuarezAutomaticAirwaySegmentation2018, jin3DConvolutionalNeural2017, garcia-ucedaAutomaticAirwaySegmentation2021}
    \item ResNet \cite{kitrungrotsakulVesselNetDeepConvolutional2019}
\end{itemize}
};
\node [below = of c31] (c32) {Advanced architectures:
\begin{itemize}
    \item DSSRN \cite{linAutomaticRetinalVessel2019}
    \item UNet-GNN \cite{garcia-ucedajuarezJoint3DUNetGraph2019}
    \item UNet-RCL \cite{Wang2019Radial}
    \item WingsNet \cite{Zheng2020WingsNet}
    \item UNet-AM \cite{qinLearningBronchioleSensitiveAirway2020,qinLearningTubuleSensitiveCNNs2021}
    \item UNet-HMRF \cite{Fan2020}
    \item TACNet \cite{chengSegmentationAirwayTree2021}
    \item BPNN \cite{tangConstructionVerificationRetinal2021}
    \item MFA-Net \cite{zhouAutomaticAirwayTree2021}
\end{itemize}
};
\node [below = of c32] (c33) {Task-based/Hybrid:
\begin{itemize}
    \item AirwayNet  \cite{qinAirwayNetVoxelConnectivityAware2019}, \item DeepVesselNet \cite{Tetteh2020}, BPD-RG \cite{wangAutomatedLabelingAirway2020}, MFN-GNN \cite{selvanGraphRefinementBased2020}
    \item 2D+3D NN \cite{zhaoBronchusSegmentationClassification2019}, Hybrid UNet \cite{yangHybridDeepSegmentation2021}
    \item UNet-CE \cite{mengTrackingSegmentationAirways2017}, CAG-Net \cite{liCascadedAttentionGuided2020}, UNet-CE-LR \cite{nadeemCTBasedAutomatedAlgorithm2021}
    \item DD-CNN \cite{ZHANG2020162}
    \item GAN \cite{zhaoHighQualityRetinal2020}, GAN-DUNet \cite{guoRetinalVesselSegmentation2020}, M-GAN \cite{parkMGANRetinalBlood2020}, SEGAN \cite{zhouRefinedEquilibriumGenerative2021}
\end{itemize}
};
\node [below = of c33] (c34) {Training strategy based:
\begin{itemize}
    \item FCN-TL \cite{jiangRetinalBloodVessel2018}
    \item OFF-eNET \cite{nazirOFFeNETOptimallyFused2020}
\end{itemize}
};
\end{scope}

\foreach \value in {1,...,4}
  \draw[->] (c1.187) |- (c1\value.west);

\foreach \value in {1,...,2}
  \draw[->] (c2.188) |- (c2\value.west);

\foreach \value in {1,...,4}
  \draw[->] (c3.188) |- (c3\value.west);
  
\end{tikzpicture}
    \caption{Taxonomy of segmentation methods}
    \label{fig:taxonomy}
\end{figure*}

\subsection{Conventional Segmentation}

\subsubsection{Basic image processing operations}
For most biomedical images, different substances have different pixel intensity values (e.g., 0 for water and -1000 for air in CT scans). Therefore, the tubular structure can be roughly segmented by certain basic image processing operations, for instance, the intensity thresholding (to provide candidate regions) \cite{xiaoyijiangAdaptiveLocalThresholding2003,jiantaopuDifferentialGeometricApproach2011}, morphology operations (to remove noise or leakages) \cite{zanaSegmentationVessellikePatterns2001,passatMagneticResonanceAngiography2006,bouraoui3DSegmentationCoronary2010}, filtering (to remove noises or enhance tubular structures) \cite{nguyenEffectiveRetinalBlood2013}, or a combination of these operations \cite{aykacSegmentationAnalysisHuman2003,camaranetoUnsupervisedCoarsetofineAlgorithm2017}.

\begin{figure}[!ht]
    \centering
    \includegraphics[width=0.4\textwidth]{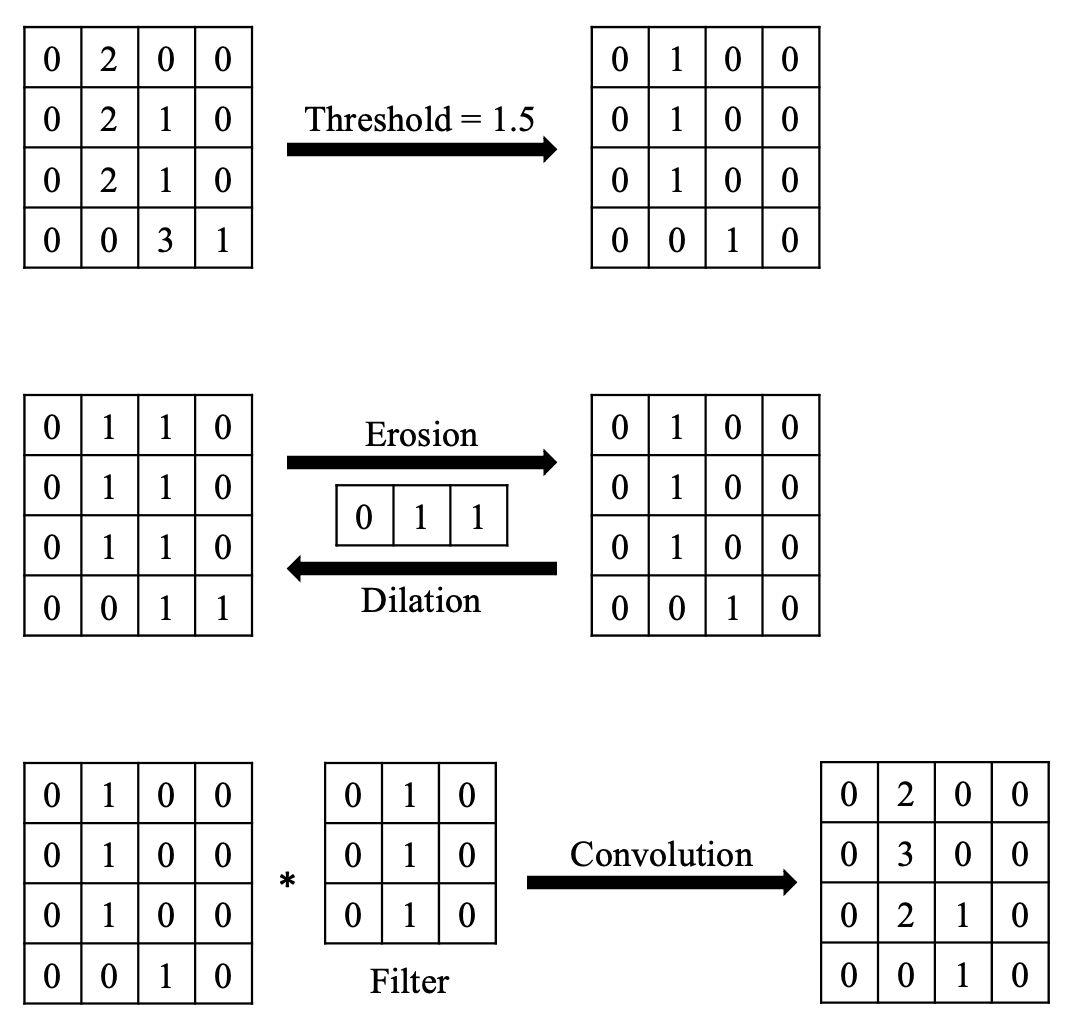}
    \caption{Basic image processing operations: thresholding, morphology operations, and filtering.}
    \label{fig:bip}
\end{figure}

\textbf{{Adaptive thresholding (AT):}} Adaptive thresholding refers to methods that use a dynamic threshold strategy for threshold-based segmentation. For instance, Jiang et al. \cite{xiaoyijiangAdaptiveLocalThresholding2003} proposed an adaptive local thresholding method based on multi-threshold probing. The airway prediction was obtained by combining the probed thresholding and non-maximum suppression post-processing. Pu et al. \cite{jiantaopuDifferentialGeometricApproach2011} relaxed the airway segmentation into the geometrical domain by analysing the principal curvature and lung anatomical structures. A progressively adaptive threshold-based modelling strategy was adopted to address the wide intensity range of airway walls. Though the dynamic threshold promotes robustness to some extent, it still faces the challenge of false-positive predictions, especially for those abnormal scans.

\textbf{{Mathematical morphology and cross-curvature evaluation (MM-CCE):}} Zana \& Klein \cite{zanaSegmentationVessellikePatterns2001} segmented vessel-like patterns using a hybrid framework of morphological filtering and cross-curvature analysis. The vessels in the monochrome retinal image can be identified by mathematical morphology methods because they are natively straight, and connected, and their curvature changes gradually along the crest line. Cross-curvature analysis is applied to evaluate which features have linearly consistent curvature.

\textbf{{Knowledge-guided mathematical morphology (MM):}} Passat et al. \cite{passatMagneticResonanceAngiography2006} proposed a preliminary approach to enhance brain vessel segmentation in Magnetic Resonance Angiography (MRA) images by incorporating high-level anatomical knowledge into the segmentation process. The first stage integrates anatomical knowledge of vessels such as density, size, and orientation into a cerebral vascular atlas extracted by a skeletonization-based algorithm, utilizing a topology-preserving non-rigid registration method. This atlas then guides MM using adaptive sets of grey-level hit-or-miss operators in the second stage. The parameters of these morphology operators are adapted to the vascular structures using anatomical knowledge modelled by the cerebral vascular atlas. A similar strategy was employed in \cite{bouraoui3DSegmentationCoronary2010} to segment coronary arteries from 3D X-ray data sequences. However, it combines a region-growing algorithm in addition to the grey-level hit-or-miss transform.

\textbf{{Multi-scale line detection (MSLD):}} Nguyen et al. \cite{nguyenEffectiveRetinalBlood2013} suggested a multi-scale line detection (MSLD) based approach. Their approach enhances pixels representing the lines of varying directions by filtering the raw image using a rotating predefined kernel. Multiple line detectors with various lengths are applied to obtain multi-scale measurement values. The final segmentation of the retinal image is gained by linearly combining multi-scale line detectors, which retain the information at varying scales and eliminates the inadequacies of single filtering. The main advantage of the approach is the fast segmentation speed due to its simplicity and scalability.

\textbf{{Thresholding and mathematical morphology (T-MM):}} Aykac et al. \cite{aykacSegmentationAnalysisHuman2003} identified the candidate airways on 2D slices by intensity thresholding and morphology. After noise reduction, the grey reconstruction (via morphological closing operation) was conducted to extract airways by identifying the grey-scale minima. The intensity thresholding then removed most of the non-airway candidates to obtain the final airway structure.

\textbf{{Coarse-to-fine AT and MM (CTF AT-MM):}} Câmara Neto et al. \cite{camaranetoUnsupervisedCoarsetofineAlgorithm2017} presented an unsupervised statistical coarse-to-fine method for detecting blood vessels in fundus pictures. After background harmonization and noise reduction, an adaptive local thresholding approach is employed to coarsely estimate the vessel tree. The threshold is computed based on the spatial dependence between pixels using the cumulative distribution function and the normalized grey-level co-occurrence matrix to evaluate the regional influence, especially in the centre and border. The coarse segmentation is refined by a morphological reconstruction using a binary curvature map as a marker to eliminate pixel mislabelling. The suggested method effectively solved major image distortion by minimizing mislabelling of central vascular reflex areas and false-positive identification of abnormal patterns.

\subsubsection{Template matching}
\textcolor{Black}{A basic template matching method involves creating an image patch (template) that is suited to a certain aspect of the search image that we want to detect. Given the intensities of pixels in the search image \textit{$I_s$} and template image \textit{$I_t$}, the template matching is normally defined by calculating the sum of absolute differences (SAD) between \textit{$I_s$} and \textit{$I_t$}. 
\begin{linenomath}
\begin{equation}
\mathcal{SAD}(x,y)= \sum_{i=0}^{T_{rows}}\sum_{j=0}^{T_{cols}}|I_s({x+i}, {y+j})-I_t(i,j)|,
\end{equation}
\end{linenomath}
where $T_{rows}$ and $T_{cols}$ denote the rows and the columns of the template image, respectively.} The segmentation performance of tubular structures heavily relies on the continual detection of identified candidates, while some pseudo-tracheal regions may occur in noisy cases, leading to leakage when dealing with small branches. To alleviate these issues, researchers designed novel filters or preset templates for image convolution to acquire the response image. Then the tubular structures can be located by analysing textual or shape features through template matching methods. Among different template matching approaches, one way is to use a preset circular mask to match the section of tubular structures \cite{hooverLocatingBloodVessels2000,zhangRetinalVesselExtraction2010,azzopardiTrainableCOSFIREFilters2015}, and another solution is to match a 3D model with multiple constraints \cite{krissianModelBasedDetectionTubular2000,worzSegmentationQuantificationHuman2007}.

\begin{figure}[!ht]
    \centering
    \includegraphics[width=0.4\textwidth]{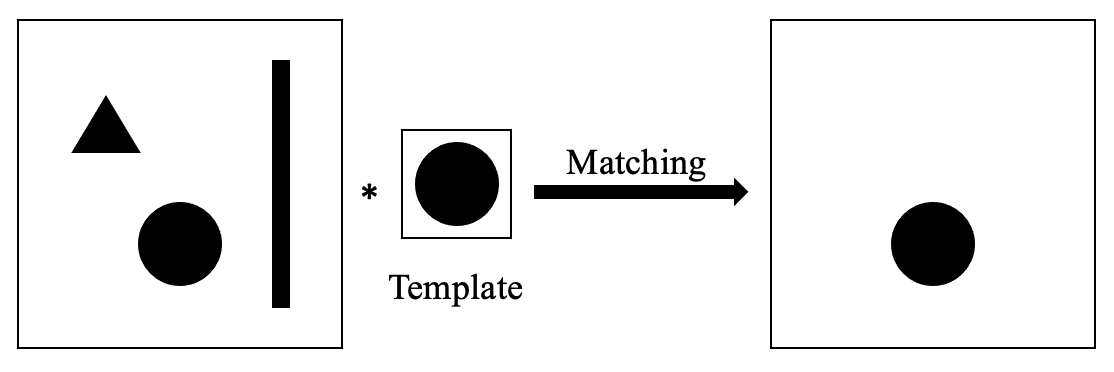}
    \caption{Template matching.}
    \label{figtm}
\end{figure}

\textbf{{Matched filtering (MF):}} MF is a classic template matching method that employs predefined convolution kernels to convolute images to acquire intense responses or features of tubular structures. Based on the assumption that the cross-section of tubular structures can be regarded as a Gaussian function, one common solution is to match different Gaussian-shaped filters for segmentation. The binary segmentation can then be obtained by thresholding the response image.  For instance, Hoover et al. \cite{hooverLocatingBloodVessels2000} proposed an adaptive thresholding method by probing an area of the matched filter response image, decreasing the threshold iteratively. However, MF approaches will obtain intense responses to both tubular structures and edges, leading to false-positive predictions.

\textbf{Matched filtering with the first-order derivative of Gaussian (MF-FDOG):} To alleviate the redundant response of MF approaches,  Zhang et al. \cite{zhangRetinalVesselExtraction2010} extended conventional MF approaches by integrating the first-order derivative of the Gaussian (FDOG). By checking the response intensity to the MF and FDOG, the non-vessel structures can be distinguished. This approach considerably reduced the false positive predictions of the original MF and discovered more small vessels.

\textbf{B-COSFIRE:} Based on the combination of shifted filter responses, Azzopardi et al. \cite{azzopardiTrainableCOSFIREFilters2015} suggested a non-linear filter named B-COSFIRE for detecting elongated patterns in images. The weighted geometric mean (of a group of different Gaussian filters) that provides orientation selectivity is calculated with the B-COSFIRE filter. Meanwhile, the rotation invariance is achieved by applying simple-shifting operations. However, these non-linear filters need to be manually adjusted to achieve optimal results on different datasets.

\textbf{{The 3D cylindrical model matching (3D CMM):}} Krissian et al. \cite{krissianModelBasedDetectionTubular2000} presented an adaptive approach for detecting three-dimensional (3D) tubular structures based on the eigenvalues and eigenvectors of the Hessian matrix. The method begins with an analytical examination of those features across a variety of models, demonstrating that a response function based on both eigenvectors and gradient is the most robust. The method then employs a cylindrical vessel model with a circular Gaussian cross-section to guide the tubular structure detection on the feature map. Because the intensity profile of the cross-section affects size estimation based on multi-scale analysis, a more accurate model with a bar-like convolved cross-section is proposed, along with an improved radius estimation algorithm based on this new model. Worz \& Rohr \cite{worzSegmentationQuantificationHuman2007} developed a new 3D cylindrical parametric intensity model for vessel segmentation which is directly fitted to image intensities using an incremental (segment-wise) Kalman filter. This mathematical model represents a cylindrical structure of a variable radius and directly depicts the image intensities of vessels and neighbouring tissues and a parameter that characterizes image blurring. Compared to previously presented Gaussian-shaped models, the new model more accurately describes a smoothed Gaussian cylinder, a more realistic representation that produces improved outcomes.

\subsubsection{Region growing (RG)}
The region growing or tracking methods segment the tubular structures by setting specific region growing rules \cite{kiralyThreeDimensionalHumanAirway2002,fetitaPulmonaryAirways3D2004,tschirrenIntrathoracicAirwayTrees2005,martinez-perezSegmentationBloodVessels2007,fabijanskaTwopassRegionGrowing2009,cetinVesselTractographyUsing2013,cetinHigherOrderTensorVessel2015} (e.g., most approaches predefine a 3D cylinder as a template to match the tubular structures), which can also be adaptive. Starting with an initial seed (given manually or automatically), the orientation and position of the tubular structure can be calculated by image-derived features such as intensities and gradients. However, due to the inhomogeneous intensity and noise in images, these approaches are prone to segmentation leakages and cannot segment small tubular structures effectively. Therefore, previous methods usually use region growing/tracking algorithms to obtain the main bronchi, then combine them with other image processing techniques, such as morphology-based methods, to segment the small trachea for further fine segmentation \cite{bartzHybridSegmentationExploration2003,mayerHybridSegmentationVirtual2004,grahamRobust3DAirway2010,Mendonca2006}. 

\begin{figure}[!ht]
    \centering
    \includegraphics[width=0.4\textwidth]{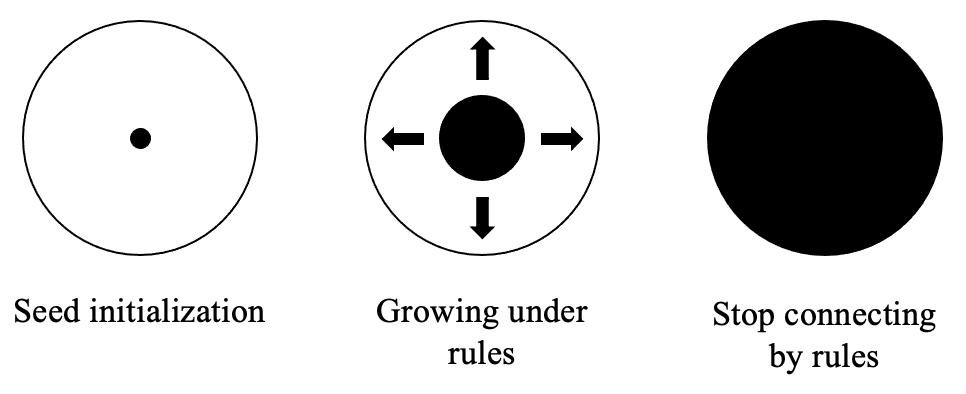}
    \caption{Region growing.}
    \label{figrg}
\end{figure}

\textbf{{Adaptive RG:}} To prevent airway leakages, Kiraly et al. \cite{kiralyThreeDimensionalHumanAirway2002} carried out a preset explosion parameter to detect leakages. Specifically, the threshold of the region growing algorithm is dynamically changed by assessing the current growth conditions (decreasing threshold while explosion occurs and vice versa). In addition, they also combined adaptive RG with morphology-based segmentation to accurately segment airway trees.

\textbf{{Energy-constrained RG:}} Fetita et al. \cite{fetitaPulmonaryAirways3D2004} proposed an energy-constrained region-growing algorithm for 3D airway segmentation. The RG is controlled by the local energy minimization exploiting the bronchial structure in terms of intensity and 3D topology. To acquire more accurate predictions of small bronchus, they also developed a mathematical morphology operator based on selective marking and depth-constrained connection costs. The experimental results indicated that their approach was robust to a large spectrum of airway pathologies, including severe stenosis.

\textbf{{Multi-seeded fuzzy connectivity (MSFC):}} Tschirren et al. \cite{tschirrenIntrathoracicAirwayTrees2005} employed a 3D cylindrical ROI (region of interest) with multi-seeded fuzzy connectivity (MSFC) to overcome image noise and gradients. The ROI for segmenting the airway adapts to its geometrical dimensions, orientation, size, and position. The MSFC grows two regions (foreground and background) simultaneously in a competitive manner, based on the similarity between input voxels and the seed voxel as fuzzy membership values. Although the prediction from MSFC is robust with fewer airway leakages, it has high computational costs, especially for large 3D volumes.

\textbf{{Geometrical RG:}} Different from the conventional region-growing approaches that perform growth through pixel/voxel intensity, the geometrical region growing utilizes geometrical features as a constraint for region growing. These geometrical features are usually used to approximate the width, size, and orientation of tubular structures. For instance, Martinez-Perez et al. \cite{martinez-perezSegmentationBloodVessels2007} calculated two geometrical features (gradient magnitude and maximum principal curvature of the Hessian matrix) based on the first and second derivatives of the intensity value. The growth is then confined to low gradient magnitude regions in the first stage, along with spatial information about the 8-neighbouring pixels. Fabijańska \cite{fabijanskaTwopassRegionGrowing2009} proposed a two-pass region growing algorithm to segment 3D airway structures, one (classic region growing) for rough airway extraction and another (based on morphological gradient) for refining segmentation. Besides, Cetin et al. \cite{cetinVesselTractographyUsing2013} utilized a second-order tensor derived from directional intensity measurements for extracting coronaries from computed tomography angiography (CTA) volumes. The anisotropic second-order tensor fit inside a vessel drives the segmentation evolution starting from a seed point. As a result, the second-order tensor effectively copes with complex structures like bifurcations. Cetin \& Unal \cite{cetinHigherOrderTensorVessel2015} utilized a similar method with higher-order tensors.

\textbf{{CTF RG:}} Due to the limitation of region-growing approaches, researchers proposed a multi-stage strategy to segment peripheral branches. The semantic predictions obtained by region growing are usually regarded as rough segmentation, followed by refinement algorithms. For instance, Bartz et al. \cite{bartzHybridSegmentationExploration2003} combined 3D region growing with 2D wave propagation and template matching to segment small branches of airway tree structures. The 2D wave propagation is presented to address the partial volume and low-resolution effects, while the 2D template matching is used to segment the small lumen further. This approach is further investigated in \cite{mayerHybridSegmentationVirtual2004}. Graham et al. \cite{grahamRobust3DAirway2010} first applied region growing for conservative segmentation of trachea and main branches, then they scanned and located all the airway cross-sections by applying an elliptic decision function on three 2D planes (transverse, coronal and sagittal planes). After that, they built a graph to connect all these airway segment candidates.  

\subsubsection{Level set segmentation}
\textcolor{Black}{The level set (LS) is a method based on partial differential equations (PDE), via moving curves and surfaces with curvature-based velocities. Given a moving speed $v$, and time $t$, the level-set function $\Phi$ satisfies
\begin{linenomath}
\begin{equation}
    \frac{\partial{\Phi}}{\partial{t}}= v|\nabla{\phi}|,
\end{equation}
\end{linenomath}
where $|\cdot|$ represents the Euclidean distance. In image world, given the pixel $(i,j)$, the motion equation can be presented as
\begin{linenomath}
\begin{align}
\begin{split}
&\frac{\Phi(i,j,t+\Delta{t})-\Phi(i,j,t)}{\Delta{t}} + max[F,0]\nabla^{+x}(i,j)\\&+min[F,0]\nabla^{-x}(i,j)=0,
\end{split}
\end{align}
\end{linenomath}
where $F$ indicates the force normal to the surface, and 
\begin{linenomath}
\begin{align}
\begin{split}
\nabla^{+x}(i,j)=&max[0, \Delta^{-x}\Phi(i,j)]^2 + \\&min[0, \Delta^{+x}\Phi(i,j)]^2, F>0
\end{split}
\end{align}
\begin{align}
\begin{split}
\nabla^{-x}(i,j)=&max[0, \Delta^{+x}\Phi(i,j)]^2 + \\&min[0, \Delta^{-x}\Phi(i,j)]^2, F<0
\end{split}
\end{align}
\end{linenomath}
It is of note that $\Delta^{+x}\Phi$ and $\Delta^{-x}\Phi$ are the left and right-side finite difference for the given pixel.} Specifically, level-set segmentation approaches are widely used in medical images with intensity inhomogeneity. By incorporating image-guided restrictions and prior clinical knowledge, level-set segmentation approaches are well-suited for segmenting tubular structures with complex architecture and size variability. The level set segmentation can be classified into edge-based \cite{hutchisonOrientedFluxSymmetry2010,lorigoCURVESCurveEvolution2001} and region-based \cite{klepaczkoSimulationMRAngiography2016}.

\begin{figure}[!ht]
    \centering
    \includegraphics[width=0.4\textwidth]{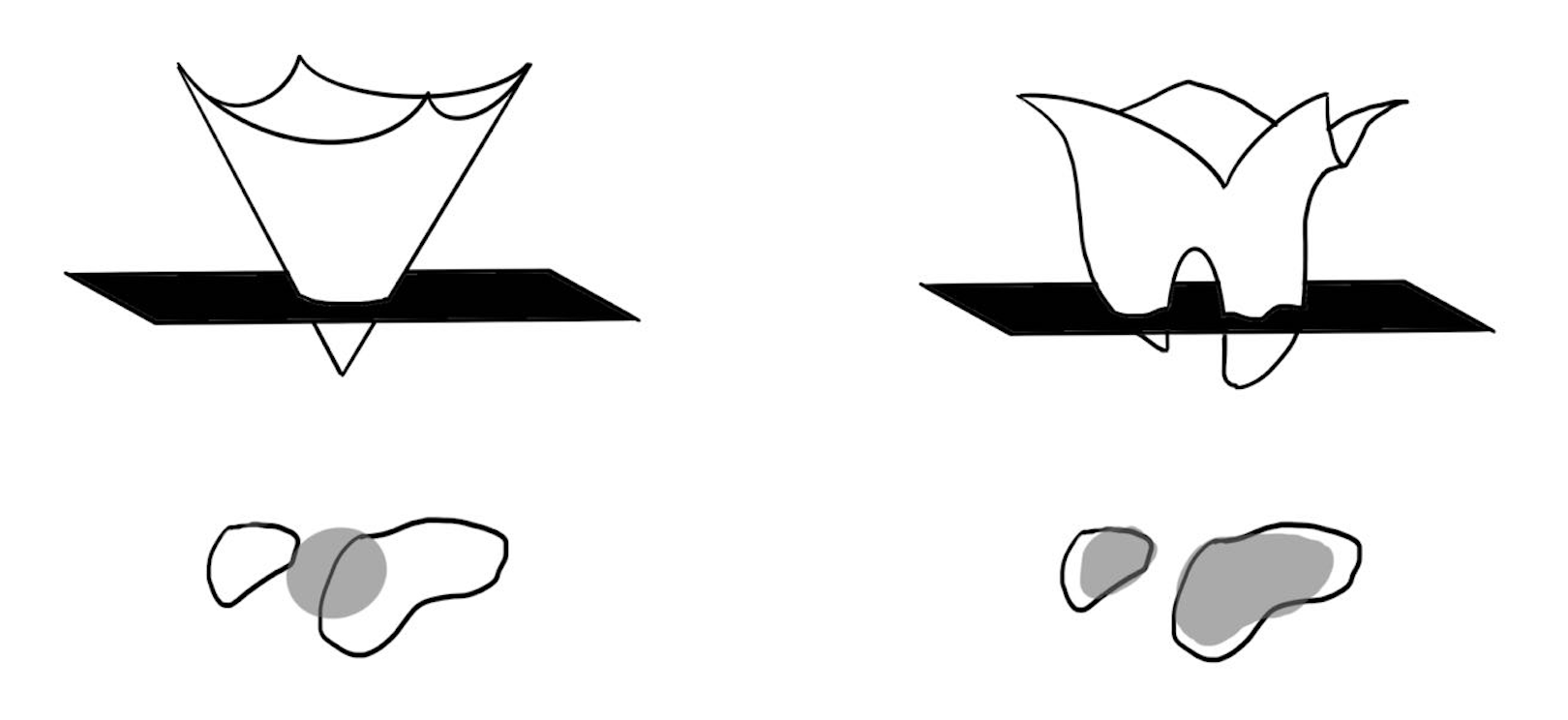}
    \caption{\textcolor{Black}{Level set segmentation.}}
    \label{figls}
\end{figure}

\textbf{{Edge-based LS:}} Edge-based level set approaches are mainly driven by intensity or gradient-derived external forces. For instance, Law et al. \cite{hutchisonOrientedFluxSymmetry2010} determined the brain vessel border orientation using gradient-flux symmetry along the vessel centrelines and gradient-flux asymmetry along vessel borders, which was then utilized to fit an active contour model by minimizing the weighted local variance. Lorigo et al. \cite{lorigoCURVESCurveEvolution2001} introduced a curve evolution-based vessel segmentation approach that models the vessel walls as manifolds that evolve iteratively to minimize an energy criterion based on image intensity and local smoothness. However, the E-LS becomes unstable and is prone to leakage when the initialized curve is out of the boundary of interest.

\textbf{{Region-based LS:}} The Region-based model was introduced to address some of the difficulties associated with edge-based deformable models, such as the leakage problem that typically occurs when coping with noise or non-uniform intensity distributions. The deformable surface was first thought to be moving under the image foreground and background regions in the region-based model. Both regions are considered statistically homogeneous, and the main distinctions in this type of algorithm pertain to how the statistics of the two regions are defined. Despite the benefits of adaptive topology, region-based deformable models have a higher computation complexity than edge-based deformable models. Klepaczko et al. \cite{klepaczkoSimulationMRAngiography2016} presented a method for validating vessel segmentation algorithms using a customed Magnetic Resonance Angiography (MRA) simulation framework. A reference model of a cerebral artery tree was created for this purpose using realistic Time-of-Flight (TOF) MRA data. Blood flow was simulated using this specified geometry, and a series of TOF pictures were created using various acquisition protocol parameters and signal-to-noise ratios. A level-set algorithm was then used to reconstruct the synthetic artery tree, which was next compared to the reference model. This validation procedure is especially beneficial for detecting minor vascular structures.

\begin{table*}[!t]
\scriptsize
\centering
\caption{Conventional segmentation methods}\label{tabc}
\begin{tabularx}{\textwidth}{lllllllc}
\toprule
\textbf{Author} & \textbf{Year} & \textbf{Region} & \textbf{Modality} & \textbf{Method} & \textbf{\begin{tabular}[t]{@{}l@{}}Validation Dataset\\ (number of cases)\end{tabular}} & \textbf{Metrics} & \multicolumn{1}{l}{\textbf{Method Class}} \\
\midrule
Jiang et al. \cite{xiaoyijiangAdaptiveLocalThresholding2003} & 2003 & RV & CFP & AT & \begin{tabular}[t]{@{}l@{}}DRIVE (40)\\ \\ STARE (20)\end{tabular} & \begin{tabular}[t]{@{}l@{}} \colorbox{acc}{\textcolor{white}{\colorbox{acc}{\textcolor{white}{Acc:}}}} 0.9212 \\ \colorbox{auroc}{\textcolor{white}{\colorbox{auroc}{\textcolor{white}{AUROC:}}}} 0.9114\\ \colorbox{acc}{\textcolor{white}{Acc:}} 0.9337 \\ \colorbox{auroc}{\textcolor{white}{\colorbox{auroc}{\textcolor{white}{AUROC:}}}} 0.8906\end{tabular} & \multirow[t]{8}{*}{\begin{tabular}[t]{@{}c@{}}Basic image\\ processing\\ operations\end{tabular}} \\
Pu et al. \cite{jiantaopuDifferentialGeometricApproach2011} & 2011 & LA & CT & AT & CT scans (75) & \textcolor{Black}{\colorbox{gc}{\textcolor{white}{GC:}} 8.2±1.2 }&  \\
Zana \& Klein \cite{zanaSegmentationVessellikePatterns2001} & 2001 & RV & CTA & MM-CCE & DRIVE (40) & \colorbox{visual}{\textcolor{white}{Visual}} &  \\
Passat et al. \cite{passatMagneticResonanceAngiography2006} & 2006 & BV & MRA & Knowledge-guided MM & PC MRA scans (30) & \begin{tabular}[t]{@{}l@{}}\colorbox{se}{\textcolor{white}{Se:}} 0.77\\ \colorbox{pre}{\textcolor{white}{Pre:}} 0.99\end{tabular} &  \\
Bouraoui et al. \cite{bouraoui3DSegmentationCoronary2010} & 2010 & CV & CTA & Knowledge-guided MM & CTA scans (60) & \colorbox{visual}{\textcolor{white}{Visual}} &  \\
Nguyen et al. \cite{nguyenEffectiveRetinalBlood2013} & 2013 & RV & CFP & MSLD & \begin{tabular}[t]{@{}l@{}}DRIVE (40)\\ STARE (20)\end{tabular} & \begin{tabular}[t]{@{}l@{}}\colorbox{acc}{\textcolor{white}{Acc:}} 0.9407\\ \colorbox{acc}{\textcolor{white}{Acc:}} 0.9324\end{tabular} &  \\
Aykac et al. \cite{aykacSegmentationAnalysisHuman2003} & 2003 & LA & CT & T-MM & CT scans (8) & \begin{tabular}[t]{@{}l@{}}\colorbox{se}{\textcolor{white}{Se:}} 0.73\\ \colorbox{gc}{\textcolor{white}{GC:}} 7-10\end{tabular} &  \\
Câmara Neto et al. \cite{camaranetoUnsupervisedCoarsetofineAlgorithm2017} & 2017 & RV & CFP & CTF AT-MM & \begin{tabular}[t]{@{}l@{}}DRIVE (40)\\ \\ STARE (20)\end{tabular} & \begin{tabular}[t]{@{}l@{}}\colorbox{se}{\textcolor{white}{Se:}} 0.7942\\ \colorbox{sp}{\textcolor{white}{Sp:}} 0.9631\\ \colorbox{se}{\textcolor{white}{Se:}} 0.7695\\ \colorbox{sp}{\textcolor{white}{Sp:}} 0.9537\end{tabular} &  \\
Hoover et al. \cite{hooverLocatingBloodVessels2000} & 2000 & RV & CFP & MF & STARE (20) & \begin{tabular}[t]{@{}l@{}}\colorbox{se}{\textcolor{white}{Se:}} 0.75\end{tabular} & \multirow[t]{5}{*}{\begin{tabular}[t]{@{}c@{}}Template\\ matching\end{tabular}} \\
Zhang et al. \cite{zhangRetinalVesselExtraction2010} & 2010 & RV & CFP & MF-FDOG & \begin{tabular}[t]{@{}l@{}}DRIVE (40)\\ \\ \\ STARE (20)\end{tabular} & \begin{tabular}[t]{@{}l@{}}\colorbox{acc}{\textcolor{white}{Acc:}} .9382\\ \colorbox{se}{\textcolor{white}{Se:}} 0.7120 \\ \colorbox{fpr}{\textcolor{white}{FPR:}} 0.0276 \\ \colorbox{acc}{\textcolor{white}{Acc:}} 0.9484\\ \colorbox{se}{\textcolor{white}{Se:}} 0.7177 \\ \colorbox{fpr}{\textcolor{white}{FPR:}} 0.0247\end{tabular} &  \\
Azzopardi et al. \cite{azzopardiTrainableCOSFIREFilters2015} & 2015 & RV & CFP & B-COSFIRE & \begin{tabular}[t]{@{}l@{}}DRIVE (40)\\ \\ \\ \\ STARE (20)\\ \\ \\ \\ CHASE-DB1 (28)\end{tabular} & \begin{tabular}[t]{@{}l@{}}\colorbox{se}{\textcolor{white}{Se:}} 0.7655\\ \colorbox{sp}{\textcolor{white}{Sp:}} 0.9704 \\ \colorbox{acc}{\textcolor{white}{Acc:}} 0.9442\\ \colorbox{auroc}{\textcolor{white}{AUROC:}} 0.9614 \\ \colorbox{se}{\textcolor{white}{Se:}} 0.7716\\ \colorbox{sp}{\textcolor{white}{Sp:}} 0.9701 \\ \colorbox{acc}{\textcolor{white}{Acc:}} 0.9497\\ \colorbox{auroc}{\textcolor{white}{AUROC:}} 0.9563\\ \colorbox{se}{\textcolor{white}{Se:}} 0.7585\\ \colorbox{sp}{\textcolor{white}{Sp:}} 0.9587 \\ \colorbox{acc}{\textcolor{white}{Acc:}} 0.9387\\ \colorbox{auroc}{\textcolor{white}{AUROC:}} 0.9487\end{tabular} &  \\
Krissian et al. \cite{krissianModelBasedDetectionTubular2000} & 2000 & BV & \begin{tabular}[t]{@{}l@{}}X-ray\\ MRA\end{tabular} & 3D CMM & \begin{tabular}[t]{@{}l@{}}X-ray scans (10)\\ MRA scans (n/a)\end{tabular} & \begin{tabular}[t]{@{}l@{}}\colorbox{visual}{\textcolor{white}{Visual}}\\ \colorbox{visual}{\textcolor{white}{Visual}}\end{tabular} &  \\
Worz \& Rohr \cite{worzSegmentationQuantificationHuman2007} & 2007 & TA & \begin{tabular}[t]{@{}l@{}}MRA\\ CTA\end{tabular} & 3D CMM & \begin{tabular}[t]{@{}l@{}}MRA scans (n/a)\\ CTA (n/a)\end{tabular} & \begin{tabular}[t]{@{}l@{}}\colorbox{visual}{\textcolor{white}{Visual}}\\ \colorbox{visual}{\textcolor{white}{Visual}}\end{tabular} &  \\
Kiraly et al. \cite{kiralyThreeDimensionalHumanAirway2002} & 2002 & LA & CT & Adaptive RG & CT scans (10) & \begin{tabular}[t]{@{}l@{}}\colorbox{gc}{\textcolor{white}{GC:}} 12\\ \colorbox{others}{\textcolor{white}{BC:}} 182\end{tabular} & \multirow[t]{10}{*}{\begin{tabular}[t]{@{}c@{}}Region\\ growing\end{tabular}} \\
Fetita et al. \cite{fetitaPulmonaryAirways3D2004} & 2004 & LA & CT & Energy-constrained RG & CT scans (30) & \begin{tabular}[t]{@{}l@{}}\colorbox{se}{\textcolor{white}{Se:}} 0.91\\ \colorbox{gc}{\textcolor{white}{GC:}} 6-7\end{tabular} &  \\
Tschirren et al. \cite{tschirrenIntrathoracicAirwayTrees2005} & 2005 & LA & CT & MSFC & LD CT scans (22) & \colorbox{others}{\textcolor{white}{NBC:}} 27.0±4.4 &  \\
Martinez-Perez et al. \cite{martinez-perezSegmentationBloodVessels2007} & 2007 & RV & CFP & Geometrical RG & \begin{tabular}[t]{@{}l@{}}DRIVE (40)\\ \\ \\ STARE (20)\end{tabular} & \begin{tabular}[t]{@{}l@{}}\colorbox{acc}{\textcolor{white}{Acc:}} 0.9344\\ \colorbox{se}{\textcolor{white}{Se:}} 0.7246 \\ \colorbox{fpr}{\textcolor{white}{FPR:}} 0.0345 \\ \colorbox{acc}{\textcolor{white}{Acc:}} 0.9410\\ \colorbox{se}{\textcolor{white}{Se:}} 0.7506 \\ \colorbox{fpr}{\textcolor{white}{FPR:}} 0.0431\end{tabular} &  \\
Fabijańska \cite{fabijanskaTwopassRegionGrowing2009} & 2009 & LA & CT & Geometrical RG & CT scans (10) & \colorbox{gc}{\textcolor{white}{GC:}} 9 &  \\
Cetin et al. \cite{cetinVesselTractographyUsing2013} & 2013 & CV & CTA & Geometrical RG & ROTTERDAM (32) & \colorbox{dsc}{\textcolor{white}{DSC:}} 0.964 &  \\
Cetin \& Unal \cite{cetinHigherOrderTensorVessel2015} & 2015 & \begin{tabular}[t]{@{}l@{}}BV\\ CV\end{tabular} & \begin{tabular}[t]{@{}l@{}}MRA\\ CTA\end{tabular} & Geometrical RG & \begin{tabular}[t]{@{}l@{}}MRA scans (50)\\ CTA (24)\end{tabular} & \begin{tabular}[t]{@{}l@{}} \textcolor{Black}{\colorbox{dsc}{\textcolor{white}{DSC:}}0.9286±0.03453}\\ \colorbox{dsc}{\textcolor{white}{DSC:}} 0.973\end{tabular} &  \\
Bartz et al. \cite{bartzHybridSegmentationExploration2003} & 2003 & LA & CT & CTF RG & CT scans (22) & \begin{tabular}[t]{@{}l@{}}\colorbox{se}{\textcolor{white}{Se:}} 0.85 \\ \colorbox{pre}{\textcolor{white}{Pre:}} 0.90\end{tabular} &  \\
Mayer et al. \cite{mayerHybridSegmentationVirtual2004} & 2004 & LA & CT & CTF RG & CT scans (22) & \colorbox{se}{\textcolor{white}{Se:}} 0.86 &  \\
Graham et al. \cite{grahamRobust3DAirway2010} & 2010 & LA & CT & CTF RG & CT scans (23) & \begin{tabular}[t]{@{}l@{}} \textcolor{Black}{\colorbox{others}{\textcolor{white}{BC:}} 253±84}\end{tabular} &  \\
Mendonca et al. \cite{Mendonca2006} & 2006 & RV & CFP & CTF RG & \begin{tabular}[t]{@{}l@{}}DRIVE (40)\\ \\ \\STARE (20)\\ \\ \end{tabular} & \begin{tabular}[t]{@{}l@{}}\textcolor{Black}{\colorbox{acc}{\textcolor{white}{Acc:}} 0.9463±0.0065}\\\colorbox{se}{\textcolor{white}{Se:}}0.7344\\\colorbox{fpr}{\textcolor{white}{FPR:}}0.7246\\\textcolor{Black}{\colorbox{acc}{\textcolor{white}{Acc:}} 0.9479±0.0123}\\\colorbox{se}{\textcolor{white}{Se:}}0.7123\\\colorbox{fpr}{\textcolor{white}{FPR:}}0.0270 \end{tabular} &  \\
Law et al. \cite{hutchisonOrientedFluxSymmetry2010} & 2010 & BV & MRA & Edge-based LS & MRA scans (4) & \colorbox{visual}{\textcolor{white}{Visual}} & \multirow[t]{3}{*}{\begin{tabular}[t]{@{}c@{}}Level set \\ segmentation\end{tabular}} \\
Lorigo et al. \cite{lorigoCURVESCurveEvolution2001} & 2001 & BV & MRA & Edge-based LS & MRA scans (n/a) & \colorbox{visual}{\textcolor{white}{Visual}} &  \\
Klepaczko et al. \cite{klepaczkoSimulationMRAngiography2016} & 2016 & BV & MRA & Region-based LS & MRA scans (n/a) & \colorbox{visual}{\textcolor{white}{Visual}} & \\
\hline
\end{tabularx}

\end{table*}

\begin{table*}
\begin{minipage}{\linewidth}\small
The regions include the retinal vessel (RV), lung airways (LA), lung vessel (LV), brain vessel (BV), coronary vessel (CV), and thoracic aorta (TA). All unnamed datasets are private and may have specific modalities such as phase-contrast (PC), low-dose (LD), standard-dose (SD), or time-of-flight (TOF). All metrics are shown as mean ± standard deviation or mean only. 
\\\\\\
\end{minipage}
\end{table*}

\subsection{Machine Learning Based Segmentation}
Most conventional methods reviewed above have numerous tuning parameters, which are hard to estimate manually. Machine learning (ML) algorithms are trained with specific features, usually based on data statistics, to automatically obtain an optimal set of parameters for a model or a classifier. There are two types of ML methods depending on their training process: unsupervised and supervised. Unsupervised learning is ideal for scenarios when gold standard (GS) labels are not available or exploratory data analysis is needed. On the other hand, supervised learning requires GS datasets to train the learning model or classifier. 

However, since ML approaches typically rely on manually extracted low-level features and criteria, they have a limited ability to learn features automatically, and segmentation accuracy is relatively low compared to deep learning approaches.

\begin{figure}[!ht]
    \centering
    \includegraphics[width=0.45\textwidth]{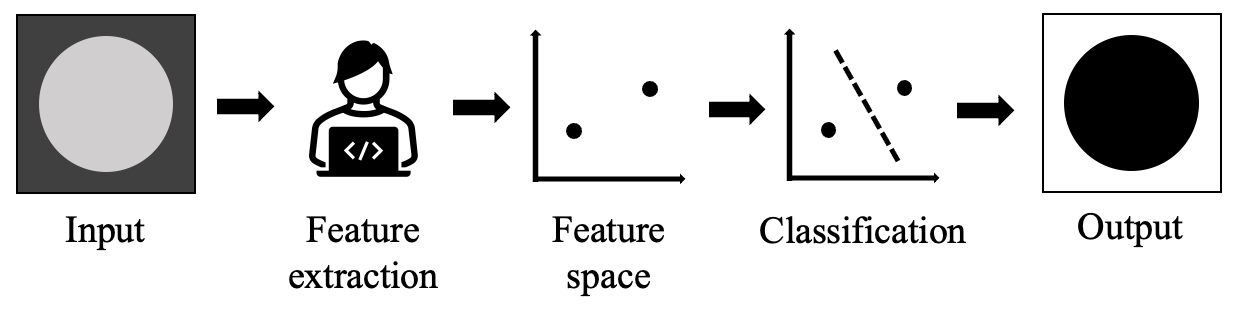}
    \caption{Machine learning based segmentation.}
    \label{figml}
\end{figure}

\subsubsection{Unsupervised}
Unsupervised ML algorithms are trained without a gold standard or ground truth. These methods discover a model or a classifier capable of describing a hidden grouping rule of input features without prior knowledge or supervision. The lack of GS can limit unsupervised segmentation performance, which is often inferior to supervised approaches. 

\textbf{{GMM:}} In the Gaussian mixture model (GMM), the image voxels are divided into backgrounds and tubular structures. Each class is characterized by a stochastic modelling process using a mixture of Gaussian or Gaussian-like distributions. \textcolor{Black}{Then the parameters of the mixture model ($\pi, \mu, \Sigma$) are estimated by maximizing the log-likelihood.
\begin{linenomath}
\begin{equation}
    \theta(\pi, \mu, \Sigma; X)=\sum_{m=1}^M{ln(\sum_{k=1}^K{\pi_k\textbf{N}(x_m|\mu_k,\Sigma_k)})}.    
\end{equation}
\end{linenomath}}
An unsupervised statistical modelling strategy using two-level GMM for segmenting the 3D cerebrovascular system from MRA data is presented in \cite{hassounaCerebrovascularSegmentationTOF2006}. Each voxel class is characterized by a low-level intensity histogram across the space and a high-level statistical dependence between adjacent voxels. The background’s low-level process is represented by two Gaussian distributions and one Rayleigh distribution, whereas the blood vessels are modelled by one Gaussian distribution. The parameters of the suggested GMM are automatically estimated utilizing an Expectation-Maximization (EM) algorithm \cite{moonExpectationmaximizationAlgorithm1996} before the Maximum Likelihood Estimation (MLE) classification. Spatial contextual information has been incorporated through the 3D Markov random field (MRF) \cite{graffigneHierarchicalMarkovRandom1995}, which serves as a prior distribution of the true label of the class of interest to avoid misclassification caused by noise and signal loss, which frequently occurs for classifiers using the pixel intensity solely.


\textbf{GMM and RG (GMM-RG):} Oliveira et al. \cite{oliveiraSegmentationLiverIts2011} also presented a method for segmenting nodules and vessels based on a GMM integrating the RG. The histogram combines three Gaussian models representing the nodules, liver parenchyma, and vasculature. The Levenberg-Marquardt (LM) optimizer is used to discover the parameter values that provide the best fit. The RG strategy allocates voxels with intensity lower than the lower threshold to nodules and is used as seeds in a process that aggregates neighbouring voxels with intensity lower than the upper threshold. The hepatic and portal veins are then identified using liver anatomical knowledge and a vein-tracking algorithm.

\textbf{Feature-based GMM:} Roychowdhury et al. \cite{roychowdhuryBloodVesselSegmentation2014} used a feature-based GMM to segment blood vessels accurately on pathological retinal images containing bright lesions, red lesions, and brightness/contrast variation while remaining computationally efficient. The GMM classifier is deployed to classify pixels as a vessel or non-vessel using a set of eight features collected from pixel neighbourhoods and first and second-order gradient images, which reduces the reliance on training data and improves the robustness. The number of pixels applied classification significantly decreases by removing major vessels overlapped by the high-pass filtered image and the morphologically (Top-hat) reconstructed image.

\textbf{{Clustering:}} Clustering is the process of discovering natural groupings in the feature space automatically. A cluster is frequently a dense area in the feature space where domain objects (e.g., image voxels) are closer to this cluster than others. \textcolor{Black}{Given the number of clusters $K$, the objective function of $N$ pixels within an image is presented as:
\begin{linenomath}
\begin{equation}
    J=\sum_{j=1}^K{\sum_{i=1}^N{{|x_i^j-c_j|}^2}},
\end{equation}
\end{linenomath}
where $c_j$ is the centroid for cluster $j$.} A centre vector represents each cluster in centroid-based clustering, which is unnecessarily a dataset member. The standard optimization objective is to locate cluster centres and assign objects with the minimized squared distance to the cluster centre.

\textbf{SWFCM:} Kande et al. \cite{kandeUnsupervisedFuzzyBased2010} presented a four-step unsupervised fuzzy-based vascular segmentation method to minimize erroneous vessel detection. First, matched filtering is applied to enhance blood vessels. The responses are then segmented using spatially weighted fuzzy c-means clustering (SWFCM) based thresholding, which preserves the spatial structure of the vascular tree. To speed up the procedure, the grey-level histogram of the image is used to determine the parameters of the FCM algorithm instead of the entire image data.

\textbf{K-means clustering (KMC):} Goceri et al. \cite{goceriVesselSegmentationAbdominal2017} proposed an unsupervised and adaptive clustering approach for segmenting portal and hepatic veins from liver magnetic resonance images. K-means clustering is employed at the initial stage to adaptively identify tissue regions and build a marker image. The results are processed iteratively with a linear contrast stretching technique in the next stage, using both edge and region-based information to generate a mask. The markers and mask images are then used to refine vascular areas using binary morphological image reconstruction \cite{vincentMorphologicalGrayscaleReconstruction1993}. The suggested methodology outperformed previous thresholding-based methods because it automatically and adaptively calculates thresholds and weights.

\textbf{{Swarm optimization:}} Swarm optimization (SO) is a bio-inspired algorithm that uses a population of candidate solutions to search for an optimal solution in the solution space iteratively. It differs from other optimization techniques in that it simply requires the objective function and is unaffected by the gradient or any derivative of the objective.

\textbf{Ant colony optimization (ACO):} Because MF alone may not be sufficient to retrieve the capillaries, Ant colony optimization (ACO) segmentation results are incorporated in \cite{cinsdikiciDetectionBloodVessels2009}. After basic preprocessing, the image is processed by the MF and the ant algorithm in parallel. In ant algorithms, several artificial ants build solutions and exchange information on the quality to find approximate solutions to the given optimization problem, using a communication mechanism similar to that used by actual ants \cite{dorigoAntColonyOptimization2006}. The result is then merged with the MF response using a simple OR gate, followed by a length filtering to acquire the entire vasculature. However, this method mistook pathological areas for vessels, and the segmented vessels are slightly thicker than manual labelling. 

\textbf{Artificial bee colony (ABC):} In \cite{hassanienRetinalBloodVessel2015}, another swarm-based optimization algorithm is utilized to develop an unsupervised retinal blood vessel segmentation approach. After enhancing the brightness, an artificial bee colony (ABC) stochastic optimization on the fuzzy c-means (FCM) compactness fitness function is used to determine cluster centres representing vessels and backgrounds. However, vessels with small diameters are usually distorted. They cannot be correctly categorized at this first level of segmentation since they are always confused with surrounding pixels, especially in pathological images. Hence, the second optimization step is employed to update the acquired cluster centres using the generalized pattern search (GPS) method \cite{torczonConvergencePatternSearch1997} to maximize the total thinness ratio of vessel segments. This method is proven to be robust to noise, exudates, haemorrhages, and variations in the pigment epithelium after deploying on DRIVE and STARE datasets. \textcolor{Black}{Su et al. \cite{SU2022} proposed CCABC which has faster convergence and can obtain higher quality solutions.}  

\textbf{Whale optimization algorithm (WOA):} Hassan \& Hassanien \cite{hassanRetinalFundusVasculature2018} further explored a novel swarm optimization method for extracting the vasculature from retinal fundus images. A hybrid model of multi-level thresholding is used in conjunction with the whale optimization algorithm (WOA) to segment the enhanced image. The WOA \cite{mirjaliliWhaleOptimizationAlgorithm2016} is utilized as an optimizer to determine thresholds that maximize between-class variance. On the DRIVE test dataset, the proposed method achieves state-of-the-art sensitivity and the results are visually identical to the ground truth.


\textbf{{MCET-HHO:}} In \cite{ramos-sotoEfficientRetinalBlood2021}, a novel unsupervised vessel segmentation approach is proposed based on optimized multi-level thresholding. The main processing stage is divided into two configurations after image smoothing. The new optimized top-hat, homomorphic filtering, and median filter were the first to segment thick vessels. The Minimum Cross-Entropy Thresholding - Harris Hawks Optimization (MCET-HHO) method \cite{rodriguez-esparzaEfficientHarrisHawksinspired2020}, a state-of-the-art multi-level image segmentation algorithm, is applied in the second configuration to segment thin vessels. Finally, an OR logic combines the thick and thin vessel masks. The suggested method outperforms most state-of-the-art methods described in terms of specificity and accuracy, while the remaining computational costs are small. \textcolor{Black}{Another multilevel thresholding method named Salp Swarm Algorithm (SSA) is optimized by Zhang et al. \cite{ZHANG2021} and has potential to be implemented in vessel segmentation.}

\subsubsection{Supervised}
In contrast, supervised learning algorithms infer a classification rule from manually annotated training pairings. The introduction of prior knowledge in the training process makes these methods usually outperform the unsupervised ones. However, that means the training time and computational cost are also increased.

\textbf{{KNN:}} The K-nearest neighbour (KNN) method is a classical supervised classification algorithm. The idea of the K-nearest neighbour method is straightforward and intuitive: if most of the K most similar (i.e., most neighbouring) samples in the feature space belong to a class, then the sample also belongs to that class.

\textbf{KNN and primitive-based method (KNN-PBM):} Staal et al. \cite{staalRidgeBasedVesselSegmentation2004} introduced a ridge-based retinal vessel segmentation methodology using a KNN classifier, taking advantage of the intrinsic feature that vessels are elongated structures rather than pixel representation. The primitive-based method (PBM) is developed to extract intrinsic features (image primitives) by grouping ridge pixels into sets that approximate straight-line elements. Each line element provides a local coordinate frame within each set, from which local characteristics are retrieved for each pixel. A sequential forward selection method is utilized to choose 27 features from convex sets and individual pixels, and then a KNN classifier is used for segmentation. This approach achieves high accuracy on the STARE dataset, but it is training data dependent, sensitive to false edges, and relatively slow due to extensive feature sets. 

\textbf{KNN and RG (KNN-RG):} Lo et al. \cite{loVesselguidedAirwayTree2010} employed KNN to classify the airway and non-airway voxels. The voxel-wise initial feature vector was constructed using the Hessian matrix and Gaussian kernel from the training data. The final airway tree segmentation is achieved using a 3D region growing method with a decision function that combines the probability map from the KNN classifier with the vessel orientation similarity evaluated by Hessian matrix analysis, which reveals how similar the orientation of a possible airway is to an adjacent vessel. By combining the KNN classifier with the vascular axial orientation constraint, the leakage is effectively limited, and the false positive rate of the segmentation results is significantly reduced. However, some fine tracheas are lost, reducing the total number of branches segmented.

\textbf{{SVM:}} A Support Vector Machine (SVM) is a class of generalized linear classifiers that performs binary classification of data in a supervised learning manner, where the decision boundary is the maximum-margin hyperplane solved for the learned samples. \textcolor{Black}{For a soft-margin SVM, the objective function $J$ is minimized to find the hyperplane. 
\begin{linenomath}
\begin{equation}
    J=\lambda{||\mathrm{\textbf{w}||}^2+[\frac{1}{n}\sum_{i=1}^n{max(0,1-y_i(\mathrm{\textbf{w}}^{T}\textbf{x}_i-b))}}],
\end{equation}
\end{linenomath}
where $y_i$ is the $i$-th target, $\mathrm{\textbf{w}}^{T}\textbf{x}_i-b$ is the $i$-th prediction. \\
}
\textbf{SVM with Line detectors (SVM-LD):} The application of line detectors as a feature vector for SVM to segment retinal vessels is proposed by Ricci \& Perfetti \cite{ricciRetinalBloodVessel2007}. The line strength response from three orthogonal line detectors is thresholded to create a simple 3D feature vector for quick supervised classification using a support vector machine (SVM). In comparison to a convolution with a 2D kernel, the suggested line detector has a lower computing cost and is less sensitive to noise due to averaging. Also, because no hypothesis is based on the cross-section profile, this detector achieves reliable detection in the presence of vessels of various sizes. However, this approach is hight dependent on the training dataset.

\textbf{SVM and graph-cut (SVM-GC):} Meng et al. \cite{mengAutomaticSegmentationAirway2017} utilized the SVM to provide an accurate supervised extraction strategy for the complicated airway tree. First, Hessian analysis and adaptive multi-scale cavity enhancement filter (CEF) enhance the areas with tube-like and cavity-like structures. SVM is then used to classify voxel candidates and remove false positives using 32D feature vectors based on local pixel intensity and structures. Finally, the candidate voxel map is refined using the graph-cut method to create the integrated airway tree. The introduction of the SVM classifier improves the accuracy, but it is highly dependent on the training data and is computationally expensive. The experimental results also showed that the lung tracheal tree segmented by this method was prone to branch breakage, particularly in the case of small tracheal branches. \textcolor{Black}{Zhai et al. \cite{Zhai16lungvessel} proposed a graph-cuts objective function combined with a Hessian based filter to segment lung vessels. They validated the method on 20 CT scans in VESSEL12 challenge and achieved a competitive performance.} 

\textbf{SVM with multi-scale filter (SVM-MSF):} Lee et al. \cite{leeHybridAirwaySegmentation2019} proposed a hybrid method integrating multi-scale filtering and an SVM classifier to segment the airway tree on 3D chest CT scans automatically. First, a modified 2D Hough transform \cite{ballardGeneralizingHoughTransform1981} was used to detect a starting seed point in the trachea. Then, a multi-scale tubular structure-based filter is employed to identify the initial candidate airway regions, including a grey-scale morphological operation and a tubular structure detection \cite{frangiMultiscaleVesselEnhancement1998,serraImageAnalysisMathematical1982}. An SVM classifier is then utilized to search for weak-contrast narrow airways and unconnected regions missed by the two enhancement filters. The fuzzy connectedness technique extracts 28-dimensional feature vectors to discover additional candidate airways while simultaneously suppressing FPR. However, the method also has shortcomings, including a high probability of false positives and the potential for leakage.


\textbf{{Trainable GMM:}} Soares et al. \cite{soaresRetinalVesselSegmentation2006} proposed a supervised learning approach for segmenting the blood vessels in retinal images using 2D Gabor wavelet. To account for vessels of various widths, the feature vector comprises the pixel’s intensity and 2D Gabor wavelet responses collected at multiple scales. The wavelet helps detect and analyse localized characteristics and singularities \cite{antoineImageAnalysisTwodimensional1993}, such as edges and blood vessels in this case. Moreover, because the Gabor wavelet can tune to specific frequencies, noise filtering, and vessel augmentation can be performed in one step. Then, a GMM and a Bayesian classifier with class-conditional probability density functions (likelihoods) are employed for fast classification. Probability distributions are computed using a training set of labelled pixels from manual segmentations. However, the training takes hours and is prone to overfit.

\textbf{{Ensemble modules:}} Based on an ensemble system of bagged and boosted decision trees, Fraz et al. \cite{frazEnsembleClassificationBasedApproach2012} proposed an accurate vessel segmentation approach for colour fundus photography. The Ensemble classification \cite{polikarEnsembleBasedSystems2006} is the technique of strategically generating and combining multiple classifiers. The decision trees are used as the classification model in this approach, and the outcomes of these weak learners are pooled using bootstrap aggregation (bagging) and boosting algorithms. The 9D feature vector comprised the inverted pixel intensity, gradient orientation maps, top-hat transform responses, line detector responses, and Gabor filter responses, which achieve high robustness. However, because of the boosting strategy with 200 decision trees, this method has a significant computational complexity.

\begin{table*}[!t]
\scriptsize
\centering
\caption{Machine learning based segmentation methods}\label{tabml}
\begin{tabularx}{\textwidth}{lllllllc}
\toprule
\textbf{Author} & \textbf{Year} & \textbf{Region} & \textbf{Modality} & \textbf{Method} & \textbf{\begin{tabular}[t]{@{}l@{}}Validation Dataset\\ (number of cases)\end{tabular}} & \textbf{Metrics} & \multicolumn{1}{l}{\textbf{Method Class}} \\
\midrule
Hassouna et al. \cite{hassounaCerebrovascularSegmentationTOF2006} & 2006 & BV & MRA & GMM & TOF MRA scans (n/a) & \colorbox{visual}{\textcolor{white}{Visual}} & \multirow[t]{9}{*}{Unsupervised} \\
Oliveira et al. \cite{oliveiraSegmentationLiverIts2011} & 2011 & LV & CT & GMM-RG & LD CT scans (15) & \colorbox{visual}{\textcolor{white}{Visual}} &  \\
Roychowdhury et al. \cite{roychowdhuryBloodVesselSegmentation2014} & 2014 & RV & CFP & Feature-based GMM & \begin{tabular}[t]{@{}l@{}}DRIVE (40)\\ \\ \\ \\ STARE (20)\\ \\ \\ \\ CHASE-DB1 (28)\end{tabular} & \begin{tabular}[t]{@{}l@{}}\textcolor{Black}{\colorbox{acc}{\textcolor{white}{Acc:}} 0.9519±0.005}\\ \textcolor{Black}{\colorbox{se}{\textcolor{white}{Se:}} 0.7249±0.0481}\\ \textcolor{Black}{\colorbox{sp}{\textcolor{white}{Sp:}} 0.983±0.0071}\\ \colorbox{auroc}{\textcolor{white}{AUROC:}} 0.962\\ \textcolor{Black}{\colorbox{acc}{\textcolor{white}{Acc:}} 0.9515±0.013}\\ \textcolor{Black}{\colorbox{se}{\textcolor{white}{Se:}} 0.7719±0.071}\\ \textcolor{Black}{\colorbox{sp}{\textcolor{white}{Sp:}} 0.9726±0.012}\\ \colorbox{auroc}{\textcolor{white}{AUROC:}} 0.9688\\ \textcolor{Black}{\colorbox{acc}{\textcolor{white}{Acc:}} 0.9530±0.005}\\ \textcolor{Black}{\colorbox{se}{\textcolor{white}{Se:}} 0.7201±0.0385}\\ \textcolor{Black}{\colorbox{sp}{\textcolor{white}{Sp:}} 0.9824±0.004}\\ \colorbox{auroc}{\textcolor{white}{AUROC:}} 0.9532\end{tabular} &  \\
Kande et al. \cite{kandeUnsupervisedFuzzyBased2010} & 2010 & RV & CFP & SWFCM & \begin{tabular}[t]{@{}l@{}}DRIVE (40)\\ \\ STARE (20)\end{tabular} & \begin{tabular}[t]{@{}l@{}}\colorbox{acc}{\textcolor{white}{Acc:}} 0.8911\\ \colorbox{auroc}{\textcolor{white}{AUROC:}} 0.9518\\ \colorbox{acc}{\textcolor{white}{Acc:}} 0.8976\\ \colorbox{auroc}{\textcolor{white}{AUROC:}} 0.9298\end{tabular} &  \\
Goceri et al. \cite{goceriVesselSegmentationAbdominal2017} & 2017 & LV & MRI & KMC & MRI scans (14) & \colorbox{acc}{\textcolor{white}{Acc:}} 0.896 &  \\
Cinsdikici \& Aydın \cite{cinsdikiciDetectionBloodVessels2009} & 2009 & RV & CFP & ACO & DRIVE (40) & \begin{tabular}[t]{@{}l@{}}\colorbox{acc}{\textcolor{white}{Acc:}} 0.9293\\ \colorbox{auroc}{\textcolor{white}{AUROC:}} 0.9407\end{tabular} &  \\
Hassanien et al. \cite{hassanienRetinalBloodVessel2015} & 2015 & RV & CFP & ABC & DRIVE (40) & \begin{tabular}[t]{@{}l@{}}\colorbox{acc}{\textcolor{white}{Acc:}} 0.9388\\ \colorbox{se}{\textcolor{white}{Se:}} 0.721\\ \colorbox{sp}{\textcolor{white}{Sp:}} 0.971 \end{tabular} &  \\
Hassan \& Hassanien \cite{hassanRetinalFundusVasculature2018} & 2018 & RV & CFP & WOA & DRIVE (40) & \begin{tabular}[t]{@{}l@{}}\colorbox{acc}{\textcolor{white}{Acc:}} 0.9793\\ \colorbox{se}{\textcolor{white}{Se:}} 0.8981\\ \colorbox{sp}{\textcolor{white}{Sp:}} 0.9883\\ \colorbox{auroc}{\textcolor{white}{AUROC:}} 0.9820\end{tabular} &  \\
Ramos-Soto et al. \cite{ramos-sotoEfficientRetinalBlood2021} & 2021 & RV & CFP & MCET-HHO & \begin{tabular}[t]{@{}l@{}}DRIVE (40)\\ \\ \\ STARE (20)\end{tabular} & \begin{tabular}[t]{@{}l@{}}\colorbox{acc}{\textcolor{white}{Acc:}} 0.9667\\ \colorbox{se}{\textcolor{white}{Se:}} 0.7578\\ \colorbox{sp}{\textcolor{white}{Sp:}} 0.9860\\ \colorbox{acc}{\textcolor{white}{Acc:}} 0.9836\\ \colorbox{se}{\textcolor{white}{Se:}} 0.7474\\ \colorbox{sp}{\textcolor{white}{Sp:}} 0.9580\end{tabular} &  \\
Staal et al. \cite{staalRidgeBasedVesselSegmentation2004} & 2004 & RV & CFP & KNN-PBM & \begin{tabular}[t]{@{}l@{}}DRIVE (40)\\ \\ \\ STARE (20)\end{tabular} & \begin{tabular}[t]{@{}l@{}}\colorbox{acc}{\textcolor{white}{Acc:}} 0.9441\\ \colorbox{auroc}{\textcolor{white}{AUROC:}} 0.9520\\ \colorbox{acc}{\textcolor{white}{Acc:}} 0.8958\\ \colorbox{auroc}{\textcolor{white}{AUROC:}} 0.9614\end{tabular} & \multirow[t]{7}{*}{Supervised} \\
Lo et al. \cite{loVesselguidedAirwayTree2010} & 2010 & LA & CT & KNN-RG & EXACT'09 (20) & \begin{tabular}[t]{@{}l@{}}\colorbox{others}{\textcolor{white}{FTR:}} 0.0011\\ \colorbox{bd}{\textcolor{white}{BD:}} 0.598\\ \colorbox{others}{\textcolor{white}{TL:}} 1184 mm\\ \colorbox{tld}{\textcolor{white}{TLD:}} 0.540\end{tabular} &  \\
Ricci \& Perfetti \cite{ricciRetinalBloodVessel2007} & 2007 & RV & CFP & SVM-LD & \begin{tabular}[t]{@{}l@{}}DRIVE (40)\\ \\ STARE (20)\end{tabular} & \begin{tabular}[t]{@{}l@{}}\colorbox{acc}{\textcolor{white}{Acc:}} 0.9595\\ \colorbox{auroc}{\textcolor{white}{AUROC:}} 0.9633\\ \colorbox{acc}{\textcolor{white}{Acc:}} 0.9646\\ \colorbox{auroc}{\textcolor{white}{AUROC:}} 0.9680\end{tabular} &  \\
Meng et al. \cite{mengAutomaticSegmentationAirway2017} & 2017 & LA & CT & SVM-GC & SD CT scans (50) & \begin{tabular}[t]{@{}l@{}}\colorbox{se}{\textcolor{white}{Se:}} 0.783\\ \colorbox{fpr}{\textcolor{white}{FPR:}} 0.10\\ \colorbox{bd}{\textcolor{white}{BD:}} 122.2\\ \colorbox{others}{\textcolor{white}{TL:}} 63.14\\ \colorbox{tld}{\textcolor{white}{TLD:}} 0.804\end{tabular} &  \\
\textcolor{Black}{Zhai et al}. \cite{Zhai16lungvessel} & 2016 & LV & CT & SVM-GC & VESSEL12 (20) & \begin{tabular}[t]{@{}l@{}}\colorbox{se}{\textcolor{white}{Se:}} 0.7331\\ \colorbox{sp}{\textcolor{white}{Sp:}} 0.7917\end{tabular} &  \\
Lee et al. \cite{leeHybridAirwaySegmentation2019} & 2018 & LA & CT & SVM-MSF & EXACT'09 (20) & \begin{tabular}[t]{@{}l@{}}\colorbox{fpr}{\textcolor{white}{FPR:}} 0.073±0.0466\\ \colorbox{tld}{\textcolor{white}{TLD:}} 0.569±0.11\end{tabular} &  \\
Soares et al. \cite{soaresRetinalVesselSegmentation2006} & 2006 & RV & CFP & Trainable GMM & \begin{tabular}[t]{@{}l@{}}DRIVE (40)\\ \\ STARE (19)\end{tabular} & \begin{tabular}[t]{@{}l@{}}\colorbox{acc}{\textcolor{white}{Acc:}} 0.9480 \\ \colorbox{auroc}{\textcolor{white}{AUROC:}} 0.9671\\ \colorbox{acc}{\textcolor{white}{Acc:}} 0.9466 \\ \colorbox{auroc}{\textcolor{white}{AUROC:}} 0.9614\end{tabular} &  \\
Fraz et al. \cite{frazEnsembleClassificationBasedApproach2012} & 2012 & RV & CFP & Ensemble modules & \begin{tabular}[t]{@{}l@{}}DRIVE (40)\\ \\ \\ \\ \\ STARE (20)\\ \\ \\ \\ \\ CHASE-DB1 (28)\end{tabular} & \begin{tabular}[t]{@{}l@{}}\colorbox{acc}{\textcolor{white}{Acc:}} 0.9534\\ \colorbox{se}{\textcolor{white}{Se:}} 0.7548\\ \colorbox{sp}{\textcolor{white}{Sp:}} 0.9763\\ \colorbox{pre}{\textcolor{white}{Pre:}} 0.7956\\ \colorbox{auroc}{\textcolor{white}{AUROC:}} 0.9768\\ \colorbox{acc}{\textcolor{white}{Acc:}} 0.9480\\ \colorbox{se}{\textcolor{white}{Se:}} 0.7406\\ \colorbox{sp}{\textcolor{white}{Sp:}} 0.9807\\ \colorbox{pre}{\textcolor{white}{Pre:}} 0.8532\\ \colorbox{auroc}{\textcolor{white}{AUROC:}} 0.9747\\ \colorbox{acc}{\textcolor{white}{Acc:}} 0.9469\\ \colorbox{se}{\textcolor{white}{Se:}} 0.7224\\ \colorbox{sp}{\textcolor{white}{Sp:}} 0.9711\\ \colorbox{pre}{\textcolor{white}{Pre:}} 0.7415\\ \colorbox{auroc}{\textcolor{white}{AUROC:}} 0.9712\end{tabular} & \\
\hline
\end{tabularx}

\end{table*}

\begin{table*}
\begin{minipage}{\linewidth}\small
The regions include the retinal vessel (RV), lung airways (LA), brain vessel (BV), coronary vessel (CV), and thoracic aorta (TA). All unnamed datasets are private and may have specific modalities such as phase-contrast (PC), low-dose (LD), standard-dose (SD), or time-of-flight (TOF). All metrics are shown as mean ± standard deviation or mean only. 
\end{minipage}
\end{table*}

\subsection{Deep Learning Based Segmentation}

Deep learning (DL) approaches automatically extract features to explore raw data without using handmade features. They can automatically learn multiple-level patterns and are not constrained by a specific application. Therefore, they have a superior generalization and recognition capabilities. DL models are composed of layers organized in a hierarchical structure called artificial neural networks (ANN) and interpret input data into meaningful output. The network’s parameters, such as kernel coefficients and fully-connected (FC) layer weights, are learned in the training process to represent the input data comprehensively. In this review, all DL models applied for treelike tubular structure segmentation are trained under supervision.

\begin{figure}[!ht]
    \centering
    \includegraphics[width=0.4\textwidth]{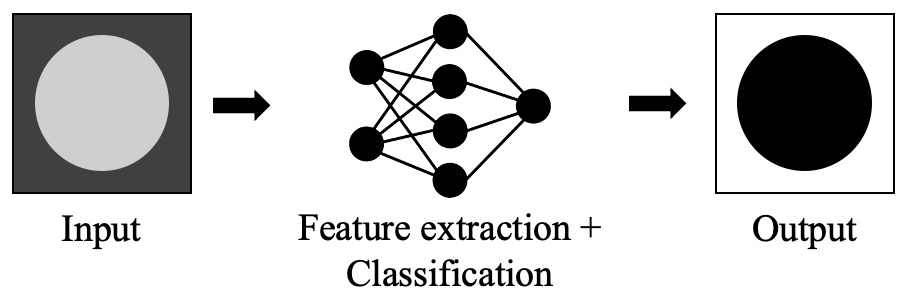}
    \caption{Deep learning based segmentation.}
    \label{figdl}
\end{figure}

\subsubsection{Classical CNNs}

Convolutional neural networks (CNNs) are a class of DL models that use convolution operations in at least one of their layers. The convolutional layers are inspired by the receptive field structure found in the animal's primary visual cortex. Hence, they are specifically designed to perform image processing tasks such as classification or pixel-wise segmentation.

\textcolor{Black}{Classical CNNs are typically comprised of multiple convolutional and pooling layers, followed by a set of fully connected layers. The convolutional layer extracts features from inputs by convolution operation, in which a convolution kernel slides throughout the input map, multiplying elementwise with a specified region called the receptive field. Then, the activation function maps inputs non-linearly to increase the visual representation ability of convolutional layers. The pooling layers are usually employed after to reduce feature position dependency and thus improve CNNs’ robustness. Besides, down-sampling along with pooling can reduce computational costs. Fully connected layers are the last block that produces a specialized semantic output, for example, foreground or background, from the feature map.}

Charbonnier et al. \cite{charbonnierImprovingAirwaySegmentation2017} presented a 2D CNN-based leakage detection approach that treats the removal of segmentation leakages as an independent classification method to improve the quality of tracheal segmentation on thoracic CT. Three 2D image patches were taken from each airway candidate, corresponding to the top, middle, and bottom layers. The airway candidate is then classified by passing three patches through three parallel stacks of convolutional and max-pooling layers separately and combining FC layers of three stacks in a final softmax layer. However, this method does not thoroughly analyse the 3D continuity of airway candidates and treelike structure. Similarly, Yun et al. \cite{yunImprovementFullyAutomated2018} presented a voxel-by-voxel airway segmentation approach for chest CT utilizing a 2.5D CNN (a pseudo-3D CNN) which takes three adjacent slices in each orthogonal direction (axial, sagittal, and coronal) at each voxel. 

\textcolor{Black}{\textbf{{UNet:}} UNet is a CNN designed for medical image segmentation \cite{ronnebergerUNetConvolutionalNetworks2015}, which has a symmetric encoder-decoder architecture with skips connections between the encoding (down-sampling) and decoding (up-sampling) pathways. UNet can process the entire image in one forward pass through its down-sampling/up-sampling architecture, resulting in a direct and faster segmentation than typical pixel-wise CNNs. Besides, using skip connections, UNets improve the accuracy of prediction by combining a low-level feature map that provides local information and a high-level feature map that represents global information. Based on the UNet architecture, Meyer et al. \cite{meyerDeepNeuralNetwork2017} proposed a retinal vessel segmentation method from Scanning Laser Ophthalmoscopy (SLO) CFP images. The 3D UNet \cite{cicek3DUNetLearning2016} based method for airway segmentation is presented in \cite{garcia-ucedajuarezAutomaticAirwaySegmentation2018}.} 

\textcolor{Black}{Jin et al. \cite{jin3DConvolutionalNeural2017} introduced a method capable of high-quality airways segmentation from incompletely labelled training datasets based on a shallow 3D UNet. This approach reduces the previous deeper 3D UNet to only two pooling layers to preserve small airways at distal sites and potentially benefit the learning of 3D structure variations. A domain-specific sampling scheme is applied to strategically select annotations from a highly-specific and moderately-sensitive conservative segmentation method \cite{xuHybridMethodAirway2015}. In addition, fuzzy-connectedness RG and skeletonization-guided leakage removal are implemented to address local discontinuities in the output. Similarly, Garcia-Uceda et al. \cite{garcia-ucedaAutomaticAirwaySegmentation2021} proposed an optimized and efficient approach for segmenting the airways based on a simple and low-memory 3D UNet architecture. This approach employs unpadded convolutions in the first three resolution levels of the UNet, which provides a 30\% reduction in memory footprint. Compared to the previous UNet-based airway segmentation algorithms, the proposed method can process larger 3D image patches, usually covering the entire lung, in a single run, resulting in an improved outcome and a faster converging of the training.}

\textbf{{ResNet:}} The residual neural network (ResNet) is a CNN that utilizes shortcuts that skip one or more layers. Adding skip connections solves the problem of gradient disappearance and avoids the increasing training error resulting from stocking more layers for higher discriminatory ability. Based on a multi-pathway ResNet, a robust 3D hepatic vessel segmentation method is proposed by Kitrungrotsakul et al. \cite{kitrungrotsakulVesselNetDeepConvolutional2019}. The proposed network structure consists of three parallel DenseNets\cite {huangDenselyConnectedConvolutional2017}, i.e., extensions of ResNet, for extracting data in the sagittal, coronal, and transverse planes, which are concatenated in the FC layer to predict the class of central voxels. In addition, a multi-scale method \cite{satoThreedimensionalMultiscaleLine1998} transforms raw images into vessel probability maps as network inputs to enhance robustness.

\textcolor{Black}{\textbf{{Generative adversarial network (GAN):}} Generative adversarial networks (GANs) \cite{GANGoodfellow} frame the segmentation task as a supervised learning problem with two components: a generator ($\mathcal{G}$), which learns the mapping from a random input $z$ with density$p_z$ to $x_g = \mathcal{G}(z)$, and a discriminator ($\mathcal{D}$), which either takes the ground truth $x_t$ with density $p_t$ or the generated data $x_g$, and outputs the probability $\mathcal{D}(x)$ of x to be true. These two components are trained adversarially to drive the generator to generate more accurate segmentation labels that can fool the discriminator. In \cite{zhaoHighQualityRetinal2020}, a retinal vessel segmentation method is developed based on the GAN. The proposed GAN takes advantage of UNet, combining high-resolution local features with up-sampled high-level features. Besides, the dilated convolutions employed by GAN effectively expand the receptive field to capture large-scale morphological vessel properties without raising the number of computations required.} 

\subsubsection{Advanced architectures}

In addition the basic CNN models, a variety of new DL architectures are developed by applying or integrating one or more advanced modules including holistically-nested edge detector (HED) \cite{linAutomaticRetinalVessel2019}, graph neural network (GNN) \cite{garcia-ucedajuarezJoint3DUNetGraph2019, selvanGraphRefinementBased2020}, mean-field network (MFN) \cite{selvanGraphRefinementBased2020}, recurrent convolutional layer (RCL) \cite{Wang2019Radial}, WingsNet\cite{Zheng2020WingsNet}, attention mechanism (AM) \cite{qinLearningBronchioleSensitiveAirway2020, qinLearningTubuleSensitiveCNNs2021, zhouRefinedEquilibriumGenerative2021, liCascadedAttentionGuided2020}, atrous convolution (AC) \cite{chengSegmentationAirwayTree2021}, back-propagation neural network (BPNN) \cite{tangConstructionVerificationRetinal2021}, multi-scale feature aggregation (MFA) \cite{zhouAutomaticAirwayTree2021, zhouRefinedEquilibriumGenerative2021}, Inception \cite{guoRetinalVesselSegmentation2020}, and multi-kernel pooling (MKP) \cite{parkMGANRetinalBlood2020}. Since the mentioned architectural improvements are also present in other types of methods, in this subsection we only detail the methods based solely on advanced modules first.

\textbf{{DSSRN:}} Lin et al. \cite{linAutomaticRetinalVessel2019} proposed a deeply supervised retinal vessel segmentation network, DSSRN, based on the holistically-nested edge detector (HED), incorporating the conditional random fields (CRFs). The proposed network begins with the HED architecture, a five-block single-stream network with multiple convolutional and ReLU layers in each block. The final predictions are formed by combining multi-scale side outputs from each hidden layer, eliminating the edge or boundary detection ambiguity. CRF provides global smoothness regularization into the system for robustness. 

\textbf{{UNet and GNN (UNet-GNN):}} Garcia-Uceda Juarez et al. \cite{garcia-ucedajuarezJoint3DUNetGraph2019} proposed an airway tree segmentation approach for chest CTs that combines a 3D UNet with graph neural network (GNN) based modules \cite{scarselliGraphNeuralNetwork2009}. This method substitutes a GNN-based module for the two deepest convolutional layers in the baseline UNet. The introduction of graph convolutions provides nodes with corresponding feature vectors containing information about graph neighbourhood connectivity. The connectivity of nodes is described by the adjacency matrix with learnable GNN filter weights. 

\textbf{{UNet with RCL (UNet-RCL):}} In \cite{Wang2019Radial}, Wang et al. incorporated 3D slice-by-slice convolutional layers called recurrent convolutional layers (RCLs) in a 3D UNet architecture for airway segmentation. Different from the conventional layer-by-layer CNN, the in-layer recurrent convolutions make the message-passing between neurons more efficient in the same layer, which helps in capturing the spatial information of elongated structures.

\textbf{{WingsNet:}}
Zheng et al. \cite{Zheng2020WingsNet} argued that the poor performance of CNN on segmenting distal small airways is attributed to the gradient erosion and dilation of the neighbourhood voxels, which is caused by the imbalance between foreground and background regions. Therefore, they proposed WingsNet with group supervision to boost the training of shallow layers. Moreover, they designed a new loss called General Union Loss to address the imbalance between large and small airways. 

\textbf{{UNet with AM (UNet-AM):}} In \cite{qinLearningBronchioleSensitiveAirway2020}, Qin et al. presented a 3D-UNet based airway segmentation method with AMs that is highly sensitive to bronchioles in the periphery. The method initially employs a feature recalibration module based on \cite{rickmannProjectExciteModules2019, zhuAnatomyNetDeepLearning2019} to make the greatest use of learned features. Moreover, inspired by \cite{houLearningLightweightLane2019, zagoruykoPayingMoreAttention2017}, an attention distillation module is introduced to promote specific structure and distribution pattern learning of airways to improve the ability to distinguish small bronchioles. The proposed method retrieved significantly more branches while maintaining comparable overall segmentation accuracy. A similar method is applied in \cite{qinLearningTubuleSensitiveCNNs2021} for both airway and artery-vein segmentation on non-contrast computed tomography (CT).

\textcolor{Black}{\textbf{{UNet with HRMF}} Fan et al. \cite{Fan2020} proposed an unsupervised UNet with hidden markov random field to segment brain vessels in TOF MRA. The method first uses a HMRF module to classify each TOF-MRA image into three classes, background, brain tissue and blodd vessels. Then the 3-D UNet takes in the output from HMRF module and segments brain vessels. }

\textbf{{TACNet:}} Cheng et al. \cite{chengSegmentationAirwayTree2021} presented the Tiny Atrous Convolutional Network (TACNet) to account for multi-scale changes in the airway. The principal convolution operation employed in the TACNet is the atrous convolution (AC), which enlarges the receptive field and captures multi-scale context information. In a conventional UNet, the feature map is downsampled numerous times, leading tiny bronchi to vanish from the feature map or become undetectable. The ACs maintain the original resolution between layers, ensuring that the feature map does not reduce in size while increasing the receptive field. 

\textbf{{Back-propagation neural network (BPNN):}} A back-propagation neural network (BPNN) for retinal vascular segmentation is proposed in \cite{tangConstructionVerificationRetinal2021} to increase the accuracy of retinal vessel segmentation. Four different green channel enhancement results are utilized to create 4D feature vectors: adaptive histogram equalization, morphological processing, Gaussian matched filtering, and Hessian matrix filtering. The feature vectors are fed into a typical three-layer BP neural network to segment blood vessels. 

\textbf{{MFA-Net:}} In \cite{zhouAutomaticAirwayTree2021}, a 3D multi-scale feature aggregation network (MFA-Net) is presented against the scale difference of various structures in airway tree segmentation. The MFA blocks apply parallel 3D atrous convolutions of different rates to capture multi-scale context information, which improves the sensitivity of MFA-Net to small bronchi segmentation to solve local discontinuities.

\subsubsection{Task-based/Hybrid approaches}


According to the morphological characteristics of human treelike tubular structures, the segmentation task of DL networks can be converted into voxel connectivity analysis \cite{qinAirwayNetVoxelConnectivityAware2019}, graph refinement \cite{selvanGraphRefinementBased2020}, or bifurcation point detection (BPD) to support conventional RG \cite{wangAutomatedLabelingAirway2020}. 

Multi-task segmentation were implemented through cascaded or hybrid modules, including separate segmentation of vertical and horizontal tubular structures \cite{zhaoBronchusSegmentationClassification2019} or thick and thin tubular structures \cite{yangHybridDeepSegmentation2021}, and coarse segmentation followed by refinement \cite{liCascadedAttentionGuided2020}.

Another strategy is to combine DL-based segmentation tasks with other tasks to improve performance using additional information, such as centreline tracking (CLT) \cite{mengTrackingSegmentationAirways2017, nadeemCTBasedAutomatedAlgorithm2021}. 

In addition, adversarial learning was also attempted, i.e., to constrain the segmentation network with discriminators. \cite{zhaoHighQualityRetinal2020, guoRetinalVesselSegmentation2020, parkMGANRetinalBlood2020, zhouRefinedEquilibriumGenerative2021}.


\textbf{{AirwayNet:}} Qin et al. \cite{qinAirwayNetVoxelConnectivityAware2019} present the AirwayNet for accurate airway segmentation that focuses on the voxel connectivity of airways. After modelling the voxel 3D connectivity using a 26-connected neighbour analysis, the binary segmentation task is converted into 26 tasks to determine whether the voxel is connected to its neighbours along a specific orientation. Therefore, the AirwayNet learns relationships between neighbouring airway voxels to distinguish the airway tree from the background. In addition, the coordinates of voxels and their distance from lung borders are fed into the AirwayNet as supplementary semantic information to take full advantage of context knowledge. The final airway tree is generated from the predicted connectivity map using fuzzy connectedness rules \cite{jin3DConvolutionalNeural2017}.

\textcolor{Black}{\textbf{{DeepVesselNet:}} Tetteh et al. \cite{qinAirwayNetVoxelConnectivityAware2019} present the DeepVesselNet based on FCN architecture for accurate 3-D angiographic feature extraction. First, they designed a 2-D orthogonal cross-hair filters to relieve computational burden. Second, a class balancing cross-entropy loss function is introduced to reduce the high false-positive rate. Finally, they generated a synthetic dataset and used this for transfer learning.}

\textbf{{Bifurcation point detection and RG (BPD-RG):}} Wang et al. \cite{wangAutomatedLabelingAirway2020} proposed an automatic airway tree labelling system using ResNets to detect bifurcation points. A hybrid of three different ResNets is constructed with cascaded residual blocks that enable the learning of large-scale spatial information. The three ResNets are trained to predict four critical bifurcation points of airway trees independently, and their predictions are fused to reduce the variance. The complete airway tree is then reconstructed using an adaptive RG approach, guided by the detected bifurcation points.

\textbf{{MFN and GNN (MFN-GNN):}} Selvan et al. \cite{selvanGraphRefinementBased2020} extracted airway trees from CT scans by first constructing a graph-based representation of the image data and then posing the tree extraction task as a graph refinement task using mean-field network (MFN) and GNN. In the first model, refinement is framed as an approximate Bayesian inference problem solved by mean-field approximation (MFA) \cite{jaakkolaImprovingMeanField1998, wainwrightGraphicalModelsExponential2007} to approximate the posterior density over various subgraphs. The optimal subgraph is represented by parameterized nodes and pair-wise probabilities. Performing the MFA updating iterations as layers in a feed-forward network enables parameter learning. The second GNN model \cite{kipfVariationalGraphAutoEncoders2016} is designed as an edge connection predictor that learns edge embeddings from an over-complete graph to produce an edge probability map. 

\textbf{{2D+3D NN:}} Zhao et al. \cite{zhaoBronchusSegmentationClassification2019} presented a 2-stage airway segmentation method based on a 2D+3D neural network (NN) and linear programming (LP) based tracking algorithm. The 2D and 3D NNs were developed to segment horizontal airways and tube-like vertical airways separately due to significant appearance differences. The 3D NN structure resembles the 3D UNet structure, and the 2D NN is a standard fully convolutional network (FCN) \cite{zhaoPyramidBasedFullyConvolutional2018}. The first stage targets the trachea and main bronchus, whereas the second stage targets the distal bronchus. The outputs from the two stages are then connected using an LP-based object tracking algorithm, which simultaneously filters out false positives. 

\textbf{{Hybrid UNet:}} A hybrid UNet-based approach for retinal vessel segmentation is proposed in \cite{yangHybridDeepSegmentation2021}, consisting of a multi-task segmentation network and a fusion network. Since the signal-to-noise ratio and contrast of thin vessels are lower than those of thick vessels, the segmentation network is built to segment thick and thin vessels separately from fundus images. The encoder extracts features corresponding to both thick and thin vessels and passes them to the two independent decoders. Afterwards, a fusion network is designed to fuse these two segmentation results to acquire the final blood vessel tree.


\textbf{{UNet with CE (UNet-CE):}} Meng et al. \cite{mengTrackingSegmentationAirways2017} developed an airway segmentation method that incorporates the 3D UNet into an adaptive volume of interest (VOI) based tracking scheme. The parent VOI is first assigned based on the coarsely segmented airway tree using the RG algorithm. The 3D extension of the UNet proposed in \cite{ronnebergerUNetConvolutionalNetworks2015} is employed to segment the airway region inside each VOI. Meanwhile, the centreline of the bronchi is retrieved by applying the gradient vector flow (GVF) magnitude and the tubular-likeness function based on the GVF. After detecting the branching point from the centreline, child VOI/VOIs can be assigned adaptively depending on the size and the travelling orientation of the bronchi. 

\textbf{{CAG-Net:}} Li et al. \cite{liCascadedAttentionGuided2020} presented a novel Cascaded Attention Guided Network (CAG-Net) for accurate retinal vascular segmentation. The CAG-Net consists of prediction and refinement modules, using Attention UNet++ (AUNet++) as the basic architecture. Because the simple 1×1 convolutional layer in \cite{zhouUNetNestedUNet2018} cannot fully fuse the decoder’s multi-scale information, attention-guided convolutional blocks (AC blocks) are employed instead to fuse the multi-scale features adaptively. In the prediction module, the AUNet++ generates an initial segmentation map that is then concatenated with the original image to create a new volume as the input of the refinement module. The final segment is formed by concatenating the outputs of the two modules in an FC layer. 

\textbf{{UNet with CE and LR (UNet-CE-LR):}} Nadeem et al. \cite{nadeemCTBasedAutomatedAlgorithm2021} integrated another tracking algorithm called freeze-and-grow (FG) with 3D UNet for airway segmentation. The FG algorithm is a traditional CT intensity-based segmentation approach that begins with conservative segmentation parameters and then iteratively captures finer details by freezing leakage-roots and relaxing parameters. In each iteration, the centreline and possible leakages are identified for propagation. A modified 3D UNet was applied to derive probability maps of the airway as inputs from the FG algorithm. Finally, a multi-task 3D UNet that classifies each pixel as airway, leak, or background is employed to remove distal leakages.

\textcolor{Black}{\textbf{{DD-CNN:}} Zhang et al. \cite{ZHANG2020162} employed a semi-supervised mixture probability model to fit the cerebrovascular intensity distribution from the sparse manual annotations and generate massive labeled pixels. They also proposed a dilated dense(DD) CNN which consists of DD blocks that use both high-level and low-level feature maps. The DD-CNN is trained on the previously generated labels.}

\textbf{GAN and dense UNet (GAN-DUNet):} In \cite{guoRetinalVesselSegmentation2020}, a dense UNet employing the Inception module is combined with GAN for accurate retinal vessel segmentation. First, the skip connections in a standard UNet are substituted with dense blocks to enable full fusing of features from shallow to deep layers, hence improving accuracy without increasing the network’s depth or width. The Inception module is then utilized to replace conventional convolutional layers to expand the receptive field and extract multi-scale vessel features from different-sized convolution kernels. Finally, a GAN is adopted in the training, where the dense UNet is considered as the generator and the loss value is backpropagated by a multilayer neural network as the discriminator.

\textbf{M-GAN:} Park et al. \cite{parkMGANRetinalBlood2020} presented a novel conditional GAN called M-GAN to perform precise retinal vessel segmentation by balancing losses of stacked deep FCNs. Based on a conditional GAN \cite{mirzaConditionalGenerativeAdversarial2014}, the M-GAN comprises a newly developed M-generator for more robust segmentation and a deeper M-discriminator for more efficient GAN model training. The M-generator consists of two stacked deep FCNs with short-term skip connections and long-term residual connections, as well as a multi-kernel pooling (MKP) block between two FCNs that supports scale-invariance of vessel segmentation of various sizes. In addition, the Lanczos resampling method \cite{duchonLanczosFilteringOne1979} is used to smooth out the segmented vessel branching and eliminate false.

\textbf{SEGAN:} Zhou et al. \cite{zhouRefinedEquilibriumGenerative2021} proposed a symmetric equilibrium GAN (SEGAN) with multi-scale features refine blocks (MSFRBs) and AMs to improve retinal vessel segmentation. SEGAN creates a symmetric adversarial architecture that eliminates the imbalance in generator-discriminator capabilities to drive the generator to produce more realistic and detailed images. Second, MSFRB is designed to optimize feature merging while retaining high-resolution and high-semantic information. In addition, AMs make the network focus on distinguishing features rather than irrelevant information. These improvements allow the proposed network to excel in extracting detailed information by maximizing the multi-scale feature representation.

\subsubsection{Training strategy based}

The training strategy of a model determines how effectively and efficiently it learns, and therefore also has a significant impact on its segmentation performance. Hence, some segmentation methods for treelike tubular structures focus on training strategies including transfer learning (TL) \cite{jiangRetinalBloodVessel2018} and cascaded training \cite{nazirOFFeNETOptimallyFused2020}.

\textbf{{FCN with TL (FCN-TL):}} In \cite{jiangRetinalBloodVessel2018}, a supervised retinal vessel segmentation method is proposed based on a pre-trained FCN \cite{shelhamerFullyConvolutionalNetworks2017}. The new training dataset was utilized to fine-tune the pre-trained FCN. Because the semantic segmentation result of retinal vessels is wider than the ground truth, and noises are added in slice merging and overlapping, additional unsupervised post-processing strategies using local Otsu’s method are applied. This paper proves the utility of transfer learning when using deep learning techniques in medical imaging.

\textbf{{OFF-eNET:}} Nazir et al. \cite{nazirOFFeNETOptimallyFused2020} trained a three-stage Optimally Fused Fully end-to-end Network (OFF-eNET) in a cascade fashion for 3D cerebral vessel segmentation. The first stage employs up-skip connections to improve information flow between the encoder and decoder and dilated convolution to preserve the spatial resolution feature map designed for thin vessels. In the second stage, residual mapping with the inception module is utilized for faster converging and learning richer representations. The cascaded training strategy is applied in the third stage to progressively achieve concrete segmentation results over three sub-stages (basic, complete, and enhanced) by utilizing transferred knowledge. During the cascaded training, the knowledge learned by the basic sub-stage is transferred to the complete sub-stage for initiation, and the enhanced sub-stage is fine-tuned based on the well-trained complete stage.

\subsubsection{Loss functions}
The commonly used loss functions are based on cross-entropy or Dice Similarity Coefficient (DSC). Since the treelike structures occupy only a small portion of the images due to their sparsity, the background pixels are usually dominant. Therefore, the global class-balancing weight is applied to the binary cross-entropy (BCE) loss function \cite{jin3DConvolutionalNeural2017,garcia-ucedajuarezAutomaticAirwaySegmentation2018,mengTrackingSegmentationAirways2017} as follows:

\begin{align}
\begin{split}
    \mathcal{L}_{wBCE}=&-w\sum\nolimits_{x=N_f}\log{p(x)}\\&-(1-w)\sum\nolimits_{x=N_b}\log{(1-p(x))},
\end{split}
\end{align}

where $p$ is the pixel-wise foreground probability. $N_f$ and $N_b$ are the number of foreground pixels and the number of background pixels, respectively. The weight $w$ is usually equal to $\frac{N_b}{N}$ where $N$ is the total number of pixels $x$.

As for the Dice loss function, a tolerance term $\epsilon$ is commonly introduced \cite{garcia-ucedajuarezAutomaticAirwaySegmentation2018,garcia-ucedajuarezJoint3DUNetGraph2019} to avoid dividing by zero in case there is no foreground pixel as defined below:

\begin{equation}
\mathcal{L}_{Dice}=-\frac{2\sum_{x=N}\log{p(x)g(x)}}{\sum_{x=N}\log{p(x)}+\sum_{x=N}\log{g(x)}+\epsilon },
\end{equation}

where $g$ is the foreground ground-truth.

Furthermore, \cite{qinLearningBronchioleSensitiveAirway2020,qinLearningTubuleSensitiveCNNs2021} combined the Dice loss with the Focal loss \cite{Lin2017} which reduces the contribution of easy examples in training and more focuses on the hard example, defined by:

\begin{align}
\begin{split}
    \mathcal{L}_{DiceFocal}=&-\frac{2\sum_{x=N}\log{p(x)g(x)}}{\sum_{x=N}\log{p(x)}+\sum_{x=N}\log{g(x)}+\epsilon }\\&-\frac{1}{N}\sum\nolimits_{x=N}{{(1-p_g(x))}^2\log{p_g(x)}},
\end{split}
\end{align}

where $p_g(x)=\ p(x)$ if $g(x)=1$. Otherwise, $p_g(x)=\ 1-p(x)$.

Wang et al. \cite{Wang2019Radial} proposed a novel loss function called radial distance loss that focuses on spatial consistency. 
\begin{equation}
\mathcal{L}_{RDL}= -\frac{1}{2}\sum_{k=0}^{1}\mathcal{W}_k\frac{2\sum_{i}^{N}p_{i,k}d_{i,k}}{\sum_{i}^{N}p_{i,k}^2+\sum_{i}^{N}d_{i,k}^2}
\end{equation}
where $p_i$ denotes the $i^{th}$ voxel predicted binary result, $d_i$ denotes the radial distance map which is defined as: 
\begin{equation}
\mathcal{D}= -\frac{1}{max(\mathcal{F})}\mathcal{F} + 1 
\end{equation}
where $\mathcal{F}$ denotes the Euclidean distance map from the ground truth centerline

In addition to the least square loss \cite{maoLeastSquaresGenerative2017} that is utilized as GAN loss and standard binary cross-entropy (BCE) loss, the generator in \cite{parkMGANRetinalBlood2020} adopts a newly designed false-negative (FN) loss function, defined as

\begin{equation}
    \mathcal{L}_{FN}=\frac{1}{N_f}\sum\nolimits_{x=N_f}{(1-p_{FN}(x))}^2,
\end{equation}

\noindent where 

\begin{equation}
    p_{FN}(x)=
    \begin{cases}
      1, & \text{if}\ p(x)\geq 0.5 \\
      p(x), & \text{otherwise}
    \end{cases}.
\end{equation}
The inclusion of the FN loss function has been shown to reduce the FNR.

\begin{table*}[!t]
\scriptsize
\centering
\caption{Deep learning based segmentation methods}\label{tabdl}
\begin{tabularx}{\textwidth}{lllllllc}
\hline
\textbf{Author} & \textbf{Year} & \textbf{Region} & \textbf{Modality} & \textbf{Method} & \textbf{\begin{tabular}[t]{@{}l@{}}Validation Dataset\\ (number of cases)\end{tabular}} & \textbf{Metrics} & \multicolumn{1}{l}{\textbf{Method Class}} \\
\hline
Charbonnier et al. \cite{charbonnierImprovingAirwaySegmentation2017} & 2016 & LA & CT & CNN & EXACT'09 (20) & \begin{tabular}[t]{@{}l@{}}\colorbox{fpr}{\textcolor{white}{FPR:}} 0.0101\\ \colorbox{tld}{\textcolor{white}{TLD:}} 0.518\end{tabular} & \multirow[t]{7}{*}{\begin{tabular}[t]{@{}c@{}}Classical\\ CNNs\end{tabular}} \\
Yun et al. \cite{yunImprovementFullyAutomated2018} & 2018 & LA & CT & CNN & EXACT'09 (20) & \begin{tabular}[t]{@{}l@{}}\textcolor{Black}{\colorbox{fpr}{\textcolor{white}{FPR:}} 0.0456±0.0373}\\ \textcolor{Black}{\colorbox{others}{\textcolor{white}{BC:}} 163.4±79.4 }\\ \textcolor{Black}{\colorbox{bd}{\textcolor{white}{BD:}} 0.657±0.131}\\\textcolor{Black}{ \colorbox{others}{\textcolor{white}{TL:}} 129.3±66.0}\\ \textcolor{Black}{\colorbox{tld}{\textcolor{white}{TLD:}} 0.601±0.119}\end{tabular} &  \\
Meyer et al. \cite{meyerDeepNeuralNetwork2017} & 2017 & RV & SLO CFP & Unet & \begin{tabular}[t]{@{}l@{}}IOSTAR (10)\\ \\ \\ \\ RC-SLO (40)\end{tabular} & \begin{tabular}[t]{@{}l@{}}\colorbox{acc}{\textcolor{white}{Acc:}} 0.9695\\ \colorbox{se}{\textcolor{white}{Se:}} 0.8038\\ \colorbox{sp}{\textcolor{white}{Sp:}} 0.9801 \\ \colorbox{auroc}{\textcolor{white}{AUROC:}} 0.9771\\ \colorbox{acc}{\textcolor{white}{Acc:}} 0.9623\\ \colorbox{se}{\textcolor{white}{Se:}} 0.8090\\ \colorbox{sp}{\textcolor{white}{Sp:}} 0.9794\\ \colorbox{auroc}{\textcolor{white}{AUROC:}} 0.9807\end{tabular} &  \\
Garcia-Uceda Juarez et al. \cite{garcia-ucedajuarezAutomaticAirwaySegmentation2018} & 2018 & LA & CT & UNet & CT scans (6) & \colorbox{dsc}{\textcolor{white}{DSC:}} 0.8 &  \\
Jin et al. \cite{jin3DConvolutionalNeural2017} & 2017 & LA & CT & UNet & EXACT'09 (20) & \colorbox{others}{\textcolor{white}{Comparative}} &  \\
Garcia-Uceda et al. \cite{garcia-ucedaAutomaticAirwaySegmentation2021} & 2021 & LA & CT & UNet & \begin{tabular}[t]{@{}l@{}}CF-CT (24)\\ DLCST (32)\\ EXACT'09 (20)\end{tabular} & \begin{tabular}[t]{@{}l@{}}\colorbox{dsc}{\textcolor{white}{DSC:}} 0.876 \\ \colorbox{dsc}{\textcolor{white}{DSC:}} 0.916\\ \colorbox{fpr}{\textcolor{white}{FPR:}} 0.0274\\ \colorbox{tld}{\textcolor{white}{TLD:}} 0.703\end{tabular} &  \\
Kitrungrotsakul et al. \cite{kitrungrotsakulVesselNetDeepConvolutional2019} & 2019 & LV & MRI & ResNet & IRCAD (20) & \begin{tabular}[t]{@{}l@{}}\colorbox{se}{\textcolor{white}{Se:}} 0.929\\ \colorbox{pre}{\textcolor{white}{Pre:}} 0.866\\ \colorbox{dsc}{\textcolor{white}{DSC:}} 0.903\\ \colorbox{others}{\textcolor{white}{VOE:}} 0.172\end{tabular} &  \\
Lin et al. \cite{linAutomaticRetinalVessel2019} & 2019 & RV & CFP & DSSRN & \begin{tabular}[t]{@{}l@{}}DRIVE (40)\\ \\ STARE (20)\\ \\ CHASE-DB1 (28)\end{tabular} & \begin{tabular}[t]{@{}l@{}}\colorbox{acc}{\textcolor{white}{Acc:}} 0.9536\\ \colorbox{se}{\textcolor{white}{Se:}} 0.7632 \\ \colorbox{acc}{\textcolor{white}{Acc:}} 0.9603\\ \colorbox{se}{\textcolor{white}{Se:}} 0.7423 \\ \colorbox{acc}{\textcolor{white}{Acc:}} 0.9587\\ \colorbox{se}{\textcolor{white}{Se:}} 0.7815\end{tabular} & \multirow[t]{8}{*}{\begin{tabular}[t]{@{}c@{}}Advanced\\ architectures\end{tabular}} \\
Garcia-Uceda Juarez et al. \cite{garcia-ucedajuarezJoint3DUNetGraph2019} & 2019 & LA & CT & UNet-GNN & LD CT scans (12) & \colorbox{dsc}{\textcolor{white}{DSC:}} 0.89 &  \\
Wang et al. \cite{Wang2019Radial} & 2019 & LA & CT & UNet-RCL & SD CT scans (38) & \begin{tabular}[t]{@{}l@{}}\textcolor{Black}{\colorbox{se}{\textcolor{white}{Se:}} 0.865±0.010}\\\textcolor{Black}{ \colorbox{dsc}{\textcolor{white}{DSC:}} 0.887±0.012}\\ \textcolor{Black}{\colorbox{others}{\textcolor{white}{CO:}} 0.766±0.060}\end{tabular} &  \\
Zheng et al. \cite{Zheng2020WingsNet} & 2020 & LA & CT & WingsNet & \begin{tabular}[t]{@{}l@{}}EXACT'09 and \\ LIDC-IDRI (20+70)\end{tabular} & \begin{tabular}[t]{@{}l@{}}\textcolor{Black}{\colorbox{pre}{\textcolor{white}{Pre:}} 0.914±0.045}\\\textcolor{Black}{ \colorbox{bd}{\textcolor{white}{BD:}} 0.887±0.079}\\ \textcolor{Black}{\colorbox{tld}{\textcolor{white}{TLD:}} 0.925±0.033}\end{tabular} &  \\
Qin et al. \cite{qinLearningBronchioleSensitiveAirway2020} & 2020 & LA & CT & UNet-AM & \begin{tabular}[t]{@{}l@{}}EXACT'09 and \\ LIDC-IDRI (20+70)\end{tabular} & \begin{tabular}[t]{@{}l@{}}\textcolor{Black}{\colorbox{se}{\textcolor{white}{Se:}} 0.936±0.050}\\ \textcolor{Black}{\colorbox{fpr}{\textcolor{white}{FPR:}} 0.035±0.014}\\ \textcolor{Black}{\colorbox{dsc}{\textcolor{white}{DSC:}} 0.925±0.020}\\\textcolor{Black}{ \colorbox{bd}{\textcolor{white}{BD:}} 0.962±0.058}\\\textcolor{Black}{ \colorbox{tld}{\textcolor{white}{TLD:}} 0.907±0.069}\end{tabular} &  \\
Qin et al. \cite{qinLearningTubuleSensitiveCNNs2021} & 2021 & \begin{tabular}[t]{@{}l@{}}LA\\ \\ \\ \\ \\ LV\end{tabular} & \begin{tabular}[t]{@{}l@{}}CT\\ \\ \\ \\ \\ CT\end{tabular} & UNet-AM & \begin{tabular}[t]{@{}l@{}}EXACT'09 and \\ LIDC-IDRI (20+70)\\ \\ \\ \\ CARVE14 (55)\end{tabular} & \begin{tabular}[t]{@{}l@{}}\colorbox{se}{\textcolor{white}{Se:}} 0.936\\ \colorbox{fpr}{\textcolor{white}{FPR:}} 0.035\\ \colorbox{dsc}{\textcolor{white}{DSC:}} 0.925\\ \colorbox{bd}{\textcolor{white}{BD:}} 0.962\\ \colorbox{tld}{\textcolor{white}{TLD:}} 0.907\\ \colorbox{acc}{\textcolor{white}{Acc:}} 0.972\\ \colorbox{se}{\textcolor{white}{Se:}} 0.971\\ \colorbox{fpr}{\textcolor{white}{FPR:}} 0.015 \\ \colorbox{dsc}{\textcolor{white}{DSC:}} 0.972\end{tabular} &  \\

\textcolor{Black}{Fan et al. \cite{Fan2020} }& 2020 & BV & TOF-MRA & UNet-HMRF & TOF-MRA (100) & \begin{tabular}[t]{@{}l@{}}\colorbox{acc}{\textcolor{white}{Acc:}} 0.9983 \\\colorbox{se}{\textcolor{white}{Se:}}0.7620\\ \colorbox{sp}{\textcolor{white}{Sp:}}0.9993\\ \colorbox{pre}{\textcolor{white}{Pre:}}0.8405 \\\colorbox{dsc}{\textcolor{white}{DSC:}} 0.7941\end{tabular} &  \\

Cheng et al. \cite{chengSegmentationAirwayTree2021} & 2021 & LA & CT & TACNet & \begin{tabular}[t]{@{}l@{}}CT scans (100)\\ \\ \\ \\ \\ \\ \\ EXACT'09 (20)\end{tabular} & \begin{tabular}[t]{@{}l@{}}\colorbox{fpr}{\textcolor{white}{FPR:}} 0.0144\\ \colorbox{dsc}{\textcolor{white}{DSC:}} 0.9032\\ \colorbox{others}{\textcolor{white}{BC:}} 215.7\\ \colorbox{bd}{\textcolor{white}{BD:}} 0.8663\\ \colorbox{others}{\textcolor{white}{TL:}} 394.9 cm\\ \colorbox{others}{\textcolor{white}{LV:}} 8207.9 $\text{mm}^3$\\ \colorbox{fpr}{\textcolor{white}{FPR:}} 0.1429\\ \colorbox{others}{\textcolor{white}{BC:}} 213.7\\ \colorbox{bd}{\textcolor{white}{BD:}} 0.849\\ \colorbox{others}{\textcolor{white}{TL:}} 186.9 cm\\ \colorbox{tld}{\textcolor{white}{TLD:}} 0.845\\ \colorbox{others}{\textcolor{white}{LC:}} 160.3\\ \colorbox{others}{\textcolor{white}{LV:}} 4396.7 $\text{mm}^3$\end{tabular} &  \\
Tang et al. \cite{tangConstructionVerificationRetinal2021} & 2021 & RV & CFP & BPNN & \begin{tabular}[t]{@{}l@{}}DRIVE (40)\\ \\ \\ STARE (20)\end{tabular} & \begin{tabular}[t]{@{}l@{}}\colorbox{acc}{\textcolor{white}{Acc:}} 0.9477\\ \colorbox{se}{\textcolor{white}{Se:}} 0.7338\\ \colorbox{sp}{\textcolor{white}{Sp:}} 0.9730\\ \colorbox{acc}{\textcolor{white}{Acc:}} 0.9498\\ \colorbox{se}{\textcolor{white}{Se:}} 0.7518\\ \colorbox{sp}{\textcolor{white}{Sp:}} 0.9734\end{tabular} &  \\
Zhou et al. \cite{zhouAutomaticAirwayTree2021} & 2021 & LA & CT & MFA-Net & CT scans (150) & \begin{tabular}[t]{@{}l@{}}\colorbox{se}{\textcolor{white}{Se:}} 0.7931\\ \colorbox{dsc}{\textcolor{white}{DSC:}} 0.8618\end{tabular} &  \\
\hline
\\
\\
\\
\end{tabularx}

\end{table*}

\begin{table*}[!t]
\scriptsize
\centering
\begin{tabularx}{\textwidth}{lllllllc}
\hline
\textbf{Author} & \textbf{Year} & \textbf{Region} & \textbf{Modality} & \textbf{Method} & \textbf{\begin{tabular}[t]{@{}l@{}}Validation Dataset\\ (number of cases)\end{tabular}} & \textbf{Metrics} & \multicolumn{1}{l}{\textbf{Method Class}} \\
\hline
Qin et al. \cite{qinAirwayNetVoxelConnectivityAware2019}~~~~~~~~~~~~~~~~~~~~ & 2019 & LA & CT & AirwayNet & CT scans (10) & \begin{tabular}[t]{@{}l@{}}\textcolor{Black}{\colorbox{se}{\textcolor{white}{Se:}} 0.847±0,049}\\\textcolor{Black}{ \colorbox{fpr}{\textcolor{white}{FPR:}} 0.011±0.008}\\\textcolor{Black}{ \colorbox{dsc}{\textcolor{white}{DSC:}} 0.902±0.028}\end{tabular} & \multirow[t]{10}{*}{Task-based/Hybrid} \\

\textcolor{Black}{Tetteh et al. \cite{Tetteh2020}} & 2020 & BV & MRA & DeepVesselNet & TOF-MRA (40) & \begin{tabular}[t]{@{}l@{}}\colorbox{se}{\textcolor{white}{Se:}} 0.8693\\ \colorbox{pre}{\textcolor{white}{Pre:}} 8644\\\colorbox{dsc}{\textcolor{white}{DSC:}} 0.8668\end{tabular} & \\

Wang et al. \cite{wangAutomatedLabelingAirway2020} & 2020 & LA & CT & BPD-RG & LUNA16 (345) & \begin{tabular}[t]{@{}l@{}}\colorbox{acc}{\textcolor{white}{Acc:}} 0.9785\\ \colorbox{dsc}{\textcolor{white}{DSC:}} 0.875\end{tabular} &  \\
Selvan et al.* \cite{selvanGraphRefinementBased2020} & 2020 & LA & CT & MFN-GNN & LD CT scans (32) & \begin{tabular}[t]{@{}l@{}}\textcolor{Black}{\colorbox{fpr}{\textcolor{white}{FPR:}} 0.078±0.046}\\\textcolor{Black}{ \colorbox{dsc}{\textcolor{white}{DSC:}} 0.848±0.033}\\ \textcolor{Black}{\colorbox{tld}{\textcolor{white}{TLD:}} 0.819±0.073}\end{tabular} &  \\
Zhao et al. \cite{zhaoBronchusSegmentationClassification2019} & 2019 & LA & CT & 2D+3D NN & CT scans (22) & \colorbox{dsc}{\textcolor{white}{DSC:}} 0.94 &  \\
Yang et al. \cite{yangHybridDeepSegmentation2021} & 2021 & RV & CFP & Hybrid UNet & \begin{tabular}[t]{@{}l@{}}DRIVE (40)\\ \\ \\ \\ STARE (20)\\ \\ \\ \\ CHASE-DB1 (28)\end{tabular} & \begin{tabular}[t]{@{}l@{}}\colorbox{acc}{\textcolor{white}{Acc:}} 0.956\\ \colorbox{se}{\textcolor{white}{Se:}} 0.813\\ \colorbox{sp}{\textcolor{white}{Sp:}} 0.976\\ \colorbox{dsc}{\textcolor{white}{DSC:}} 0.819\\ \colorbox{acc}{\textcolor{white}{Acc:}} 0.963 \\ \colorbox{se}{\textcolor{white}{Se:}} 0.976\\ \colorbox{sp}{\textcolor{white}{Sp:}} .982\\ \colorbox{dsc}{\textcolor{white}{DSC:}} 0.8155\\ \colorbox{acc}{\textcolor{white}{Acc:}} 0.9632\\ \colorbox{se}{\textcolor{white}{Se:}} 0.817\\ \colorbox{sp}{\textcolor{white}{Sp:}} 0.9776 \\ \colorbox{dsc}{\textcolor{white}{DSC:}} 0.7997\end{tabular} &  \\
Meng et al. \cite{mengTrackingSegmentationAirways2017} & 2017 & LA & CT & UNet-CE & SD CT scans (50) & \begin{tabular}[t]{@{}l@{}}\colorbox{se}{\textcolor{white}{Se:}} 0.796\\ \colorbox{fpr}{\textcolor{white}{FPR:}} 0.001\\ \colorbox{dsc}{\textcolor{white}{DSC:}} 0.866\end{tabular} &  \\
Li et al.* \cite{liCascadedAttentionGuided2020} & 2020 & RV & CFP & CAG-Net & \begin{tabular}[t]{@{}l@{}}DRIVE (40)\\ \\ \\ \\ \\ STARE (20)\\ \\ \\ \\ \\ CHASE-DB1 (28)\end{tabular} & \begin{tabular}[t]{@{}l@{}}\colorbox{acc}{\textcolor{white}{Acc:}} 0.9700\\ \colorbox{se}{\textcolor{white}{Se:}} 0.8397\\ \colorbox{sp}{\textcolor{white}{Sp:}} 0.9827\\ \colorbox{dsc}{\textcolor{white}{DSC:}} 0.8298\\ \colorbox{auroc}{\textcolor{white}{AUROC:}} 0.9867 \\ \colorbox{acc}{\textcolor{white}{Acc:}} 0.9791\\ \colorbox{se}{\textcolor{white}{Se:}} 0.8217\\ \colorbox{sp}{\textcolor{white}{Sp:}} 0.9901\\ \colorbox{dsc}{\textcolor{white}{DSC:}} 0.8254\\ \colorbox{auroc}{\textcolor{white}{AUROC:}} 0.9894 \\ \colorbox{acc}{\textcolor{white}{Acc:}} 0.9768\\ \colorbox{se}{\textcolor{white}{Se:}} 0.8520\\ \colorbox{sp}{\textcolor{white}{Sp:}} 0.9853\\ \colorbox{dsc}{\textcolor{white}{DSC:}} 0.8227\\ \colorbox{auroc}{\textcolor{white}{AUROC:}} 0.9880\end{tabular} &  \\
Nadeem et al. \cite{nadeemCTBasedAutomatedAlgorithm2021} & 2021 & LA & CT & UNet-CE-LR & \begin{tabular}[t]{@{}l@{}}SD CT scans (120)\\ LD CT scans (40)\end{tabular} & \begin{tabular}[t]{@{}l@{}}\colorbox{bd}{\textcolor{white}{BD:}} 0.952\\ \colorbox{bd}{\textcolor{white}{BD:}} 0.987\end{tabular} &  \\
\textcolor{Black}{Zhang et al.} \cite{ZHANG2020162} & 2019 & BV & TOF-MRA & DD-CNN & MIDAS21 & \begin{tabular}[t]{@{}l@{}}\colorbox{acc}{\textcolor{white}{Acc:}}0.9756\\\colorbox{se}{\textcolor{white}{Se:}} 0.9622\\ \colorbox{dsc}{\textcolor{white}{DSC:}} 0.9747\end{tabular} &  \\
Zhao et al. \cite{zhaoHighQualityRetinal2020} & 2020 & RV & CFP & GAN & \begin{tabular}[t]{@{}l@{}}DRIVE (40)\\ \\ \\ \\ \\ STARE (20)\end{tabular} & \begin{tabular}[t]{@{}l@{}}\colorbox{acc}{\textcolor{white}{Acc:}} 0.9563\\ \colorbox{se}{\textcolor{white}{Se:}} 0.8390\\ \colorbox{sp}{\textcolor{white}{Sp:}} 0.9736\\ \colorbox{dsc}{\textcolor{white}{DSC:}} 0.8299 \\ \colorbox{auroc}{\textcolor{white}{AUROC:}} 0.9812\\ \colorbox{acc}{\textcolor{white}{Acc:}} 0.9684\\ \colorbox{se}{\textcolor{white}{Se:}} 0.8390\\ \colorbox{sp}{\textcolor{white}{Sp:}} 0.9736\\ \colorbox{dsc}{\textcolor{white}{DSC:}} 0.8465\\ \colorbox{auroc}{\textcolor{white}{AUROC:}} 0.9853\end{tabular} &  \\
Park et al.* \cite{parkMGANRetinalBlood2020} & 2020 & RV & CFP & M-GAN & \begin{tabular}[t]{@{}l@{}}DRIVE (40)\\ \\ \\ \\ \\ \\ \\ \\ STARE (20)\\ \\ \\ \\ \\ \\ \\ \\ CHASE-DB1 (28)\\ \\ \\ \\ HRF (45)\end{tabular} & \begin{tabular}[t]{@{}l@{}}\colorbox{acc}{\textcolor{white}{Acc:}} 0.9706\\ \colorbox{se}{\textcolor{white}{Se:}} 0.8346\\ \colorbox{sp}{\textcolor{white}{Sp:}} 0.9836\\ \colorbox{pre}{\textcolor{white}{Pre:}} 0.8302 \\ \colorbox{dsc}{\textcolor{white}{DSC:}} 0.8324\\ \colorbox{others}{\textcolor{white}{IoU:}} 0.7129\\ \colorbox{others}{\textcolor{white}{MCC:}} 0.8163\\ \colorbox{auroc}{\textcolor{white}{AUROC:}} 0.9868\\ \colorbox{acc}{\textcolor{white}{Acc:}} 0.9876\\ \colorbox{se}{\textcolor{white}{Se:}} 0.8324\\ \colorbox{sp}{\textcolor{white}{Sp:}} 0.9938\\ \colorbox{pre}{\textcolor{white}{Pre:}} 0.8417 \\ \colorbox{dsc}{\textcolor{white}{DSC:}} 0.8370\\ \colorbox{others}{\textcolor{white}{IoU:}} 0.7198\\ \colorbox{others}{\textcolor{white}{MCC:}} 0.8306\\ \colorbox{auroc}{\textcolor{white}{AUROC:}} 0.9873\\ \colorbox{acc}{\textcolor{white}{Acc:}} 0.9736\\ \colorbox{dsc}{\textcolor{white}{DSC:}} 0.8110\\ \colorbox{others}{\textcolor{white}{MCC:}} 0.7979\\ \colorbox{auroc}{\textcolor{white}{AUROC:}} 0.9859\\ \colorbox{acc}{\textcolor{white}{Acc:}} 0.9761\\ \colorbox{dsc}{\textcolor{white}{DSC:}} 0.7972\\ \colorbox{others}{\textcolor{white}{MCC:}} 0.7845\\ \colorbox{auroc}{\textcolor{white}{AUROC:}} 0.9852\end{tabular} & \\
\hline
\end{tabularx}

\end{table*}

\begin{table*}[!t]
\scriptsize
\centering
\begin{tabularx}{\textwidth}{lllllllc}
\hline
\textbf{Author} & \textbf{Year} & \textbf{Region} & \textbf{Modality} & \textbf{Method} & \textbf{\begin{tabular}[t]{@{}l@{}}Validation Dataset\\ (number of cases)\end{tabular}} & \textbf{Metrics} & \multicolumn{1}{l}{\textbf{Method Class}} \\
\hline
Guo et al.* \cite{guoRetinalVesselSegmentation2020} & 2020 & RV & CFP & GAN-DUNet & DRIVE (40) & \begin{tabular}[t]{@{}l@{}}\colorbox{acc}{\textcolor{white}{Acc:}} 0.9542\\ \colorbox{se}{\textcolor{white}{Se:}} 0.8283 \\ \colorbox{sp}{\textcolor{white}{Sp:}} 0.9726\\ \colorbox{dsc}{\textcolor{white}{DSC:}} 0.8215\\ \colorbox{auroc}{\textcolor{white}{AUROC:}} 0.9772\\ \colorbox{others}{\textcolor{white}{AUPR:}} 0.9058\end{tabular} & \multirow[t]{2}{*}{Task-based/Hybrid}\\
Zhou et al.* \cite{zhouRefinedEquilibriumGenerative2021}~~~~~~~~~~~~~~~~~~ & 2021 & RV & CFP & SEGAN & \begin{tabular}[t]{@{}l@{}}DRIVE (40)\\ \\ \\ \\ \\ \\ \\ STARE (20)\\ \\ \\ \\ \\ \\ \\ CHASE-DB1 (28)\\ \\ \\ \\ \\ \\ \\ HRF (45)\end{tabular} & \begin{tabular}[t]{@{}l@{}}\colorbox{acc}{\textcolor{white}{Acc:}} 0.9563\\ \colorbox{se}{\textcolor{white}{Se:}} 0.8294\\ \colorbox{sp}{\textcolor{white}{Sp:}} 0.9812 \\ \colorbox{pre}{\textcolor{white}{Pre:}} 0.8397\\ \colorbox{dsc}{\textcolor{white}{DSC:}} 0.8345\\ \colorbox{others}{\textcolor{white}{G:}} 0.9021\\ \colorbox{auroc}{\textcolor{white}{AUROC:}} 0.9830\\ \colorbox{acc}{\textcolor{white}{Acc:}} 0.9671\\ \colorbox{se}{\textcolor{white}{Se:}} 0.8812\\ \colorbox{sp}{\textcolor{white}{Sp:}} 0.9781\\ \colorbox{pre}{\textcolor{white}{Pre:}} 0.7952\\ \colorbox{dsc}{\textcolor{white}{DSC:}} 0.8359\\ \colorbox{others}{\textcolor{white}{G:}} 0.9283\\ \colorbox{auroc}{\textcolor{white}{AUROC:}} 0.9863\\ \colorbox{acc}{\textcolor{white}{Acc:}} 0.9630\\ \colorbox{se}{\textcolor{white}{Se:}} 0.8435\\ \colorbox{sp}{\textcolor{white}{Sp:}} 0.9783\\ \colorbox{pre}{\textcolor{white}{Pre:}} 0.8013  \\ \colorbox{dsc}{\textcolor{white}{DSC:}} 0.8218\\ \colorbox{others}{\textcolor{white}{G:}} 0.9083\\ \colorbox{auroc}{\textcolor{white}{AUROC:}} 0.9872\\ \colorbox{acc}{\textcolor{white}{Acc:}} 0.9559\\ \colorbox{se}{\textcolor{white}{Se:}} 0.8310\\ \colorbox{sp}{\textcolor{white}{Sp:}} 0.9730\\ \colorbox{pre}{\textcolor{white}{Pre:}} 0.8115\\ \colorbox{dsc}{\textcolor{white}{DSC:}} 0.8211\\ \colorbox{others}{\textcolor{white}{G:}} 0.8992\\ \colorbox{auroc}{\textcolor{white}{AUROC:}} 0.9693\end{tabular} & \\
Jiang et al. \cite{jiangRetinalBloodVessel2018} & 2018 & RV & CFP & FCN-TL & \begin{tabular}[t]{@{}l@{}}DRIVE (40)\\ \\ \\ \\ STARE (20)\\ \\ \\ \\ CHASE-DB1 (28)\\ \\ \\ \\ HRF (45)\end{tabular} & \begin{tabular}[t]{@{}l@{}}\textcolor{Black}{\colorbox{acc}{\textcolor{white}{Acc:}} 0.9593±0.0073}\\ \textcolor{Black}{\colorbox{se}{\textcolor{white}{Se:}} 0.7121±0.0548} \\ \textcolor{Black}{\colorbox{sp}{\textcolor{white}{Sp:}} 0.9832±0.0069}\\ \colorbox{auroc}{\textcolor{white}{AUROC:}} 0.9580\\ \textcolor{Black}{\colorbox{acc}{\textcolor{white}{Acc:}} 0.9653±0.0107}\\\textcolor{Black}{ \colorbox{se}{\textcolor{white}{Se:}} 0.7820±0.1085}\\ \textcolor{Black}{\colorbox{sp}{\textcolor{white}{Sp:}} 0.9798±0.0117}\\ \colorbox{auroc}{\textcolor{white}{AUROC:}} 0.9857\\ \textcolor{Black}{\colorbox{acc}{\textcolor{white}{Acc:}} 0.9591±0.0065}\\ \textcolor{Black}{\colorbox{se}{\textcolor{white}{Se:}} 0.7217±0.0821}\\ \textcolor{Black}{\colorbox{sp}{\textcolor{white}{Sp:}} 0.9770±0.0037}\\ \colorbox{auroc}{\textcolor{white}{AUROC:}} 0.9580\\ \textcolor{Black}{\colorbox{acc}{\textcolor{white}{Acc:}} 0.9662±0.0052}\\ \textcolor{Black}{\colorbox{se}{\textcolor{white}{Se:}} 0.7686±0.0378}\\ \textcolor{Black}{\colorbox{sp}{\textcolor{white}{Sp:}} 0.9826±0.0057}\\ \colorbox{auroc}{\textcolor{white}{AUROC:}} 0.9770\end{tabular} & \multirow[t]{2}{*}{\begin{tabular}[t]{@{}c@{}}Training strategy\\ based\end{tabular}} \\
Nazir et al. \cite{nazirOFFeNETOptimallyFused2020} & 2020 & BV & CTA & OFF-eNET & CTA scans (70) & \begin{tabular}[t]{@{}l@{}}\textcolor{Black}{\colorbox{pre}{\textcolor{white}{Pre:}} 0.8956±0.0020}\\ \textcolor{Black}{\colorbox{dsc}{\textcolor{white}{DSC:}} 0.9075±0.0030}\\ \textcolor{Black}{\colorbox{others}{\textcolor{white}{HD:}} 5.01±1.05 mm}\\ \textcolor{Black}{\colorbox{others}{\textcolor{white}{AVD:}} 0.8986±0.0010}\end{tabular} & 
\\
\hline
\end{tabularx}
\vspace{2ex}

\begin{minipage}{\linewidth}\small
The regions include the retinal vessel (RV), lung airways (LA), brain vessel (BV), coronary vessel (CV), and thoracic aorta (TA). All unnamed datasets are private and may have specific modalities such as phase-contrast (PC), low-dose (LD), standard-dose (SD), or time-of-flight (TOF). *: methods that also use advanced architectures. All metrics are shown as mean ± standard deviation or mean only.
\end{minipage}
\end{table*}

\subsection{Preprocessing}
Preprocessing techniques are commonly employed as the first steps of segmentation methods, which involve intensity normalization, noise reduction, and contrast enhancement. Different combinations of these techniques must be used according to variable imaging modalities, conditions, and targets. 

The intensity normalization can reduce the variance of the image data caused by different imaging conditions and improve the computational efficiency using uniform intensity scales. Many segmentation methods preprocessed the images to a standard intensity scale using uniformity transformations. For deep learning methods, rescaling intensities also promotes a faster convergence in the optimization stage. Besides, spacing resampling is often adopted to standardize the 3D image data. 

After data normalization, outliers and noise can be eliminated using the intensity histogram or histogram-based function such as LogSig \cite{cinsdikiciDetectionBloodVessels2009}. Filter-based techniques such as anisotropic diffusion \cite{peronaScalespaceEdgeDetection1990} are also applied by some methods to reduce the noise. As for the image with nonuniform lighting conditions, filters calculating the local mean can be used for brightness correction \cite{hassanienRetinalBloodVessel2015}. 

Many techniques are available for tubular structure enhancement and are usually followed by data re-normalization. Nonlinear intensity transformations such as squaring intensity value \cite{roychowdhuryBloodVesselSegmentation2014} and gamma enhancement \cite{cinsdikiciDetectionBloodVessels2009} can directly improve the contrast. Mathematical morphology operations, including Top-hot transform, can extract small tubular structures using structuring elements that fit the largest diameter \cite{roychowdhuryBloodVesselSegmentation2014}. Using a filter like the unsharp mask filter can significantly reduce the noise and enhance the walls of tubular structures \cite{mengAutomaticSegmentationAirway2017}. For Coloured Fundus Photography (CFP) images, the green channel shows the highest contrast between vessel and background \cite{zanaSegmentationVessellikePatterns2001}. Hence, most methods only adopt and convert the green channel into greyscale. However, matching histograms of the green channel and the red channel can further enhance this contrast \cite{kandeUnsupervisedFuzzyBased2010}.

Furthermore, most deep learning methods adopt typical data augmentation techniques in order to manually enlarge the training dataset to improve its robustness, including rotation, mirroring, scaling, cropping, rigid transformations, and elastic deformations.

\subsection{Postprocessing}
A postprocessing step may be applied after the segmentation to rejoin disconnected segments or to remove small, segmented regions usually corresponding to noise or abnormalities in the image.

The connected component analysis (CCA) \cite{fiorioTwoLinearTime1996} is a common postprocessing technique to isolate individual components of the segmentation output using connected neighbourhood and label propagation. In order to filter unwanted and disconnected small segments under a threshold, some methods define a specific number of pixels \cite{cinsdikiciDetectionBloodVessels2009,kandeUnsupervisedFuzzyBased2010,chengSegmentationAirwayTree2021,garcia-ucedaAutomaticAirwaySegmentation2021} and some use a thinness ratio \cite{hassanienRetinalBloodVessel2015}. Besides, some methods have developed new algorithms to refine the output using information from neighbour pixels \cite{kitrungrotsakulVesselNetDeepConvolutional2019}. Moreover, the morphological reconstruction is also adopted in the postprocessing to refine the output \cite{camaranetoUnsupervisedCoarsetofineAlgorithm2017}. 

For the CNN-based method, the border effects due to zero-paddings make the output less reliable toward the boundaries \cite{garcia-ucedaAutomaticAirwaySegmentation2021}. Hence, a quadratic polynomial decrease towards the borders or a Gaussian-distributed weight map can be applied to the probability map, accompanied by an overlapping slide-window sampling strategy. The final segmentation can be obtained by placing together all patches and averaging the overlapped regions.
\subsection{Open-source Implementations}
\textcolor{Black}{
To promote the usage of the state-of-the-art (SOTA) models and help future researchers generate comparable segmentation results, we recommend some open-source implementations of aforementioned algorithms here.}
\begin{itemize}
    \item MONAI: Project MONAI was originally started by NVIDIA and King's College London. It is a freely available, PyTorch-based deep learning framework for healthcare imaging with comprehensive tutorials. The framework provides the user with SOTA, end-to-end training workflows and many other features such as interactive data labelling. More information can be found here: \url{https://monai.io/index.html}
    \item Alleviating Class-wise Gradient Imbalance for Pulmonary Airway Segmentation [\url{https://github.com/haozheng-sjtu/3d-airway-segmentation}] \cite{Zheng2020WingsNet}
    \item Automatic airway segmentation from computed tomography using robust and efficient 3-D convolutional neural networks [\url{https://github.com/antonioguj/bronchinet}]  \cite{garcia-ucedaAutomaticAirwaySegmentation2021}
    \item Lung-Vessel-Segmentation-Using-Graph-cuts [\url{https://github.com/Zhiwei-Zhai/Lung-Vessel-Segmentation-Using-Graph-cuts}] \cite{Zhai16lungvessel}

\end{itemize}
\section{Dataset}
To compare the performance of different segmentation models on the same scale, several datasets were made public and summarized in the table below. 
\subsection{Airways}
Currently, there is a very limited number of public datasets for airway segmentation which hinders the development in this area of research. 
To our best, we summarized four commonly seen datasets in Table \ref{table:2}. EXACT'09 \cite{loExtractionAirwaysCT2012} dataset consists of 40  volumetric chest CT scans obtained using different acquisition protocols. The dataset has large intra-variation because it ranges from clinical dose to ultra-low dose scans, from healthy volunteers to patients with severe lung disease, and from full inspiration to full expiration. 
LIDC-IDRI dataset \cite{LIDC} was created by seven academic centres and eight medical imaging companies, containing 1018 lung cancer screening thoracic low-dose CT scans. It is worth noting that the original dataset does not contain airway annotations. Qin et al. \cite{qinAirwayNetSESimpleYetEffectiveApproach2020} annotated the airway of 40 CT scans from LIDC-IDRI and can be download from \href{https://geronsushi.github.io/lung.html}{here}
COPDGene study contains 400 CT scans taken from subjects who had a minimum of 10 pack-year smoking history. Volunteers suffered from a range of COPD. Detailed inclusion and exclusion criteria and acquisition protocols can be found in the publication. 
KOLD dataset contains 477 volumetric lung CT scans from patients with chronic obstructive pulmonary disease (COPD), asthma, or other unclassified obstructive lung diseases. Detailed inclusion criteria can be found in \cite{Lee2008}\cite{Chae2010}.

\begin{table*}[!t]
\begin{center}
\caption{Airway Dataset}
\label{table:2}
{
\begin{tabularx}{\textwidth}{XXXXXXX}
\toprule
Name & Imaging Modality & Number of Scans & Image Dimension(h*w) & Year Published & Publicly Available & Link\\
\midrule
DLCST & CT & 32 & 512*512 & 2009 & No & \href{https://doi.org/10.1097/JTO.0b013e3181a0d98f}{Publication}\\

COPDGene & CT & 400 & 512*512 & 2010 & No & \href{http://www.copdgene.org/imaging\#Diagnosis}{Website} \href{https://www.tandfonline.com/doi/full/10.3109/15412550903499522}{Publication}\\

EXACT'09 & CT & 40  & 512*512 & 2012 & Yes & \href{http://image.diku.dk/exact/}{Website} \href{https://ieeexplore.ieee.org/document/6249784}{Publication} \\

LIDC-IDRI & CT & 1018 & 512*512 & 2015 & Yes & \href{https://geronsushi.github.io/lung.html}{Website} \href{https://aapm.onlinelibrary.wiley.com/doi/abs/10.1118/1.3528204}{Publication} \\

CF-CT & CT & 24 & 512*512 & 2017 & No & \href{https://doi.org/10.1007/s00330-017-4819-7}{Publication}\\

KOLD & CT & 65 & 512*512 & 2019 & No & \href{https://www.e-trd.org/journal/view.php?doi=10.4046/trd.2014.76.4.169}{Publication}\\

Subset of LUNA16 & CT & 345 & 512*512 & 2020 & No & \href{https://doi.org/10.1007/s11517-020-02184-y}{Publication}\\
\bottomrule
\end{tabularx}
}
\end{center}
\end{table*}

\subsection{Blood Vessels}
\subsubsection{Retinal blood vessels}
Fundus imaging refers to the projection of 3-D retinal tissues onto a 2-D imaging plane through reflected light, which encompasses a wide range of modalities. Detailed review on retinal imaging and fundus photography has been carefully done by \cite{Abramoff2010} and  \cite{Panwar2016}
To stimulate the development of more and more sophisticated methods of fundus image analysis, many publicly available datasets with expert annotation for vessel segmentation have been established which greatly stimulates the community of research. Table \ref{table:3} summarizes brief information of these datasets.

\begin{table*}[!t]
\begin{center}
\caption{Retinal Blood Vessel Dataset}
\label{table:3}
{
\begin{tabularx}{\textwidth}{XXXXXXX}
\toprule
Name & Imaging Modality & Number of Scans & Image Dimension(h*w) & Year Published & Publicly Available & Link\\
\midrule
STARE & CFO & 20 & 605*700 & 2000 & Yes & \href{https://cecas.clemson.edu/~ahoover/stare/probing/index.html}{Website} \href{https://ieeexplore.ieee.org/document/845178}{Publication} \\
DRIVE & CFO & 40 & 565*584 & 2004 & Yes & \href{https://drive.grand-challenge.org/}{Website} \href{https://ieeexplore.ieee.org/document/1282003}{Publication} \\
ARIA & CFO & 143 & 768*576 & 2006 & Yes & \href{https://www.researchgate.net/post/How_can_I_find_the_ARIA_Automatic_Retinal_Image_Analysis_Dataset}{Website} \href{https://doi.org/10.1016/j.jfranklin.2008.04.009}{Publication}\\
DIARETDB1 & CFO & 89 & 1500*1152 & 2007 & Yes & \href{https://www.it.lut.fi/project/imageret/diaretdb1/}{Website} \href{http://www.bmva.org/bmvc/2007/papers/paper-60.html}{Publication}\\
REVIEW & CFO & 16 & N/A & 2008 & Yes & \href{http://reviewdb.lincoln.ac.uk/}{Website} \href{https://ieeexplore.ieee.org/document/4649647}{Publication}\\
CHASE-DB1 & CFO & 28 & 999*960 & 2009 & Yes & \href{https://blogs.kingston.ac.uk/retinal/chasedb1/}{Website} \href{https://iovs.arvojournals.org/article.aspx?articleid=2126623}{Publication}\\
ROC & CFO & 100 & 1389*1383 & 2010 & Yes & \href{http://webeye.ophth.uiowa.edu/ROC/}{Website} \href{https://ieeexplore.ieee.org/document/5282586}{Publication}\\
VICAVR & CFO & 58 & 768*584 & 2010 & Yes & \href{http://www.varpa.es/research/ophtalmology.html\#vicavr}{Website}\\
HRF & CFO & 45 & 3504*2336 & 2013 & Yes &\href{https://www5.cs.fau.de/research/data/fundus-images/}{Website} \href{https://digital-library.theiet.org/content/journals/10.1049/iet-ipr.2012.0455}{Publication}\\
Messidor  & CFO & 1748 & 1440*960 & 2014 & Yes & \href{https://www.adcis.net/en/third-party/messidor2/}{Website} \href{https://www.ias-iss.org/ojs/IAS/article/view/1155/959}{Publication}\\
RC-SLO & SLO CFO & 40 & 360*320 & 2015 & Yes & \href{http://www.retinacheck.org/datasets}{Website} \href{https://ieeexplore.ieee.org/document/7530915}{Publication}\\
IOSTAR & SLO CFO & 30 & 1024*1024 & 2016 & No & \href{https://link.springer.com/chapter/10.1007/978-3-319-20801-5_35}{Publication}\\
PREVENT & OCTA & 11 & 91*91 & 2020 & Yes & \href{https://datashare.ed.ac.uk/handle/10283/3528}{Website} \href{https://www.ncbi.nlm.nih.gov/pmc/articles/PMC7718823/}{Publication}\\
\bottomrule
\end{tabularx}
}
\end{center}
\end{table*}

\subsubsection{Others}
Blood vessels have vital roles in many other anatomical regions as well. The segmentation of hepatic vessels, brain vessels and coronary vessels is often seen in the area of research and has great implications in many clinical fields such as oncology and surgery. Table \ref{table:4} summarizes commonly used datasets.
\begin{table*}[!t]
\begin{center}
\caption{Others}
\label{table:4}
{
\begin{tabularx}{\textwidth}{YYYYYYYY}
\toprule
Name & Anatomy & \makecell{Imaging \\ Modality} & \makecell{Number \\of Scans} & \makecell{Image \\ Dimension(h*w)} & Year Published & \makecell{Publicly \\Available} & Link\\
\midrule
IRCAD & liver & CT & 20 & 512*512 & 2009 & Yes & \href{https://www.ircad.fr/research/data-sets/liver-segmentation-3d-ircadb-01/}{Website}\\
VASCUSYNTH & liver & Algorithm synthesized & 10 & 101*101 & 2010 & Yes & \href{https://vascusynth.cs.sfu.ca/Data.html}{Website} \href{https://www.sciencedirect.com/science/article/pii/S0895611110000534\#sec14}{Publication} \\
OSMSC & Coronary and lung & MRA and CT & 93 & N/A & 2013 & Yes &
\href{https://www.vascularmodel.com/dataset.html}{Website} \href{https://www.ncbi.nlm.nih.gov/pmc/articles/PMC4023857/}{Publication}\\
ROTTERDAM & Coronary & CTA & 32 & N/A & 2009 & No & \href{https://www.sciencedirect.com/science/article/pii/S1361841509000474}{Publication}\\
CASDQEF & Coronary & CTA & 48 & N/A & 2013 & No &  \href{https://pubmed.ncbi.nlm.nih.gov/23837963/}{Publication}\\
VESSEL12 & Lung & CT & 20 & 512*512 & 2012 & Yes & \href{https://vessel12.grand-challenge.org/}{Website} \href{https://doi.org/10.1016/j.media.2014.07.003}{Publication}\\
MSD & Liver & CT & 443 & 512*512 & 2021 & Yes & \href{http://medicaldecathlon.com/}{Website} \href{https://arxiv.org/abs/2106.05735}{Publication}\\
MIDAS21 & Brain & TOF-MRA & 109 & 448*448 & 2005 & Yes & \href{https://www.insight-journal.org/midas/community/view/21}{Website} \href{https://www.academicradiology.org/article/S1076-6332(05)00564-7/fulltext}{Publication}\\
\bottomrule
\end{tabularx}
}
\end{center}
\end{table*}
\section{Evaluation Metrics}

Since the metric is an essential basis for model evaluation and selection, it is meaningful and necessary to investigate the appropriate metric. In general, pixels are classified as foreground pixels (positive) or background pixels (negative). On this basis, there are four basic pixel measures compared with GS segmentation, namely TP (true positive), FP (false positive), FN (false negative), and TN (true negative). Some general metrics to evaluate the performance of segmentation networks are presented in Table~\ref{tab1}.

\begin{table*}[!t]
\renewcommand{\arraystretch}{1.6}
\centering
\caption{Evaluation metrics}\label{tab1}
\begin{tabularx}{\textwidth}{ll}
\hline
Metric ~~~~~~~~~~~~~~~~~~~~~~~~~~~~~~~~~~~~~~~~~~~~~~~~~~ & Definition\\
\hline
Accuracy (Acc) & $\frac{TP+TN}{TP+TN+FP+FN}$ \\
Sensitivity (Se) & $\frac{TP}{TP+FN}$ \\
Specificity (Sp) & $\frac{TN}{TN+FP}$ \\
Precision (Pre) & $\frac{TP}{TP+FP}$ \\
False positive rate (FPR) & $\frac{FP}{TP+FP}$ \\
True negative rate (TNR) & $\frac{TN}{TN+FN}$ \\
Dice similarity coefficient (DSC) & $\frac{2TP}{TN+FN+2TP}$ \\
Jaccard index (IoU) & $\frac{TP}{TN+FN+TP}$ \\
Volumetric overlap error (VOE) & $1-\frac{TP}{TN+FN+TP}$ \\
Geometric mean (G) & $\sqrt{Se\times Sp}$ \\
Matthews correlation coefficient (MCC) & $\frac{(TP\cdot TN)-(FP\cdot TN)}{\sqrt{(TP+FP)(TP+FN)(TN+FP)(TN+FN)}}$ \\
Generation count (GC) & Number of correctly detected generations \\
Branch count (BC) & Number of correctly detected branches excluding the trachea \\
Branches detected (BD) & Percentage of correctly detected branches \\
Named Branch Count (NBC) & Number of correctly detected named branches\\
Tree length (TL) & Length of correctly detected branches excluding the trachea \\
Tree length detected (TLD) & Percentage of correctly detected tree length \\
Leakage count (LC) & Number of disconnected leakage \\
Leakage volume (LV) & Volume of leakage \\
Centreline overlap (CO) & Percentage of centreline overlap \\
\hline
\end{tabularx}
\end{table*}

In addition, the Hausdorff distance (HD), across the result and the GT, measures the maximum distance between two label boundaries. The absolute volumetric difference (AVD) represents the percentage ratio of the absolute difference between the GT volume and the segmented volume, to the GT volume. 

The receiver operating characteristic (ROC) curve is a graph that illustrates the trade-off between the Pre and FPR of the model at different thresholds. The closer the curve is to the upper left corner, the better the performance of the system. The most common performance metric extracted from the ROC curve is the area under the curve (AUROC), which is used to compare different models at the same threshold, with an AUROC of 1 for an optimal system. Similarly, the precision-recall (PR) curve demonstrates the trade-off between the Pre and Se, where the area under the curve (AUPR) is also considered a performance metric.

\section{Discussion}
\subsection{\textcolor{Black}{Research Trends}}
\begin{figure}[!t]
\centering
\includegraphics[width=3.4in]{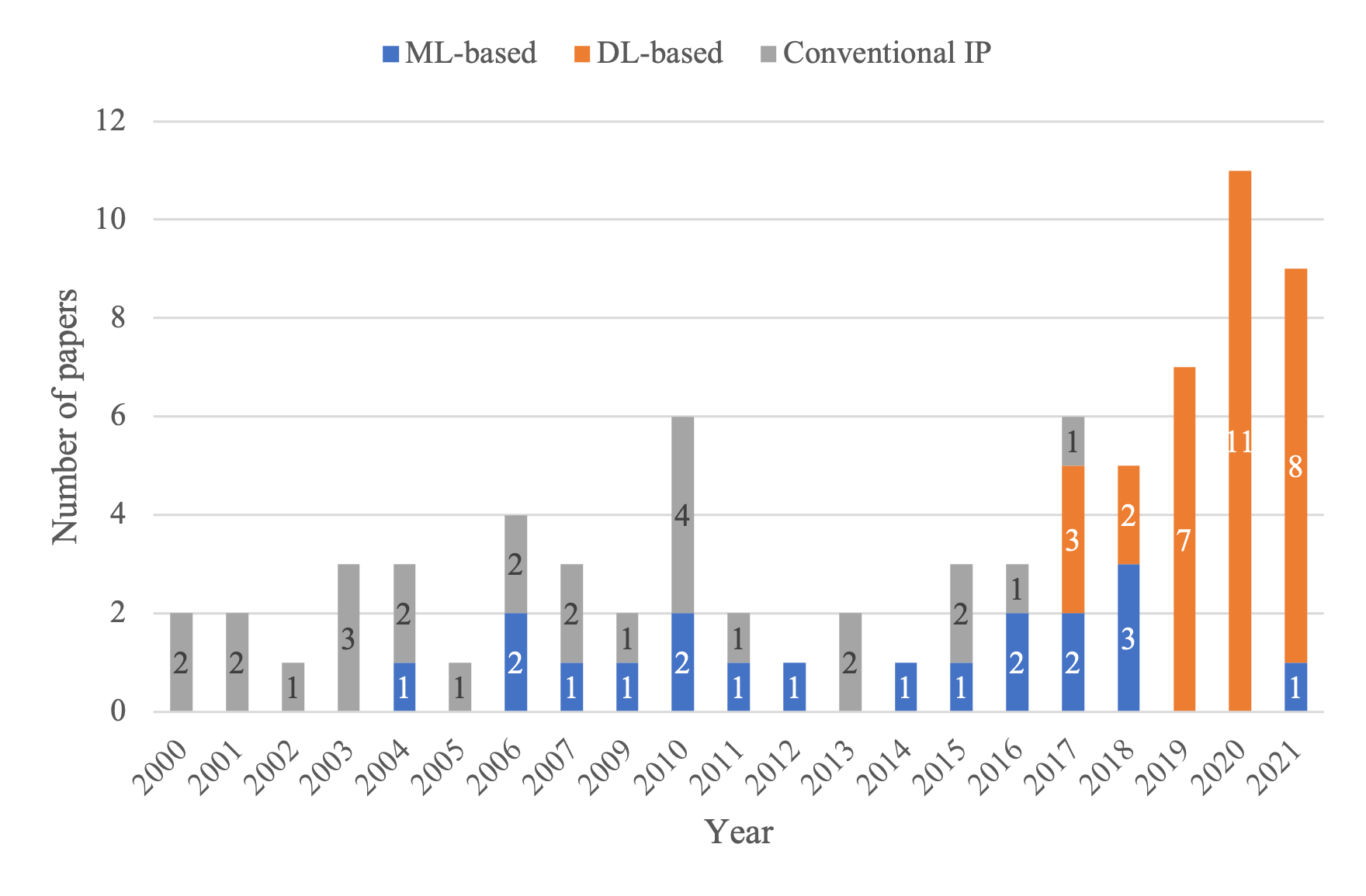}
\caption{\textcolor{Black}{Number of publications and years in terms of segmentation methods.}}
\label{yrs}
\end{figure}

The segmentation of human treelike tubular structures has been a heavily researched area in recent years. As shown in Figure~\ref{yrs}, the number of publications shows an overall upward trend from 2000 to 2021. However, the recent development of interest in this research area is mainly attributed to the emergence of DL algorithms. According to the publication of CNNs and the UNet that specifically for medical image segmentation \cite{ronnebergerUNetConvolutionalNetworks2015}, the number of DL-based methods for human treelike tubular structure segmentation started to increase remarkably in 2017 and became dominant in the last 3 years. In contrast, the number of new conventional methods has been decreasing since the first decade and became zero in 2018. Although new ML-based methods have been proposed since 2004, they have hardly dominated due to the need to extract features manually and the higher computational cost than conventional methods. Therefore, they were rapidly replaced by DL-based methods after ANNs were proposed and high-performance GPUs became available. 

According to the trend analysis, the utilization of the DL algorithm for human treelike tubular structure segmentation is the main research direction recently and in the future. Compared with the conventional method and ML-based method, the DL-based method has the following advantages. First, conventional and unsupervised ML-based methods may suffer from performance shortcomings because they do not benefit from the manually labelled ground truth. Moreover, traditional methods require special templates or rules designed for segmentation targets, and ML algorithms need to extract meaningful features from images, both requiring deep domain expertise. DL algorithms, on the other hand, can directly extract suitable internal representations of images for automatic learning. Meanwhile, manual feature selection is application-specific and thus lacks the ability to generalize and learn new features. DL algorithms, on the contrary, can automatically learn features at various levels without application-specific restrictions, and hence have higher generalizability and robustness.

\subsection{\textcolor{Black}{Segmentation Methods}}
\begin{figure}[!t]
\centering
\includegraphics[width=3.4in]{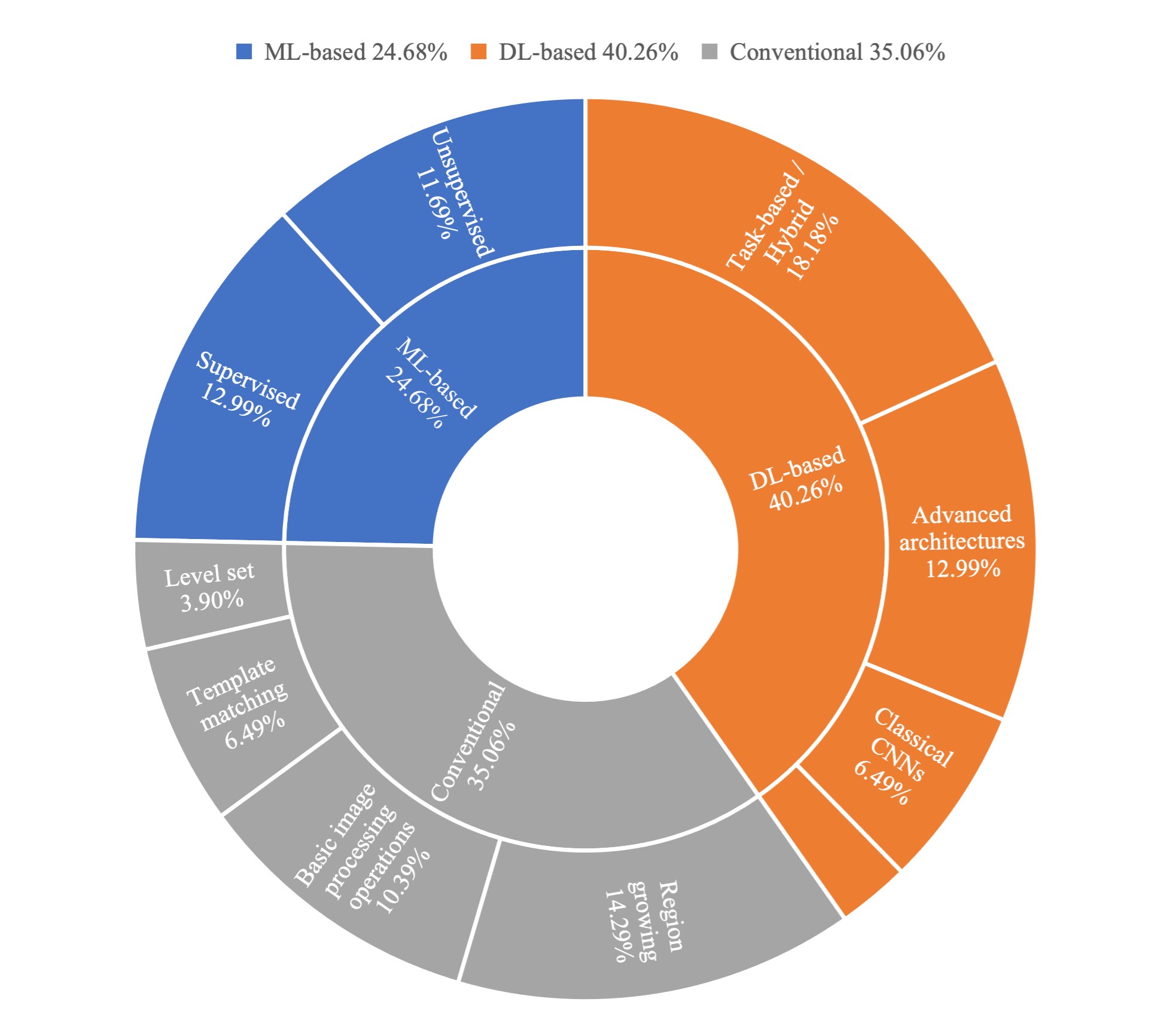}
\caption{\textcolor{Black}{Percentage of publications in terms of segmentation methods.}}
\label{methods}
\end{figure}

From the perspective of the method classification, \textcolor{Black}{40.26\% (31) of the 77 papers} reviewed combined DL algorithms, conventional methods accounted for \textcolor{Black}{35.06\%}, while ML-based methods had the lowest percentage. A more detailed breakdown of the percentage of method categories is shown in Figure~\ref{methods}. For the conventional methods, the RG method was the most widely used, probably due to its efficiency in 3D tree segmentation. Supervised and unsupervised ML algorithms were equally studied for human treelike tubular structure segmentation.

\textcolor{Black}{As for DL-based segmentation methods, task-based/hybrid algorithms were the most dominant, followed by architecture-based improvements. These methods had made targeted modifications to the segmentation task or model architecture for tree-like tubular structures. For example, some methods performed multi-task segmentation of thick and thin tubular structures or introduced GNN for tree-like connectivity. Classical CNNs were not popular because they are more generic and therefore did not achieve optimal segmentation performance on tree-like structures. Practically, for most task-based/hybrid segmentation, advanced architecture was a must.} Therefore, a large portion of task-based methods also developed novel DL models at the same time, such as combining GNN in order to transform segmentation into graph refinement \cite{selvanGraphRefinementBased2020}. In addition, the percentage of DL-based methods based on training strategies was only \textcolor{Black}{2.60\%}. It is also worth noting that all DL-based methods currently applied to human treelike tubular structure segmentation were supervised learning, in contrast to ML-based methods.

\subsection{\textcolor{Black}{Evaluation Metrics and Quantitative Analysis}}
\begin{figure}[!t]
\centering
\includegraphics[width=3.2in]{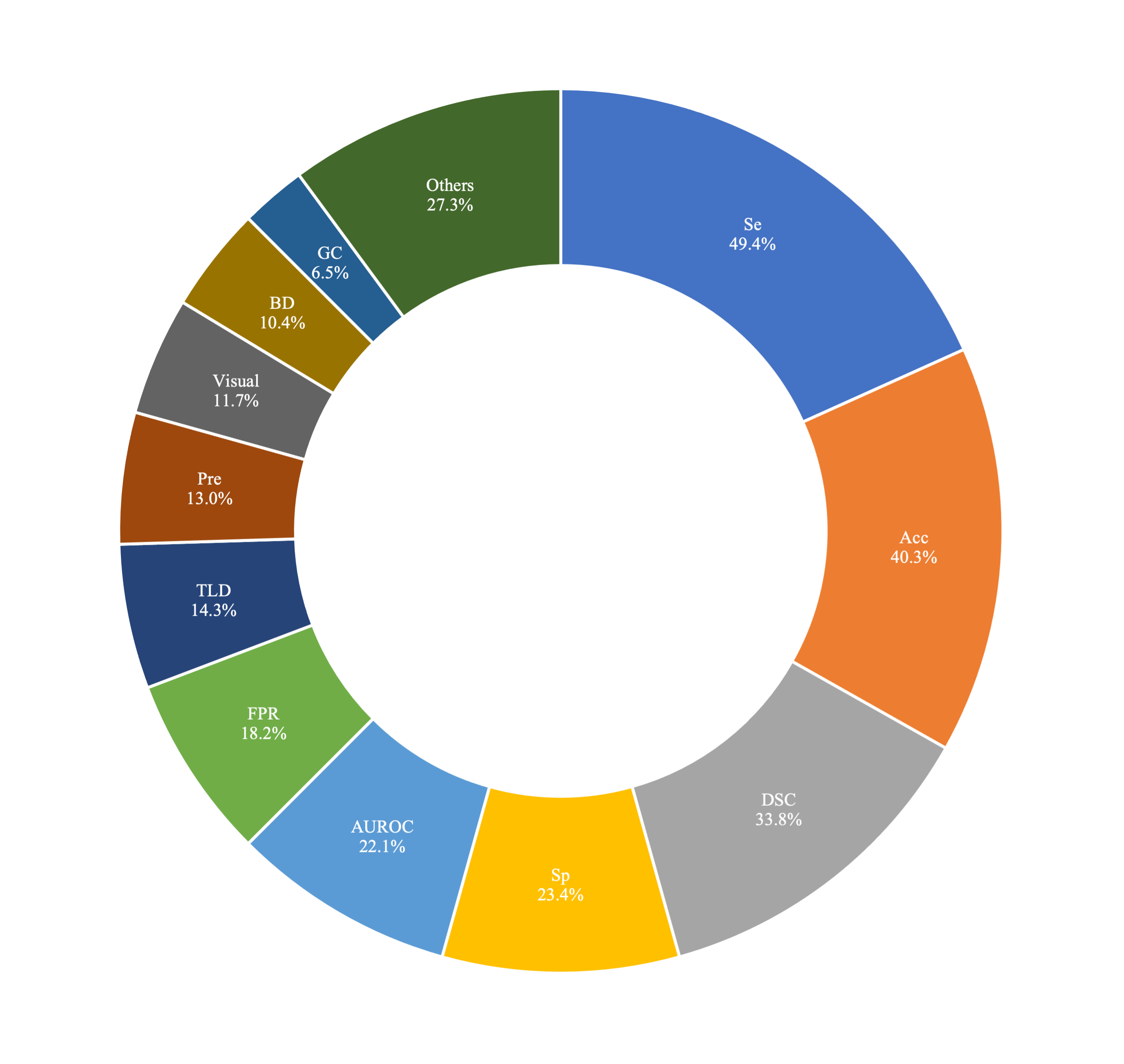}
\caption{\textcolor{Black}{Percentages of different evaluation metrics used in publications.}}
\label{metrics}
\end{figure}

\textcolor{Black}{Se is the most common metric to evaluate the human treelike tubular structure segmentation, which is adopted by nearly half of the reviewed papers, as shown in Figure~\ref{metrics}. By comparing with Acc, the second frequently used metric, Se can help the model avoid classifying all candidates as foreground or background. DSC is also a common choice that reflects the similarity of segmentation results and GT. In addition, Sp and FPR are often used to reflect the segmentation leakage, while AUROC depicts the trade-off between Se and FPR. As for airway segmentation, TLD and BD are the most commonly applied metrics.}

\textcolor{Black}{Quantitative analysis of segmentation performance for all reviewed methods is limited as they were evaluated on different datasets with various modalities and protocols. Therefore, the quantitative analysis was conducted representatively on two public datasets adopted by most methods, DRIVE and EXACT'09, for blood vessel segmentation and airway segmentation, respectively.}

\subsubsection{\textcolor{Black}{Blood vessel segmentation}}

\begin{figure*}[!t]
\centering
\begin{subfigure}{0.3\textwidth}
    \centering
    \includegraphics[width=\textwidth]{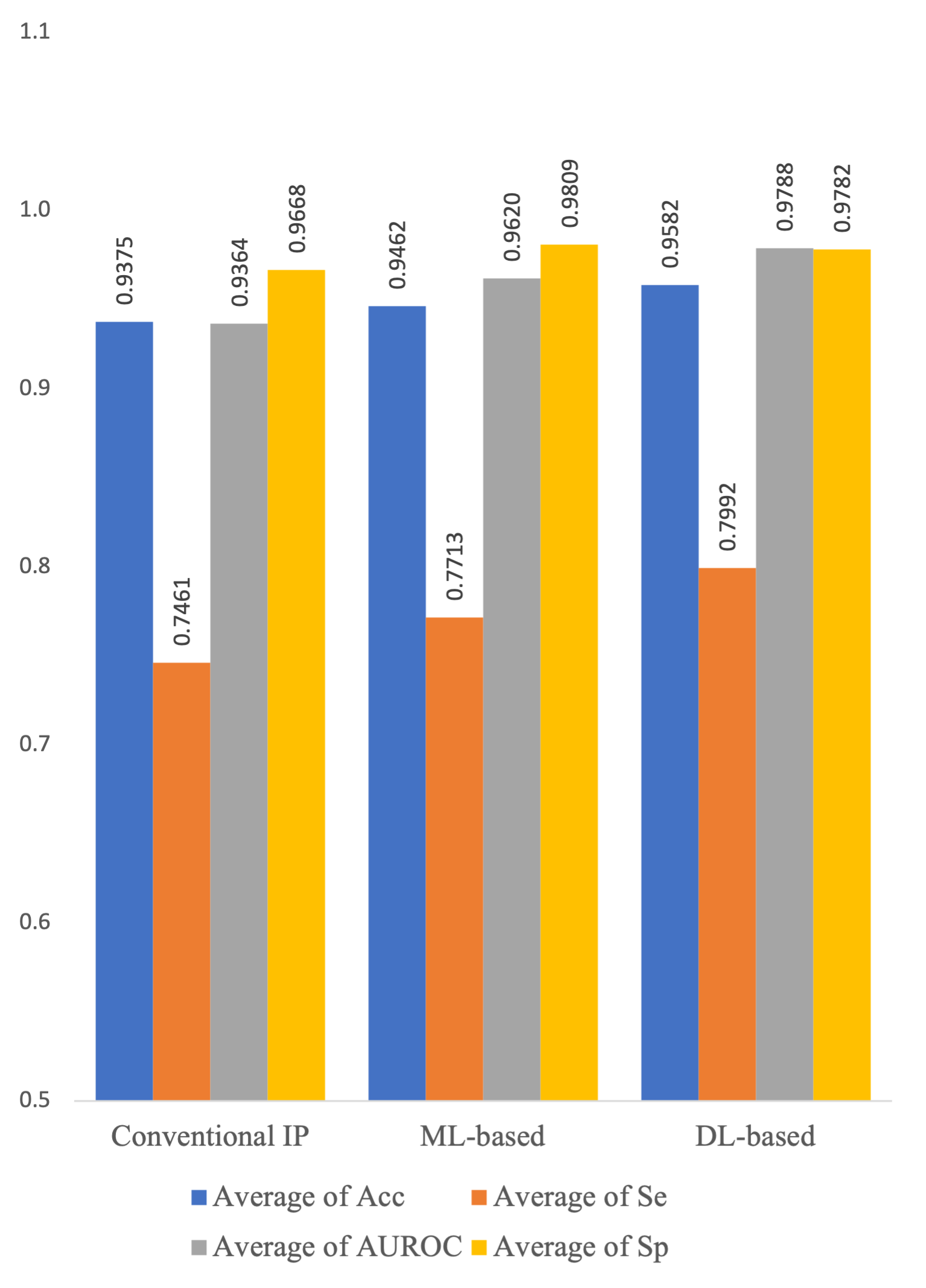}
    \caption{\textcolor{Black}{At category level.}}
    \label{DRIVE1}
\end{subfigure}
\hfill
\begin{subfigure}{0.69\textwidth}
    \centering
    \includegraphics[width=\textwidth]{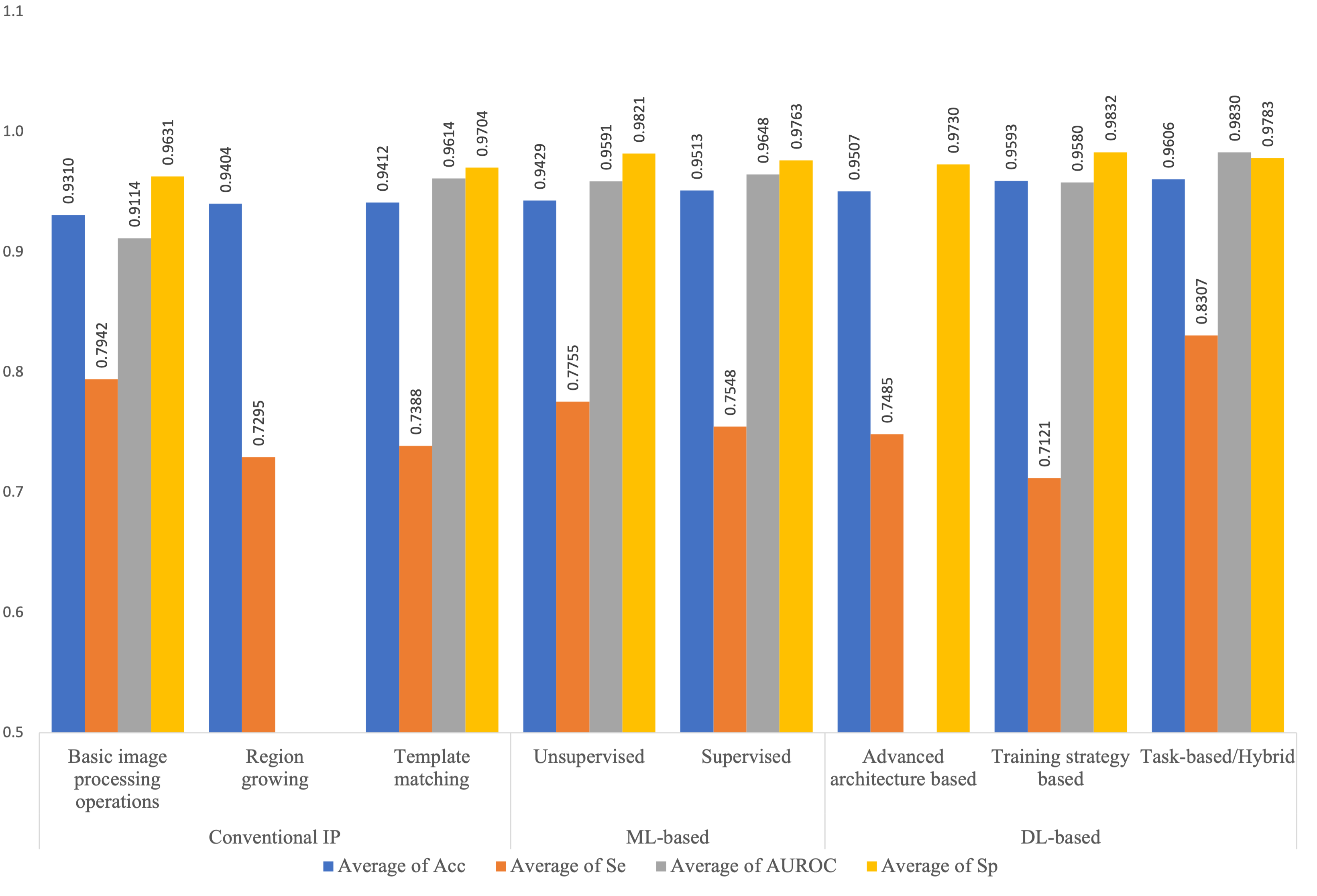}
    \caption{\textcolor{Black}{At subcategory level.}}
    \label{DRIVE2}
\end{subfigure}
\hfill
\begin{subfigure}{\textwidth}
    \centering
    \includegraphics[width=\textwidth]{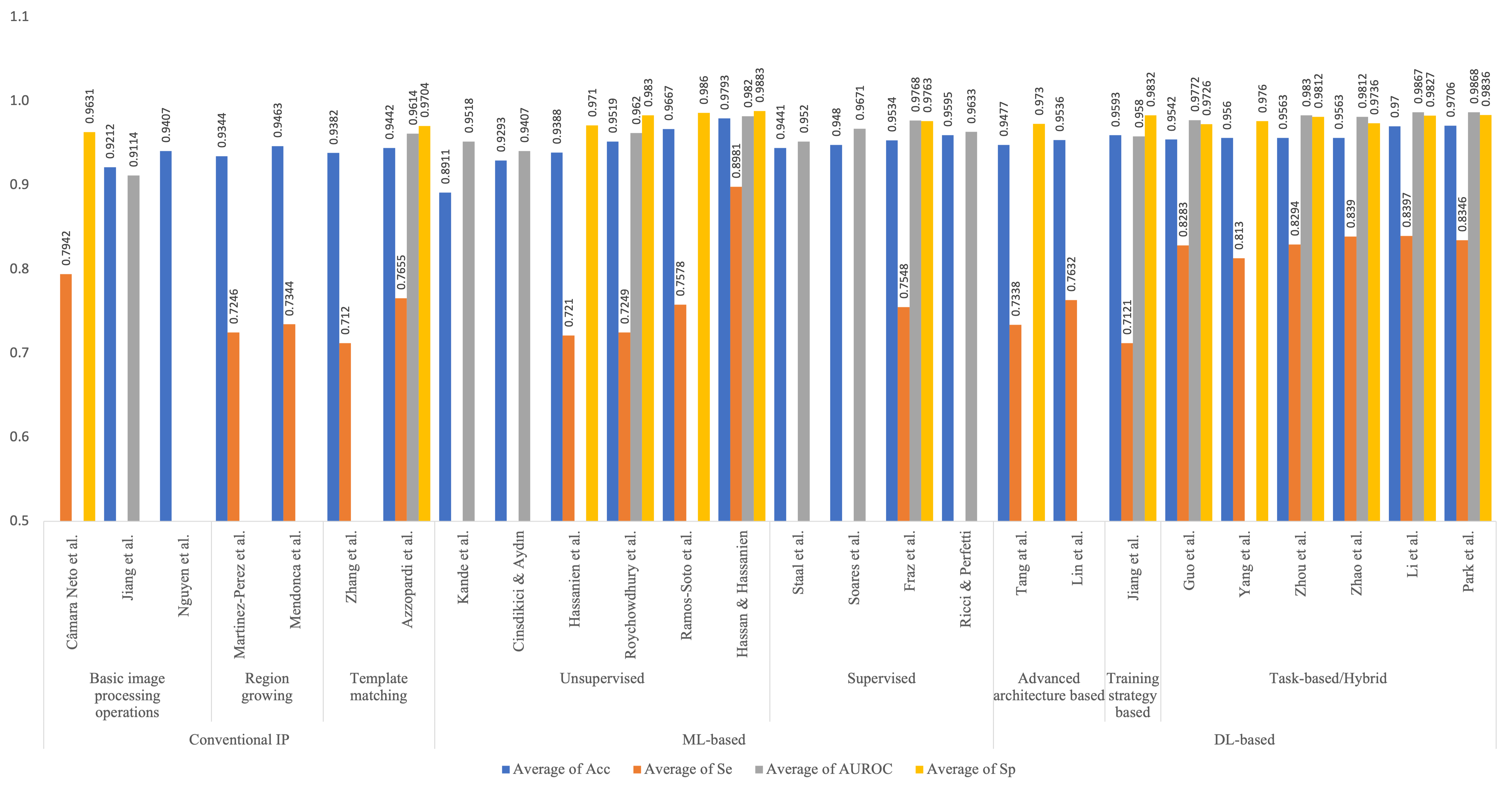}
    \caption{\textcolor{Black}{At method level.}}
    \label{DRIVE3}
\end{subfigure}
\hfill
\caption{\textcolor{Black}{Quantitative segmentation metrics comparison of methods evaluated on DRIVE dataset.}}
\label{DRIVE}
\end{figure*}

\textcolor{Black}{27 of the 77 reviewed papers validated their performance for segmenting blood vessels on the DRIVE dataset. We quantitatively compared means of the four most used metrics (Acc, Se, AUROC, and Sp), but excluded a method using Visual evaluation \cite{zanaSegmentationVessellikePatterns2001}, as shown in Figure~\ref{DRIVE}. It's worth noting that not every method used all four metrics, so our mean comparisons were only based on methods that used a particular metric.}

\textcolor{Black}{Overall, DL-based methods have the best segmentation performance, followed by ML-based methods, and finally Conventional methods. For the Conventional method, basic image processing operations are most sensitive to vascular structure, but template matching has higher segmentation accuracy and specificity with fewer false positives according to AUROC. RG has moderate segmentation accuracy, but the sensitivity is not satisfactory. Similarly, unsupervised ML-based methods are more sensitive than supervised methods but slightly inferior in other segmentation metrics. On the other hand, task-based/hybrid DL-based methods achieve the best results on all metrics. DL-based methods based solely on advanced architecture or training strategies are lower in segmentation sensitivity even than Conventional methods, which is an obvious problem. Specifically, as shown in Figure~\ref{DRIVE3}, all task-based/hybrid DL-based methods can achieve good segmentation performance. However, among all unsupervised ML-based methods, Hassan \& Hassanien's WOA method performs exceptionally well, reaching the state-of-the-art.}

\subsubsection{\textcolor{Black}{Airway segmentation}}
\begin{figure}[!t]
\centering
\includegraphics[width=0.5\textwidth]{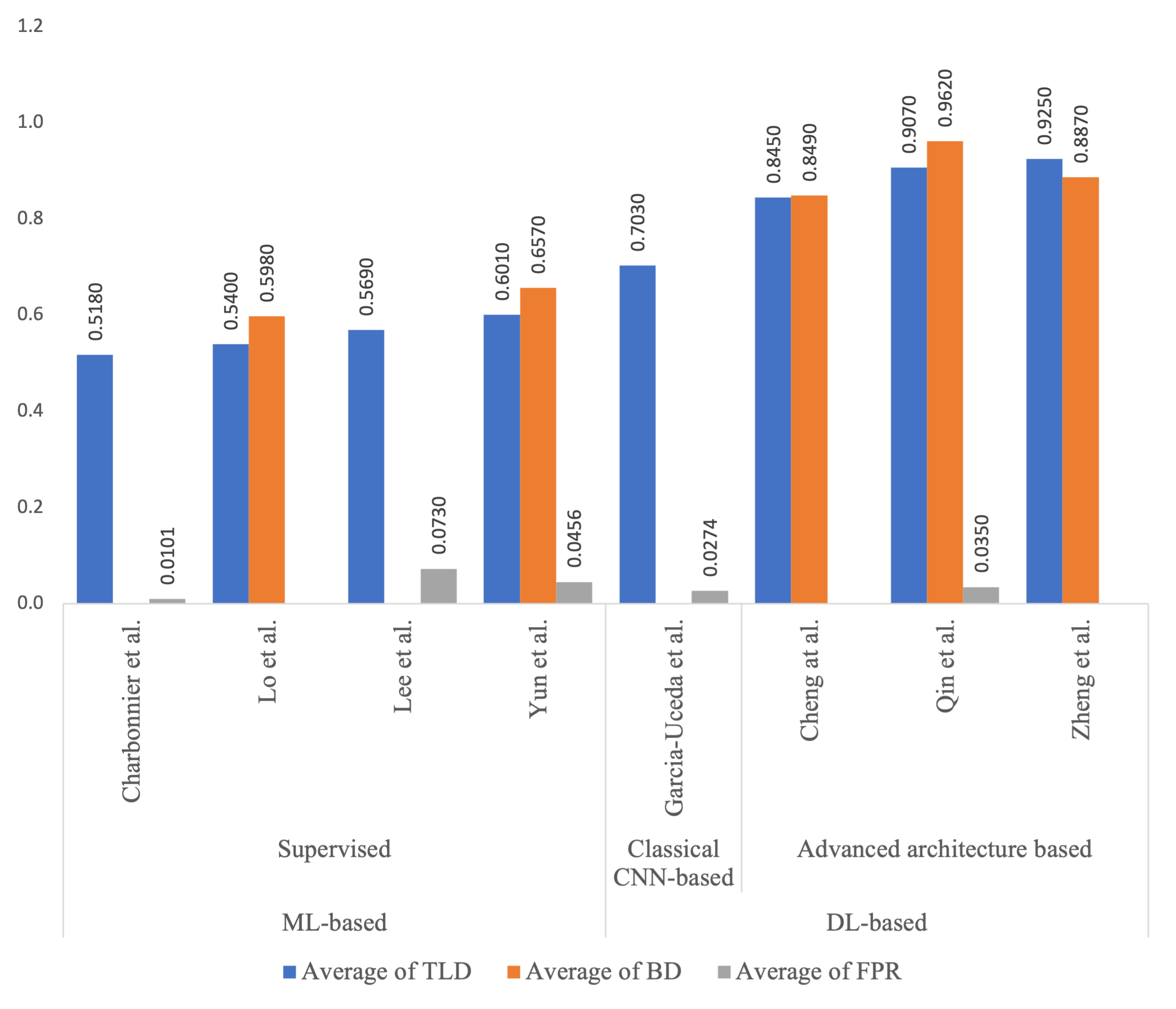}
\caption{\textcolor{Black}{Quantitative segmentation metrics comparison of methods evaluated on EXACT'09 dataset.}}
\label{EXACT09}
\end{figure}

\textcolor{Black}{For airway segmentation, a total of 10 papers were evaluated in EXACT'09. After excluding one paper using duplicate methods \cite{qinLearningTubuleSensitiveCNNs2021} and one paper doing comparative analysis \cite{jin3DConvolutionalNeural2017}, the remaining 8 methods only covered part of the subcategories of ML-based and DL-based segmentation algorithms, as shown in Figure~\ref{EXACT09}. We used mean values of the three most commonly applied metrics for quantitative comparison (TLD, BD, and FPR). Similarly, not every method used all three metrics.}

\textcolor{Black}{Overall, the segmentation performance of DL-based methods is significantly higher than that of ML-based methods. Supervised ML-based methods are limited by the difficulty of finding suitable features and rules to maximize airway detection rate while minimizing segmentation leakage. Clearly, the method of Charbonnier et al. \cite{charbonnierImprovingAirwaySegmentation2017} has the lowest detection rate but the least leakage, while the method of Yun et al. \cite{yunImprovementFullyAutomated2018} has a higher detection rate but also increased FPR. On the other hand, the advent of CNN enabled more suitable CT image features to be extracted, thus achieving a better balance between detection rate and FPR. We can see that by improving the classic CNN architecture, the detection rate of tracheal segmentation is further improved by more than 14\% while maintaining a low FPR.}

\subsection{\textcolor{Black}{Open Challenges}}
\textcolor{Black}{Conventional segmentation methods suffer from incomplete prediction and leakage (shown in Fig. \ref{fig:airway_leakage}). EXACT'09 challenge has summarized performance of them on airway segmentation. On average, no traditional method achieves over 74\% in TLR.
Despite the recent sprout of deep learning based methods, the problem of missing thin terminal bronchioles in the final prediction still persists as shown in Fig. \ref{fig:airway_break} and Fig. \ref{fig:retinal_break}. Commonly used cost functions such as Binary cross entropy loss and Dice loss failed to address this problem. Results with high Dice coefficient can still have many disconnected peripheral branches which can be reflected by the low tree length detected ratio and branch detected ratio (often less than 70\%). This problem is more predominant in the 3D cases than the 2D cases. 
In addition, model performance on pathological cases should be addressed in the future. morphological changes in structure can pose a further challenge to the robustness of the model. 
\begin{figure}[!t]
    \centering
    \includegraphics[width=0.4\textwidth]{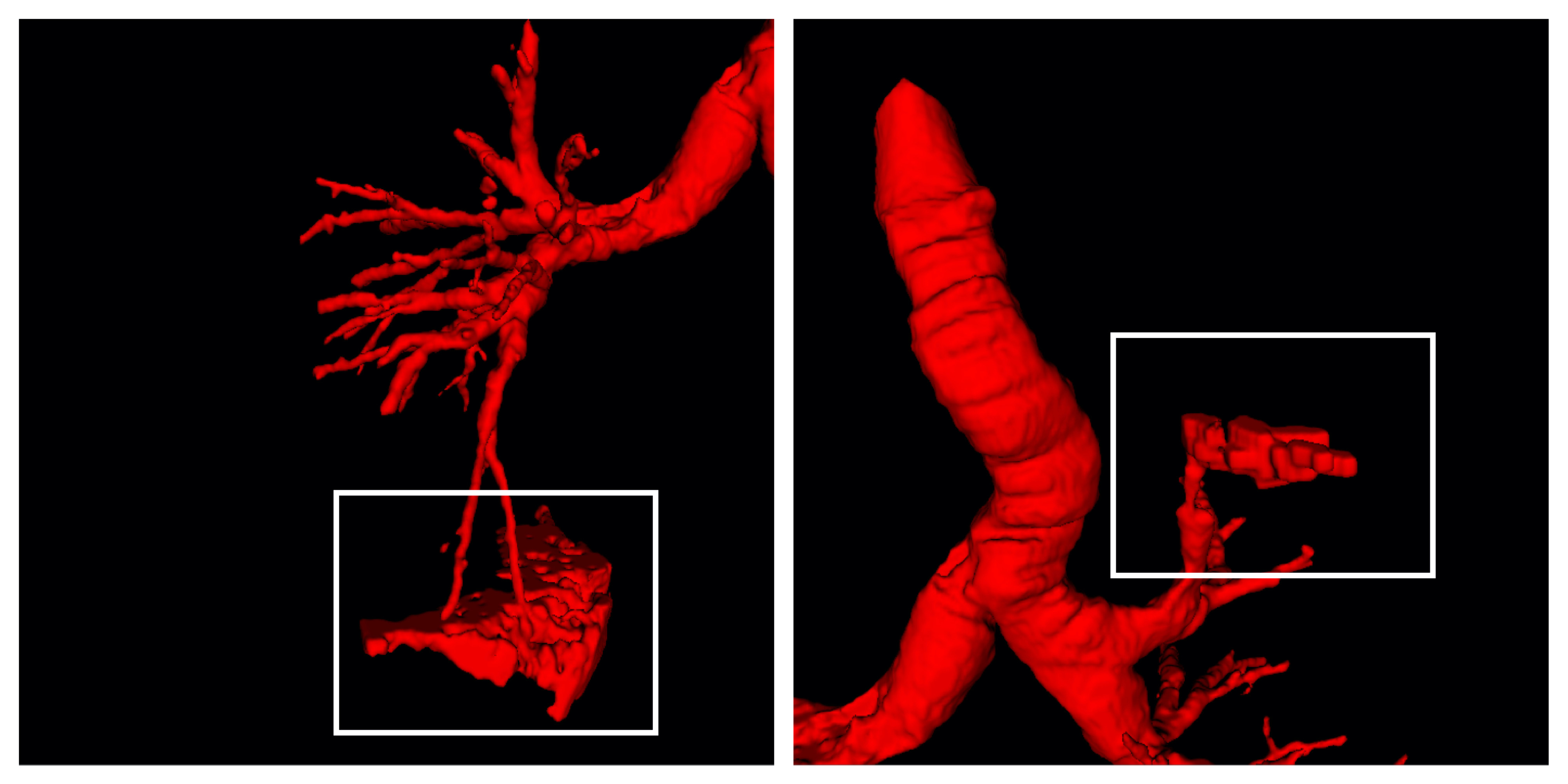}
    \caption{Visual Illustration of leakage at terminal bronchioles from CASE14 from EXACT'09;}
    \label{fig:airway_leakage}
\end{figure}
\begin{figure}[!t]
    \centering
    \includegraphics[width=0.4\textwidth]{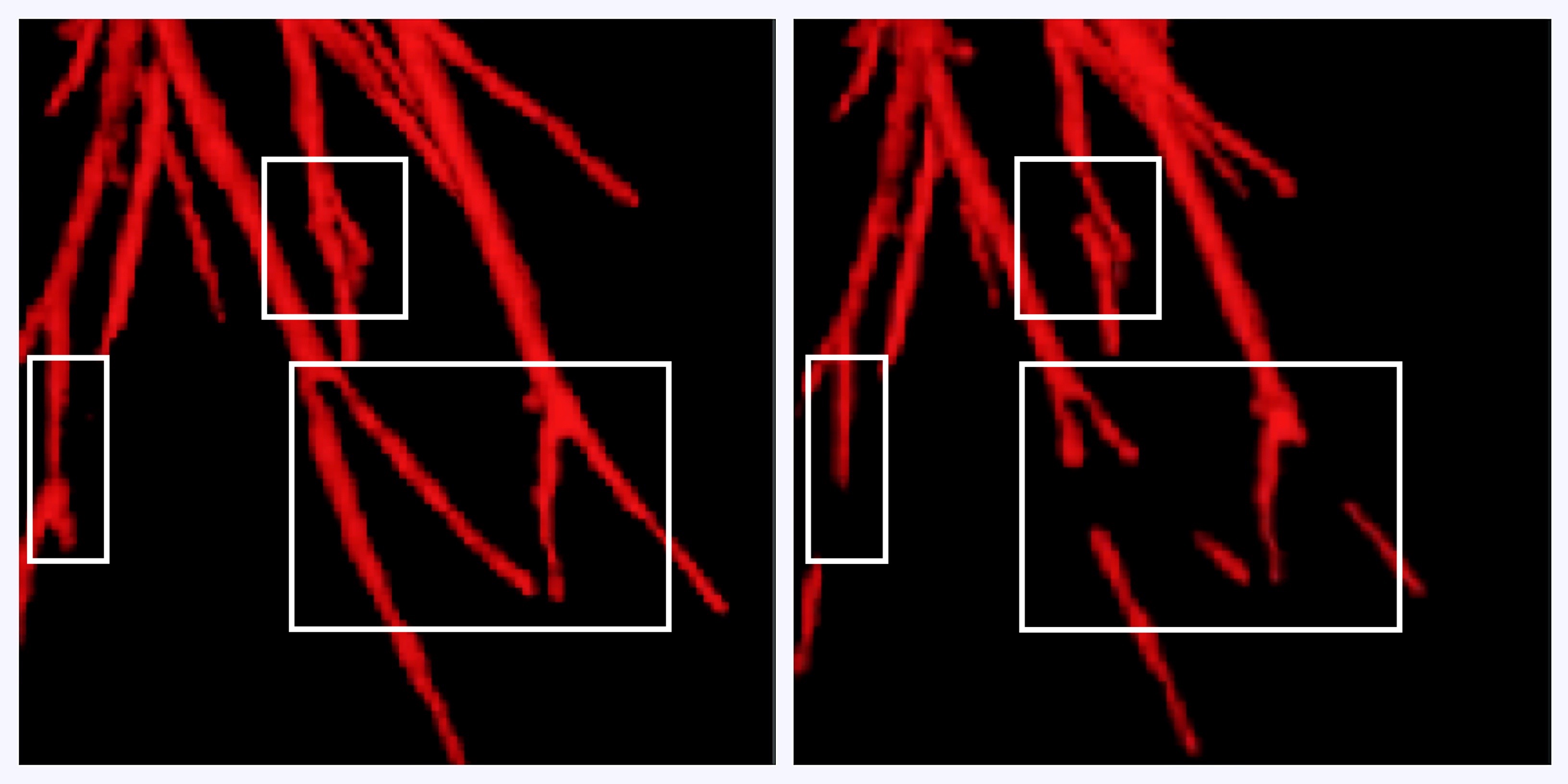}
    \caption{Visual Illustration of discontinuity at terminal bronchioles from CASE2 from EXACT'09; On the left is the ground truth and on the right is the prediction}
    \label{fig:airway_break}
\end{figure}
\begin{figure}[!t]
    \centering
    \includegraphics[width=0.4\textwidth]{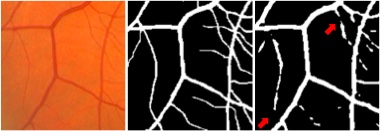}
    \caption{Visual Illustration of discontinuity of retinal blood vessels in DRIVE dataset from \cite{Shit2020}; From left to right is the real image, the ground truth and the segmentation result from U-Net trained by Dice loss}
    \label{fig:retinal_break}
\end{figure}
}

\begin{figure*}[t!]
    \centering
    \includegraphics[width = 1.0\textwidth]{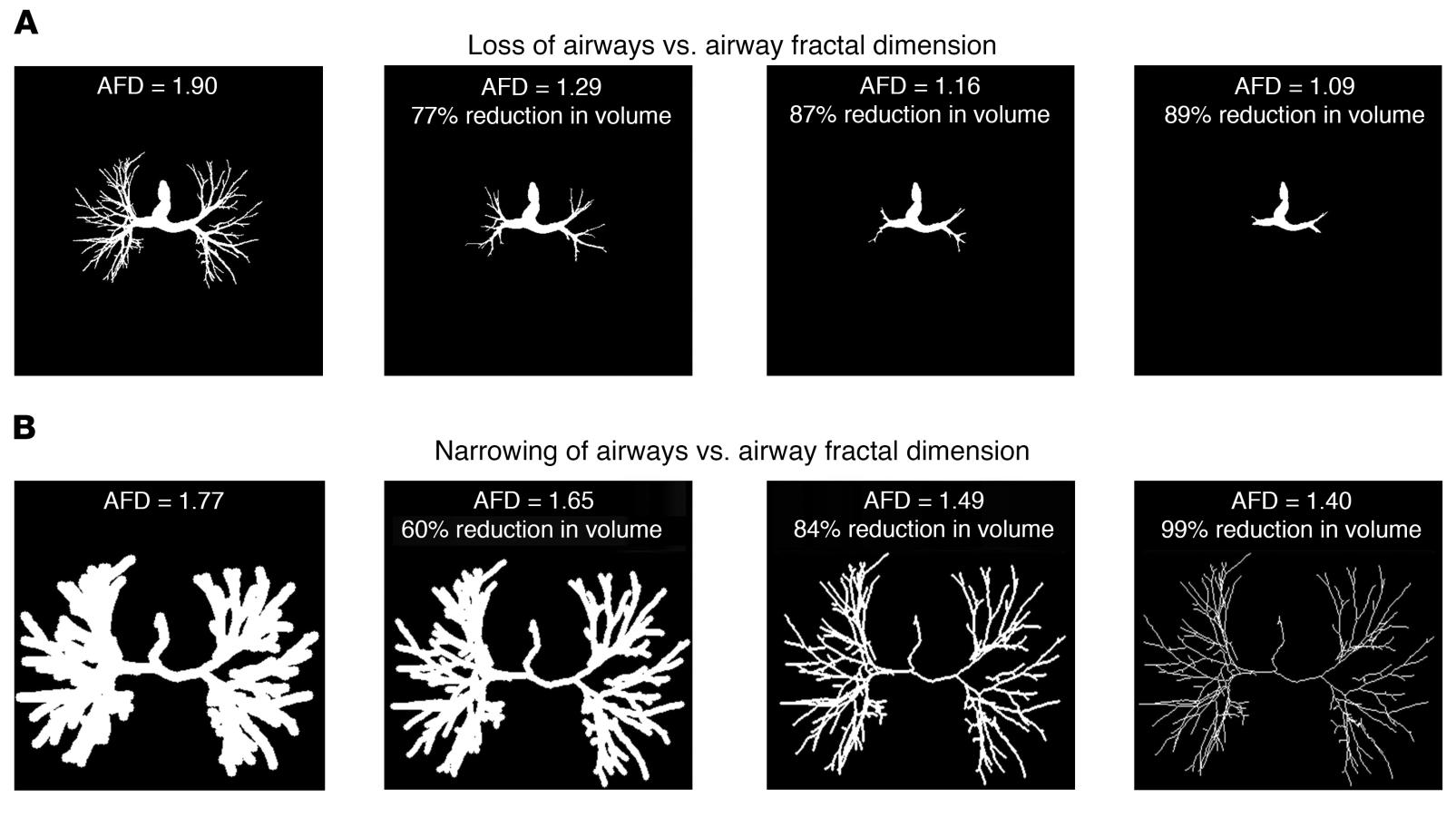}
    \caption{Visual Illustration of AFD from \cite{Bodduluri2018}}
    \label{fig:fractal}
\end{figure*}

Another problem is the lack of metrics to quantify the morphological shape of the airway which often contains useful information for lung-related disease diagnosis and prognosis. \cite{Orlandi2005,Achenbach2008,deJong2005,Hasegawa2006} measure the airway thickness and show a strong correlation between it and reduced air flow of the lung. Apart from the airway wall thickness, very few studies focused on other morphological parameters such as branching patterns. This is partly because there is no mathematical way to quantify the spatial patterns of the human airway. \cite{Bodduluri2018} proposed a new metric named Airway Fractal Dimension(AFD) to help predict respiratory morbidity and mortality in COPD. AFD is calculated using the Minkowski-Bougliand box-counting dimension which essentially measures the self-similarity of a graph. In the case of airway structures, the more branch it has the more detailed it is considered, A formal mathematical definition of the box-counting dimension is shown below:
\begin{equation*}
    D = \lim_{\epsilon \to \infty}\frac{\log N(\epsilon)}{-\log \epsilon }
\end{equation*}
wehre $D$ is the fractal dimension, $N(\epsilon)$ is the number of boxes required to cover the complete fractal structure at a given scale $\epsilon$.
They analysed 8,135 participants enrolled in the COPDGene cohort and found a strong correlation ($p<0.001$) between AFD and forced expiratory volume. They also did a multivariate analysis and confirmed that AFD can be used in addition to airway thickness to give better prognosis. An intuitive visual representation is also provided by them as shown in Fig. \ref{fig:fractal}.
A review on computing the fractal dimension in medical imaging has been done by Konatar I. et al. \cite{Konatar2020}

There hasn't been any release of new datasets with airway annotation in the community for years. 
Only 90 CT scans (20 from EXACT'09 and 70 from LIDC-IDRI) are widely publicly available, and they come from patients with a wide range of physical conditions. 

Most proposed models cannot extend to more complex clinical settings because the algorithms would fail to predict airway regions when encountering pathological cases such as fibrosis and COVID-19. Moreover, the variations between CT scans acquired by different protocols also pose a challenge to the generalization of the model.

To summarize, there are still some open challenges for accurate and robust human treelike tubular structure segmentation, which are as follows:

\begin{itemize}
    
    \item DL-based approaches through unsupervised or semi-supervised learning is a topic that has not been thoroughly investigated yet. Unsupervised training can overcome the current dilemma of lacking properly annotated datasets by domain experts.
    
    \item In terms of training strategies, the DL algorithm may have great room for improvement. For instance, applying hard sampling mining to improve the segmentation performance in pathological cases.
    
    \item There is a lack of a standard set of evaluation metrics and a systematic evaluation process for human tree tubular structure segmentation methods. The inconsistent metrics and the fact that most of the datasets are not publicly available make the performance comparison between algorithms very difficult. As mentioned above, using sole metrics such as accuracy, IoU and Dice cannot fully access the model on a spatial scale. Therefore, an open evaluation scheme across datasets is urgently needed.
    

    \item Binary cross entropy loss and Dice loss are not able to capture the discontinuity in peripheral bronchioles. New loss functions, such as clDice \cite{Shit2020}, that emphasize connectivity should also be incorporated in training the model.
    
    \item \textcolor{Black}{For conventional image processing methods, their results are fully explainable. However, deep neural networks are normally considered as 'black boxes', whereas their interpret ability for tubular structure segmentation remains to be explored. Researchers should focus more on explainable AI solutions \cite{yang2022unbox}. This could be achieved by sensitivity analysis \cite{adebayo2018sanity}, gradient-based methods \cite{doshi2018considerations}, and data harmonisation strategies for multi-centre studies} \cite{nan2022data}.
\end{itemize}

\section{Conclusion}
This review summarizes the algorithms, datasets, and evaluation metrics for human treelike structure segmentation methods in the literature. We present a systematic classification of the different algorithms, supported by tables reporting the anatomical regions of interest, the datasets used, and performance metrics, which can help researchers to better understand the available options and methods. Based on the analysis of the literature, deep learning based segmentation methods have become dominant with their advantages of capturing hidden information in complex structures. In this regard, this review proposes feasible research directions based on deep-learning algorithms, evaluation metrics, and loss functions to accelerate the development and improvement of human treelike tubular structure segmentation methods.

\printcredits

\bibliographystyle{model1-num-names}

\bibliography{treelike_Vessels_Seg}

\bio{}
\endbio

\endbio

\end{document}